\documentclass[twoside,twocolumn,english,superscriptaddres,showpacs,longbibliography,aps,prb,preprintnumbers]{revtex4-2}

\usepackage[T1]{fontenc}
\usepackage[utf8]{inputenc}
\usepackage[english]{babel}
\usepackage{bm}
\usepackage{amsmath,physics}
\usepackage{amssymb}
\usepackage{graphicx}
\usepackage{blkarray}
\usepackage[unicode=true, pdfusetitle, bookmarks=true,bookmarksnumbered=true,bookmarksopen=true,bookmarksopenlevel=1,
breaklinks=false,pdfborder={0 0 0},pdfborderstyle={},backref=false,color links, citecolor=blue,linkcolor=red,urlcolor=blue] {hyperref}
\usepackage[nameinlink]{cleveref}
\Crefname{figure}{Fig.}{}

\usepackage{comment}
\usepackage{caption}
\usepackage{here}

\makeatletter
\def\set@firstnote#1{%
 \@ifnum{\firstnote@num=#1\relax}{}{%
  \class@warn@end{Endnote numbers changed: rerun LaTeX}%
 }%
 \immediate\write\@mainaux{%
   \global\mathchardef\string\firstnote@num#1\relax
 }%
}%

\usepackage{braket}
\usepackage{bm}
\usepackage{tikz}
\usepackage{pgfplots}
\usepackage{pgf}
\usepackage{subfigure}
\usepackage{nicematrix}
\usepackage{diagbox}
\usepackage{orcidlink}

\renewcommand{\selectlanguage}[1]{}

\makeatother

\pgfplotsset{compat=1.18}
\begin{document}

\title{Organizing transitions and their cascades: Generalized symmetry enforcement in massless flows or Higgs transitions}
\author{Yoshiki Fukusumi}
\affiliation{Physics Division, National Center for Theoretical Sciences, National Taiwan University, Taipei 106319, Taiwan}
\author{Yuma Furuta}
\affiliation{Institute for Advanced Study, Kyushu University, Fukuoka 819-0395, Japan}
\pacs{73.43.Lp, 71.10.Pm}
\preprint{KYUSHU-HET-367}
\date{\today}
\begin{abstract}
We study the role of generalized symmetry in massless renormalization group flows or Higgs transitions. In particular, we revisit the massless renormalization group flows in unitary minimal models, $\mathbf{M}(p,p+1) \rightarrow \overline{\mathbf{M}(p-1,p)}$ preserving the fusion ring symmetry $\text{FR} (SU(2)_{p-2})\subset \mathbf{M}(p,p+1)$ where $p$ is an integer satisfying $p>3$. In this series of flows, we demonstrate that the unbroken fusion ring symmetry $\text{FR} (SU(2)_{p-2})$ eliminates all relevant perturbations in the $\overline{\mathbf{M}(p-1,p)}$ model. Hence, the infrared theory $\overline{\mathbf{M}(p-1,p)}$ is stable at the level of the scaling analysis and can be interpreted as a (weak-)symmetry-enforced gapless phase in contemporary theoretical physics. Phenomenologically, by the folding trick, the unbroken fusion ring symmetry corresponds to a (half-)integer spin nonsimple current, a variant of the Cooper pair involving nonabelian anyons generated from the coset or level-rank duality structures. Moreover, we demonstrate that the structure of (half-)integer spin nonsimple current plays a fundamental role in causing the resonance effect of relevant and dangerously irrelevant perturbations. This resonance effect may result in the cascade of phase transitions (or the system flows to unexplored fixed points), and the symmetry can be a stopper of such unconventional flows.

\end{abstract}

\maketitle

\section{introduction}
\label{introduction}

\subsection{Background}

The generalized symmetry is a rapidly developing research field in contemporary physics and mathematics\cite{Cobanera:2009as,Cobanera:2011wn,Cobanera:2012dc,Gaiotto:2014kfa,McGreevy:2022oyu,Cordova:2022ruw,Bhardwaj:2023kri}. From a traditional viewpoint, it is a set of integrals of motion forming a ring satisying well-organized properties\cite{Petkova:2000ip}, a natural generalization of a group. An important point in this research direction is that one can assign its correspondence to defects\cite{Petkova:2000ip} or boundary conditions\cite{Cardy:1989ir,Behrend:1999bn} (with a close connection to anyons\cite{Leinaas:1977fm,Goldin:1979ki}), and the related points have been noticed in subfactor theory\cite{article,Bockenhauer:1999wt} (see a review \cite{Kawahigashi:2021hds}). However, it should be noted that the rigorous categorical formulation of this correspondence narrows the scope of the frameworks at the present stage, especially in studying hierarchical structure by massless renormalization group (RG) flows in conformal field theories (CFTs)\cite{Zamolodchikov:1987ti,Zamolodchikov:1987jf,Zamolodchikov:1989hfa} or related gapped domain walls in topological quantum field theories (TQFTs)\cite{Kitaev:2011dxc,Lan:2014uaa}. These classifications of the CFTs and TQFTs should be related consistently when the paradigm of the bulk-edge correspondence or CFT/TQFT correspondence holds\cite{Laughlin:1983fy,Witten:1988hf,Moore:1991ks}, but their connections contain still unclear points\cite{Hung:2015hfa,Wan:2016php,Zeev:2022cnv,Fukusumi_2022_c,Fukusumi:2024ejk,Antinucci:2025fjp}. To establish their connections, one needs to study $\mathbb{C}$-linear formulation of fusion rings and the corresponding category\cite{Kong:2013aya}\footnote{$\mathbb{C}$ is the field of complex number. In this manuscript, we also use $\mathbb{Z}$ as the set of integers.}. However, the exact construction of $\mathbb{C}$-linear tensor functors representing gapped domain walls is fundamental, but the corresponding studies have been  limitted\cite{Zhao:2023wtg,Fukusumi:2024ejk,Fukusumi:2025clr,Fukusumi:2025ljx,Fukusumi:2025xrj,Fukusumi:2025fir,Fukusumi:2025fvb}. We highlight a related reasearch direction which has a closer connection to the present manuscript, quantum field theoretic formulations of RG domain walls\cite{Brunner:2007ur,Gaiotto:2012np,Stanishkov:2016pvi,Stanishkov:2016rgv,Poghosyan:2022mfw,Poghosyan:2022ecv}, a class of conformal interfaces which are equivalent to massless RG flows.

More traditionally, the corresponding transitions can be regarded as a variant of the Higgs transitions\cite{Higgs:1964pj} involving the projection\cite{Brunner:2007ur,Klos:2019axh,Klos:2020upw} $\rho: \mathcal{H}_{\mathbf{A}}\rightarrow \mathcal{H}_{\mathbf{A'}}$, where $\mathcal{H}$ is the Hilbert space and $\mathbf{A}$ and $\mathbf{A'}$ are the fusion ring of the ultraviolet (UV) and infrared (IR) theories respectively\cite{Frohlich:2003hg,Frohlich:2003hm,Cordova:2025eim,Fukusumi:2025clr,Fukusumi:2025xrj}. By applying the folding trick\cite{Wong:1994np} to these models related by a projection, one can obtain nontrivial currents or symmetries of coupled models through ``condensation\cite{Bais:2008ni}" (or proliferation in a recent reference\cite{Cheng:2026qax}) from the unbroken symmetries $\mathbf{A}_{\text{ub}}\subset \mathbf{A}$ satisfying $\rho(\mathbf{A}_{\text{ub}})=\mathbf{A'}_{\text{ub}}$. The composite structure $\mathbf{A}_{\text{ub}}\boxtimes \overline{\mathbf{A'}_{\text{ub}}}$ as we introduce later has a close connection to the extension of the coupled model\cite{Fukusumi_2022_c,Fukusumi:2024ejk,Fukusumi:2025ljx}. By combining older terminologies in high energy physics theory, this structure, which provides extension of the coupled models, can be called (half-)integer spin nonsimple currents\cite{Vafa:1986wx,Dijkgraaf:1989hb,Schellekens:1989uf,Schellekens:1990ys,Gato-Rivera:1990lxi,Kreuzer:1993tf,Fuchs:2000gv} satisfying the (half-)integer spin constraint for the particles $s_{\alpha_{\text{ub}}\boxtimes \overline{\alpha'_{\text{ub}}}}=\mathbb{Z}/2$ where $\alpha_{\text{ub}}$ ($\alpha'_{\text{ub}}$) is the anyons in $\mathbf{A}_{\text{ub}}$ ($\mathbf{A'}_{\text{ub}}$). When restricting our attention to systems with (half-)integer spin simple currents, the construction of a generalized electron operator in this context can be seen widely in the studies of fractional quantum Hall states\cite{Cappelli:2010jv,Schoutens:2015uia,Hansson_2017}. They provide a building block of single-valued wavefunctions of topologically ordered states (or TQFTs) because of the (half-)integral spin conditions of the objects. More recently, it was called connected \'{e}tale algebra\cite{DavydovMugerNikshychOstrik+2013+135+177} or condensable algebra\cite{Kong:2013aya}, but to emphasize the role of extension (or proliferation) in the earlier references, we rather use the term (half-)integer spin nonsimple current. In condensed matter, one can see its analogy (or correspondence) to the composite particle to the BCS pair creations\cite{Bardeen:1957mv}, and related aspects have been studied in \cite{Bais:2008ni,PhysRevB.84.125434,Kong:2013aya,Kaidi:2021gbs,Huston:2022utd,Fukusumi:2022xxe,Fukusumi_2022_c,Zhang:2024bye,Kikuchi:2024ibt,Fukusumi:2024ejk,Kikuchi:2024cjd,Fukusumi:2025ljx}, for example. More mathematically, the variant of the Witt equivalence\cite{davydov2011structure,DavydovMugerNikshychOstrik+2013+135+177} is the key structure in studying such phenomena, and this equivalence naturally selects the fundamental parts of a theory, constructing the well-defined wavefunctions of TQFTs or correlation functions in CFTs. This selection of organized parts of a theory can be interpreted as an algebraic analog of the conformal bootstrap\cite{Polyakov:1974gs}.

By unfolding the coupled model $\mathbf{A}\boxtimes \overline{\mathbf{A'}}$ to $\rho:\mathcal{H}_{\mathbf{A}}\rightarrow \mathcal{H}_{\mathbf{A'}}$, the (half-)integer spin nonsimple current $\mathbf{A}_{\text{ub}}\boxtimes \overline{\mathbf{A'}_{\text{ub}}}$ corresponds to the preserved or unbroken symmetry structure through the transition between models. More traditionally, the structure can be represented by using the relative tensor product, or \emph{pairing}, $\mathcal{H}_{\mathbf{A}}\otimes_{\mathbf{A}_{\text{ub}}} \overline{\mathcal{H}_{\mathbf{A'}}}$ which is a $\mathbf{A}_{\text{ub}}$ balanced $\mathbf{A}-\overline{\mathbf{A'}}$ bimodule\footnote{However, we stress that the exact and detailed construction of this relative tensor product structure between theories has not been studied sufficiently. Moreover the resultant model can be outside of the conventional models such as modular tensor categories (MTCs) because of the field identification problems even when taking $\mathbf{A}$ and $\mathbf{A'}$ as MTCs. Inversely, to obtain the resultant bimodules as a MTC, it can be necessary to take $\mathbf{A}$ and $\mathbf{A'}$ outside of MTCs. For example, in the case with $\mathbf{A}_{\text{ub}}=\mathbb{Z}_{2}$, the underlying models $\mathbf{A}$ and $\mathbf{A'}$ can be fermionic models outside of MTCs. This modification seems to correspond to a variant of Karoubi envelop in mathematics and parton construction in condensed matter. We study this aspect elsewhere.}. In other words, the pairing in mathematics and BCS pair creation in physics have a straightforward connection through the folding trick and (half-)integer spin statistics. This transparent structure in the transitions is the main focus of the present manuscript. The unbroken symmetry $\{ \mathcal{Q}_{\alpha_{\text{ub}}}\}=\mathbf{A}_{\text{ub}}$, which is a symmetry unbroken under the (massless) RG flow, is fundamental in studying RG between QFTs and their lattice realization. Hence, we provide a general argument based on the quantum Hamiltonian formalism of CFTs and their RGs (or Hamiltonian truncation)\cite{Cardy:1986ie,Yurov:1989yu,Yurov:1991my,Hogervorst:2014rta,James_2018}, focusing on the phenomenology at the IR. Many puzzling (and interesting) problems exist in the interpolation between UV and IR, but we rather focus on the IR or the consequences of RG, except for the unbroken symmetry structure $\mathbf{A}_{\text{ub}}$ persistent in the transition.

First, let us assume a UV theory is deformed by some perturbation $H_{\text{pert}}$. Then, the UV Hamiltonian $H$ with unbroken symmetry $\{ \mathcal{Q}_{\alpha_{\text{ub}}}\}=\mathbf{A}_{\text{ub}}(\subset \mathbf{A})$ deformed from a CFT Hamiltonian $H_{\text{CFT}}$ with the \emph{fusion ring symmetry}\cite{Fukusumi:2025clr,Fukusumi:2025xrj,Fukusumi:2025fir,Fukusumi:2025fvb} $\{ \mathcal{Q}_{\alpha}\}=\mathbf{A}$ is given as follows;
\begin{equation}
H=H_{\text{CFT}}+H_{\text{pert}}
\end{equation}
where the commutation relations $[H_{\text{CFT}},\mathcal{Q}_{\alpha}]=0$ and $[H,\mathcal{Q}_{\alpha_{\text{ub}}}]=0$ hold. These commutation relations provide a fundamental aspect of the generalized symmetries as conserved charges\cite{Petkova:2000ip}. One can see examples in \cite{Grimm:1990ch,Grimm:2001dr,Belletete:2018eua,Belletete:2020gst}.

Next, in the same way as in the UV case, we assume the IR Hamiltonian $H_{\text{CFT}}'$ with the (emergent) fusion ring symmetry $\{ \mathcal{Q}_{\alpha'}\}=\mathbf{A'}$ can be written as follows;
\begin{equation}
H'=H'_{\text{CFT}}+H'_{\text{pert}}
\end{equation}
where the commutation relations $[H'_{\text{CFT}},\mathcal{Q}_{\alpha'}]=0$ and $[H',\mathcal{Q}_{\alpha'_{\text{ub}}}]=0$ holds, and $\mathbf{A'}_{\text{ub}}=\{ \mathcal{Q}_{\alpha'_{\text{ub}}} \}$ is the unbroken symmetry satisying the algebraic equivalence $\{ \mathcal{Q}_{\alpha_{\text{ub}}}\}=\mathbf{A'}_{\text{ub}}=\mathbf{A}_{\text{ub}}$. In this manuscript, we distinguish the IR objects from the UV objects by the prime symbol '. The RG in Hamiltonian formalism is a process relating $H$ to $H'$, and the corresponding procedure has been studied in the truncation methods\cite{Yurov:1989yu,Yurov:1991my}. There exist some subtle points in the process\cite{Giokas:2011ix}, but the definition of symmetry is (or should be) compatible.

On the other hand, there exists a more straightforward (or nonperturbative) way of studying the transition property of $\mathbf{A}$ and $\mathbf{A'}$. The renormalization group domain wall\cite{Brunner:2007ur,Gaiotto:2012np} provides the process of RG as a projection $\rho$ from a UV Hilbert space $\mathcal{H}_{\mathbf{A}}$ to an IR Hilbert space$\mathcal{H}_{\mathbf{A'}}$\cite{Klos:2019axh,Klos:2020upw},
\begin{equation}
\rho:\mathcal{H}_{\mathbf{A}}\rightarrow \mathcal{H}_{\mathbf{A'}},
\end{equation}
and $\rho$ provides the mapping $H_{\text{CFT}}\rightarrow H'_{\text{CFT}}$.
Hence, when restricting our attention to CFT parts or transitions of universalities, there exist several theoretical techniques applicable to the setting. In this formalism, the unbroken symmetry $\mathbf{A}_{\text{ub}}$ is defined by the subsymmetry $\mathbf{A}_{\text{ub}}\subset \mathbf{A}$ satisfying the following relation
\begin{equation}
\rho(\mathbf{A}_{\text{ub}})=\mathbf{A'}_{\text{ub}}
\end{equation}
However, we stress that the correspondence between $\mathbf{A}_{\text{ub}}$ and $\mathbf{A'}_{\text{ub}}$ in the categorical language requires more careful arguments, involving (half-)integer spin nonsimple currents. Whereas there exist subtle points in formulating the corresponding category theories, this algebraic method by $\rho$ enables one to study how UV states, operators, and symmetries transform to those in IR. This clear and intuitive description of transition properties is a fundamental benefit in understanding the corresponding quantum phase transitions.

\subsection{Problem and summary of results}
The algebraic method in recent studies is useful in understanding the consequences of the (conjectural) RG flow compared with other analytical methods. However, it should be stressed that the algebraic construction of $\rho: H_{\text{CFT}}\rightarrow H'_{\text{CFT}}$ itself does not ensure the stability of the RG flow or the irrelevance of $H'_{\text{pert}}$ straightforwardly, because the construction is different from the perturbative formalism. In other words, the method is intrinsically nonperturbative and complementary to the perturbative method. The algebraic method usually (implicitly) requires one to \emph{assume} the existence of the RG flow, and this contrasts with the analytical method, which usually attempts to \emph{demonstrate} the existence and the resultant stability of the RG flow.

The points can be itemized as follows;
\begin{itemize}
\item{Algebraic method: $\rho$ determines the transformation laws of states, operators, and symmetries. However, it does not ensure the stability of the process, i.e., irrelevance of $H'_{\text{pert}}$.}
\item{Perturbative method: $H_{\text{pert}}$ should result in $H'_{\text{pert}}$, and this process provides stability analysis of $H'_{\text{pert}}$. However, it does not provide transformation laws of quantum states or operators without higher-order calculations\cite{Poghosyan:2013qta,Poghossian:2013fda}, i. e. $\rho$.}
\end{itemize}
Hence, it is necessary to bridge the algebraic construction of $\rho$ and analysis for $H_{\text{pert}}$ and $H'_{\text{pert}}$. For this purpose, we attempt to provide arguments answering the following question;
\begin{equation}
\text{\emph{Is $H'_{\text{pert}}$ derived from $\rho$ irrelevant?}}
\end{equation}

In this manuscript, we focus on the IR unbroken symmetry $\mathbf{A'}_{\text{ub}}=\rho(\mathbf{A}_{\text{ub}})$ which commutes with IR (and UV) Hamiltonians, and study their restriction on $H'_{\text{pert}}$ (or the consequence of action of $\rho$ on $H_{\text{pert}}$). Because the structure of the unbroken symmetry is common to both the perturbative and algebraic methods, the arguments in this manuscript apply to a general system. Moreover, because we focus on the IR theory, the technical difficulty of $\rho$ in the algebraic formulation and the subtlety of the truncation method in the perturbative formulation\cite{Giokas:2011ix} do not affect most of our analysis. In particular, in the massless flows between minimal CFTs\cite{Belavin:1984vu}, $\mathbf{M}(p,p+1)\rightarrow \overline{\mathbf{M}(p-1,p)}$ preserving the fusion ring symmetry $\text{FR}(SU(2)_{p-2})$, we demonstrate that
\begin{equation}
\begin{split}
&\text{Under the unbroken symmetry $\text{FR}(SU(2)_{p-2})$,} \\
&\text{$H'_{\text{pert}}$ is irrelevant.}
\end{split}
\label{main_claim}
\end{equation}
This establishes the stability of the flow $\mathbf{M}(p,p+1)\rightarrow \overline{\mathbf{M}(p-1,p)}$ under the unbroken symmetry $\text{FR}(SU(2)_{p-2})$.
In other words, only by assuming the existence of the RG flows and unbroken symmetries, the RG flows are robust to the detailed form of the (unnecessary) perturbations. This stabilization of the IR fixed point can be regarded as (weak-) symmetry-enforced gaplessness in recent studies\cite{Wang:2014lca,Wang:2016gqj,Sodemann:2016mib,Wang:2017txt,Kim:2025tzc}, and our analysis clarifies the sequence of the massless RG flow as (weak-) symmetry-enforced gaplessness by the fusion ring symmetry. This aspect will be important when studying the realization of massless RGs in lattice models, because the lattice models can include unnecessary perturbations without fine-tuning and/or symmetry constraints.

We also explain the role of unbroken symmetry in the coset representation \cite{Goddard:1984vk,Goddard:1984hg,Goddard:1986ee,Goddard:1988md} and the level-rank duality\cite{Kuniba:1990im,Kuniba:1990zh,Nakanishi:1990hj,Altschuler:1990th,Naculich:1990hg,Aharony:2016jvv,Hsin:2016blu} of CFTs. It should be stressed that our phenomenology has a close connection to the Higgs phenomena\cite{Higgs:1964pj}, and it implies a unifying relationship between classification by generalized symmetry (or finite ring theory) and coset structure of Lie groups. Connections between ring theory and group representation theory have been studied in (pure) mathematics, and we attempt to emphasize that their coset or level-rank duality structures will be useful in the context of RG in physics. In particular, we propose the following coset representation containing the level-rank duality structure generating the (half-)integer spin nonsimple currents $\text{FR}(SU(2)_{p-2})$,
\begin{equation}
\begin{split}
&\mathbf{M}(p,p+1)\boxtimes \mathbf{M}(p-1,p) \\
&=\frac{SU(p-1)_{2}\otimes SU(2)_{p-3}\otimes SU(2)_{1}}{SU(2)_{p-2}\boxtimes \left(U(1)_{p-2}\otimes SU(p-2)_{2} \right) }
\end{split}
\end{equation}
The coset representation in \cite{Blumenhagen:1994ik,Blumenhagen:1994wg} plays the fundamental role. Comparing the corresponding coset in the pioneering works\cite{Crnkovic:1989ug,Gaiotto:2012np}, the structure of the (half-)integer spin nonsimple current, the $SU(2)_{p-2}$ level-rank duality structure in the denominator, can be seen more straightforwardly in this expression. The same structure of the coset representation containing the level-rank duality structure holds for $W_{N}-$symmetric models more generally\cite{Martins:1991hi,Dunning:2002cu}, only by replacing $SU(2)$ with $SU(N)$ and so on\footnote{We thank Yunqin Zheng for the corresponding discussion.}. Related observations can be seen in the concluding remark of \cite{Ambrosino:2026umb}.

Moreover, by revisiting the role of symmetry at UV, we study the nontrivial relationship between the (half-)integer spin nonsimple current structure and the nontrivial cascade of phase transitions $H\rightarrow H'\rightarrow H''\rightarrow ...$. This cascade can be triggered by the coexistence of a relevant perturbation and other irrelevant perturbations constructed from the (half-)integer spin nonsimple currents, whereas the relevant perturbation triggers only the first flow $H\rightarrow H'$. Whereas the irrelevant perturbations themselves do not trigger RG flow, the resonance effect with a relevant perturbation is nontrivial away from the established flow $H\rightarrow H'$ by a relevant perturbation. In this context, we \emph{rediscover} the following point in the context of generalized symmetry:
\begin{equation}
\begin{split}
&\text{One cannot neglect irrelevant perturbations in general} \\
&\text{when there exists a relevant perturbation.}
\end{split}
\label{cascade_claim}
\end{equation}
Hence, in realizing the massless RG $H\rightarrow H'$ in a lattice model, the exclusion of unnecessary irrelevant perturbations triggering undesired flow $H\rightarrow H''$ is fundamental. In this context, symmetry plays a fundamental role in excluding such unnecessary perturbations. To our knowledge, this resonance effect from the irrelevant perturbations has been studied as an effect of a ``dangerously irrelevant perturbation'' in quantum field theory. However, the meaning of a dangerously irrelevant perturbation is different from that in statistical mechanics. To avoid confusion, we rather identify the phenomenon as a resonance phenomenon. We demonstrate that this resonance phenomenon (and its consequent cascades) are ubiquitous in the massless RG (or the Higgs transitions\cite{Cordova:2025eim}), and provide the spin-$2$ chiral-chiral nonsimple current (or ``phantom" current in references\cite{Antinucci:2025uvj,Furuta:2025ahl,Zhang:2026gqp}) as a building block constructing a series of cascades. In particular, we propose the following cascade of flows,
\begin{equation}
\mathbf{M}(p,p+1)\rightarrow \overline{\mathbf{M}(p-1,p)} \rightarrow \mathbf{M}(p-2,p-1)
\end{equation}
triggered by the coexistence of the relevant perturbation $\Phi_{|1,3|}$ and irrelevant perturbation $\Phi_{|3,1|}$ in $\mathbf{M}(p,p+1)$. Very interestingly, it has been proposed in \cite{Gukov:2015qea} that there exists an alternative scenario where the system flows to another (unexplored) fixed point, and further studies on the resonance effect are an interesting problem.

Historically, the corresponding phenomena have been studied in the supersymmetric theories\cite{Seiberg:1994pq,Kutasov:1995np,Kutasov:1995ve,Intriligator:1995ff,Kutasov:1995ss,Leigh:1996ds} and weak-first-order transition \cite{Gorbenko:2018ncu}. General aspects have been studied in \cite{Gukov:2015qea,Gukov:2016tnp} by using a variant of the Morse theory. The earlier examples and related classification schemes have been summarized in \cite{Amit:1982az}. In this context, what we want to emphasize is
\begin{equation}
\begin{split}
&\text{Existence of ``dangerously irrelevant operators"} \\
&\text{or resonance phenomena is ubiquitous.}
\end{split}
\end{equation}
This observation is based on the existence of the integer spin nonsimple current structure in the coset or level-rank duality structure of many series of conformal field theories, with a close connection to the hierarchical structure of topological orders.

\begin{figure}[htbp]
\begin{center}
\includegraphics[width=0.5\textwidth]{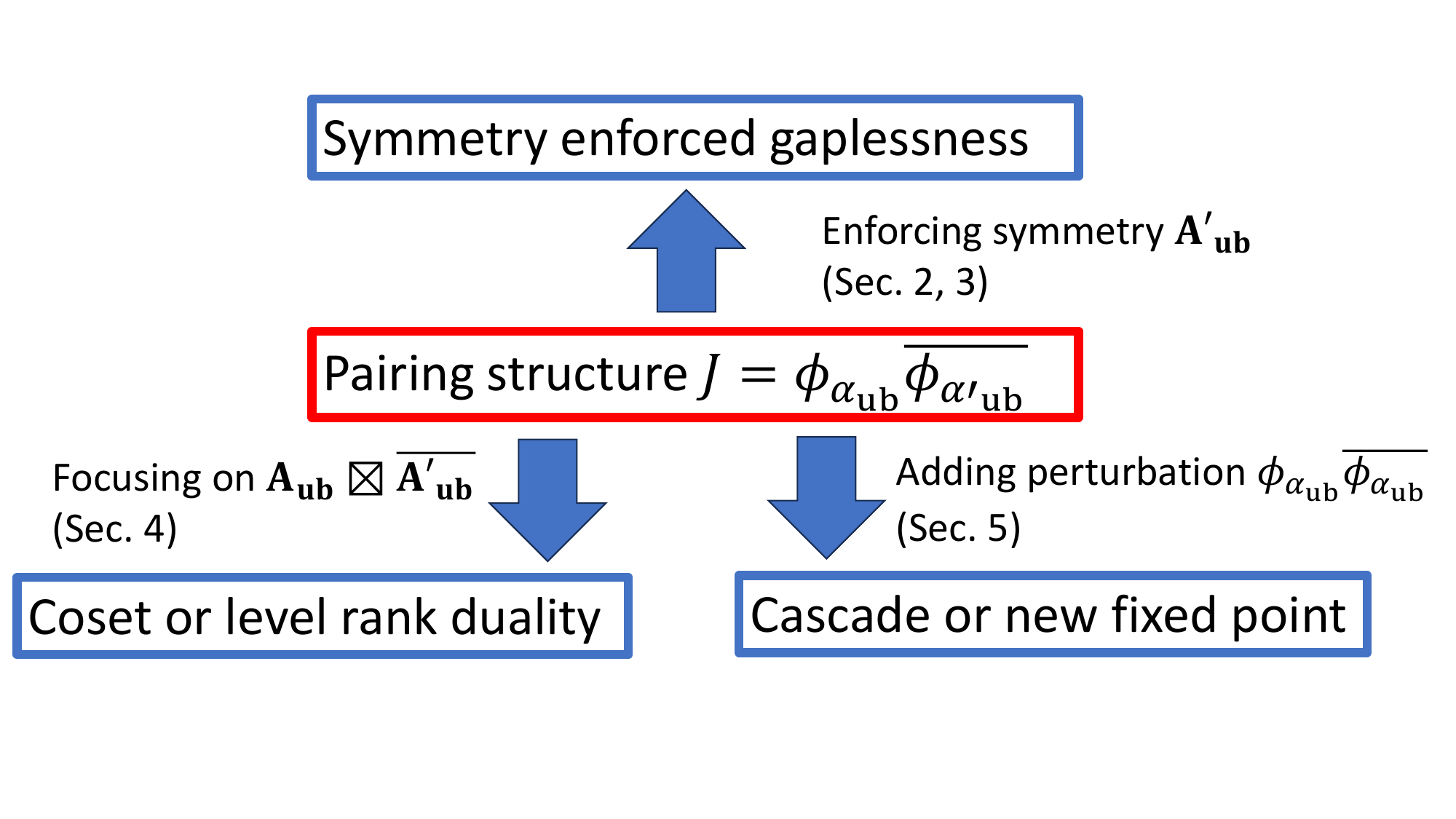}
\caption{Summary of nontrivial phenomena from the (half-)integer spin simple current. $J=\phi_{\alpha_{\text{ub}}}\phi_{\alpha'_{\text{ub}}}\in \mathbf{A}_{\text{ub}}\boxtimes \overline{\mathbf{A'}_{\text{ub}}}$ is the particle satisfying the spin constraint $s_{J}=\mathbb{Z}/2$ and we call (half-)integer spin nonsimple current. The same object is also called connected \'{e}tale algebra or condensable algebra\cite{DavydovMugerNikshychOstrik+2013+135+177,Kong:2013aya}. We interpret this object as symmetry operators in CFTs, electron operators in TQFTs, and perturbations triggering RG flows.}
\label{fig:nonsimple}
\end{center}
\end{figure}

The rest of the manuscript is organized as follows (also see Figure \ref{fig:nonsimple}). Sec. \ref{section_main} is the main part of the present manuscript. We provide the rigorous demonstration of the statement, Eq. \eqref{main_claim}. Sec. \ref{section_example} is the application of our method to simple flows: the flow from the tricritical Ising CFT to the Ising CFT and that from the tetracritical Ising CFT to the tricritical Ising CFT. In Sec. \ref{section_Higgs}, we study the phenomenological relationship between the unbroken symmetry and the coset representation or level-rank duality. This section can be regarded as an improvement of Appendix B in \cite{Fukusumi:2025clr}, and we clarify the role of the (half-)integer spin nonsimple currents. In Sec. \ref{section_cascade}, we revisit the implications of the unbroken symmetry in the UV theory by applying the arguments in the previous sections. We provide a symmetry-based argument demonstrating the statement Eq. \eqref{cascade_claim}, and clarify the spin-$2$ chiral-chiral nonsimple current as a source of the cascades. The main points are summarized in Eq. \eqref{cascade_example} and \eqref{cascade_general}. Sec. \ref{section_conclusion} contains the concluding remarks of the present manuscript. We comment on the status of the ideas in the present manuscript and the possible future research directions. We include elementary parts of techniques in the CFTs and their topological symmetries in the Appendix, and it will be helpful for the reader unfamiliar with CFTs. We also apply our method to more general flows in \cite{Nakayama:2024msv}, but the stabilization or symmetry-enforced gaplessness cannot be achieved without some additional assumptions.

\subsection{Comment on fusion category or fusion ring symmetry}

For a technical reason, we do not use the term ``fusion category symmetry", such as $\text{Rep}(SU(2)_{p-2})$. The fusion category symmetry often fixes many specific categorical data such as the central charge and braidings, and this limitation is inconvenient when studying relationships between different theories. Moreover, in some of the references, fusion category symmetries do not admit $\mathbb{C}$-linearlity (or linear algebraic structure), which is fundamental in studying quantum systems\footnote{This incompatibility with $\mathbb{C}$-linear structure in category theories is fundamental in studying boundary conditions and defects, but this is sometimes inconvenient in studying \emph{operators} and their \emph{algebra}.}. For readers interested in this problem of fusion category symmetry, we note recent works by the first author \cite{Fukusumi:2025clr,Fukusumi:2025xrj,Fukusumi:2025fir,Fukusumi:2025fvb}. However, we also note that the role of $\mathbb{C}$-linearity in the present manuscript is limited because we focus on the particular type of unbroken symmetries in the massless RG flows between minimal models, which has been studied in \cite{Gaiotto:2012np}. In this setting, the distinction between defects (without $\mathbb{C}$-linearlity) and topological symmetry operators (with $\mathbb{C}$-linearity) becomes less relevant to some extent. However, this distinction still plays an important role at several technical points. Our discussion is based on symmetry analysis in linear algebra and the standard scaling analysis on the stability of RGs(see \cite{Cardy:1986ie,Cardy:1996xt}, for example). In particular, the action of symmetry operators on bulk fields is fundamental, and corresponding arguments can be seen in \cite{Buican:2017rxc,Nakayama:2024msv,Gaberdiel:2026sfg,Ambrosino:2026umb,Benedetti:2026drn}.

We also note that the (conjectural) relationship between the massless RG flow and the gapped domain wall (or $\mathbb{C}$-linear tensor functor)\cite{Wan:2016php,Zeev:2022cnv,Zhao:2023wtg,Fukusumi:2024ejk,Fukusumi:2025clr} is not the main scope of the present manuscript except for the latter part of Sec. \ref{section_Higgs}, and this point contrasts with previous works by the author(s)\cite{Fukusumi:2025clr,Fukusumi:2025xrj,Fukusumi:2025fir,Fukusumi:2025fvb}. Related to this point, we also note that \emph{we focus on the role of symmetry at IR} in the present manuscript, whereas the author(s) focused on that at UV (and their reduction or interpolation to low energy sectors of the IR) in the previous works\cite{Fukusumi:2025clr,Fukusumi:2025xrj,Fukusumi:2025fir,Fukusumi:2025fvb}\footnote{After constructing a series of homomorphisms or monoidal functors, we noticed that the connection between homomorphisms and the massless RGs is not straightforward. There exist many homomorphisms by fixing a UV and IR theory, and these homomorphisms should be interpreted as interpolation (or breaking process) of symmetries from the mixing by relevant perturbations. In TQFTs, they provide classifications of fusion rules, but there exist other classifications by the profunctor, as we study in the present manuscript.}.

\section{Stability for Zamolodchikov renormalization group flows}
\label{section_main}

In this section, we present a general stability analysis for the massless RG flow\cite{Zamolodchikov:1987jf,Zamolodchikov:1987ti,Zamolodchikov:1989hfa}. The main result is the statement Eq. \eqref{statement_minimal}. We summarize basic knowledge of CFTs and their topological symmetry in the Appendix  \ref{section_review}, and we focus on the algebraic and scaling analysis relevant in the main proposal. We note \cite{DiFrancesco:1997nk,Ginsparg:1988ui,Ribault:2016sla} as fundamental references in general aspects of CFTs and \cite{Petkova:2000ip,Graham:2003nc,Recknagel:2013uja,Northe:2024tnm} as references on defect and boundary CFT. We study the flows between minimal conformal field theories\cite{Belavin:1984vu},
\begin{equation}
\mathbf{M}(p,p+1)\to \overline{\mathbf{M}(p-1,p)}.
\end{equation}
To distinguish this flow from the usual flow from the coset structure, we have denoted the IR theory as $\overline{\mathbf{M}(p-1,p)}$. We study this point in Sec. \ref{section_Higgs}, and the readers can forget about this point of the chirality in this section. We only make some historical remarks for the readers who care about this subtle point. These nontrivial changes of the chirality in the massless RG flow can be observed in \cite{Gaiotto:2012np} based on the nontrivial structure of the corresponding folded model in \cite{Crnkovic:1989ug}. The related analysis on the conformal spin of objects between theories have been studied in \cite{Nakayama:2024msv,Kikuchi:2024cjd}, and their relation to the Lieb-Schulz-Mattice type classification\cite{Lieb:1961fr,Schultz:1964fv} of CFTs has been studied in \cite{Fukusumi_2022_c,Fukusumi:2024ejk,Fukusumi:2025clr,Fukusumi:2025ljx,Fukusumi:2025xrj}. We also note related earlier works on the CFTs with discrete $\mathbb{Z}_{N}$ symmetry \cite{Furuya:2015coa,Lecheminant:2015iga,Numasawa:2017crf,Yao:2018kel,Tanizaki:2018xto}. When focusing on the $\mathbb{Z}_{N}$ symmetry of the $SU(N)_{K}$ Wess-Zumino-Witten model\cite{Wess:1971yu}, the model should provide the anomaly classification with $K$ modulo $N$. In this context, when coupling two different $SU(N)_{K}$ and $SU(N)_{K'}$, the theory should provide different connectivity depending on $K-K'=0, \ \text{mod.}N$ or $K+K'=0, \ \text{mod.}N$. These two signs distinguish the $\mathbb{Z}_{N}$ preserving flows $SU(N)_{K}\rightarrow SU(N)_{K'}$ and $SU(N)_{K}\rightarrow \overline{SU(N)_{K'}}$. This research direction has been explored in the context of the Haldane conjecture\cite{Haldane:1981zza,Haldane:1983ru,Wamer:2019oge} or symmetry-protected topological phases\cite{MOshikawa_1992,Pollmann:2009ryx,Pollmann:2009mhk}, and one can observe their connections to gapless symmetry-protected topological phases\cite{Scaffidi:2017ppg}.

\subsection{Main proposal and fundamental relations}

First, let us assume that the primary fields in the IR theory are labelled by $a'$, and they are denoted as $\Phi_{a'}=\phi_{a'}\boxtimes \overline{\phi_{a'}}$ where $\phi$ and $\overline{\phi}$ are the chiral or antichiral primary fields respectively and $\boxtimes$ is called the Deligne product\footnote{We drop space-time coordinates and spatial integrals to simplify the notations.}.  The $\boxtimes$ is fundamental in constructing bulk CFTs consistent with the bootstrap technique, and this results in the ring isomorphism $\{\phi_{\alpha}\}=\{\Phi_{\alpha}\}=\mathbf{A}$ known as the Moore-Seiberg data\cite{Moore:1988ss,Moore:1988qv,Moore:1989vd,Fuchs:2002cm}. Hence, we can focus on the common algebraic structure $\{\alpha\}=\mathbf{A}$. The linear sum of $\Phi_{a'}$ will provide the IR perturbation $H'_{\text{pert}}$ from the IR CFT $\overline{\mathbf{M}(p-1,p)}$. 

\begin{figure}[htbp]
\begin{center}
\includegraphics[width=0.5\textwidth]{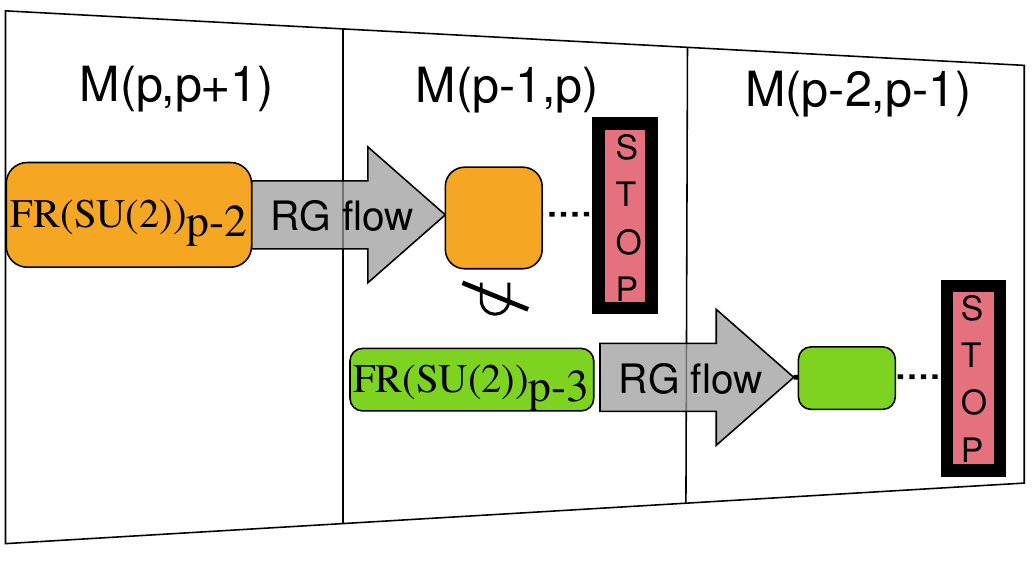}
\caption{The conceptual picture of the RG flow between unitary minimal models. The unbroken symmetry is $\text{FR(SU(2))}_{p-2}$ through the RG flow, and as stated in section \ref{subsec:derivation}, this symmetry is stabilized under the flow. Then the flow will be 
stopped and will not go further to $\textbf{M}(p-2,p-1)$. However, an emergent symmetry $\text{FR}(SU(2)_{p-3})$,which is not a subsymmetry of $\text{FR}(SU(2)_{p-2})$, allows $\overline{\textbf{M}(p-1,p)}$ to flow to $\textbf{M}(p-2,p-1)$.}
\label{fig:RG_flow_emargent}
\end{center}
\end{figure}

We also introduce the unbroken symmetry at the IR theory $\mathcal{Q}_{\alpha'_{\text{ub}}}\in \mathbf{A'}_{\text{ub}}=\mathrm{FR}(SU(2)_{p-2})$\cite{Gaiotto:2012np}. To emphasize the preservation of the fusion ring under the flow, we denoted $\mathbf{A'}_{\text{ub}}=\mathrm{FR}(SU(2)_{p-2})$, not $\mathbf{A'}_{\text{ub}}=\mathrm{FR'}(SU(2)_{p-2})$. We focus on the primary fields $\Phi_{a'}$ constrained by the relation 
\begin{equation}
[\mathcal{Q}_{\alpha'_{\text{ub}}}, \Phi_{a'}]=0 
\end{equation}
for all $\mathcal{Q}_{\alpha'_{\text{ub}}}\in \mathbf{A'}_{\text{ub}}=\mathrm{FR}(SU(2)_{p-2})$. This relation constrains the form of the perturbations very strongly, and we obtain the following claim.
\begin{equation}
\text{$\Phi_{a'}$ preserving $\mathrm{FR}(SU(2)_{p-2})$ is irrelevant, i. e. $h_{a'}>1$.} 
\label{statement_minimal}
\end{equation}
where $h_{a'}$ is the chiral conformal dimension of $\phi_{a'}$. This phenomenon can be depicted as in Figure \ref{fig:RG_flow_emargent}. Hence, for this flow, $H'_{\text{pert}}$ contains only irrelevant perturbations, and at the level of this scaling analysis,
\begin{equation}
\text{The IR theory $\overline{\mathbf{M}(p-1,p)}$ is stable in $\mathrm{FR}(SU(2)_{p-2})$.}
\end{equation}
In other words, \emph{the massless RG flow can be identified as a sequence of (weak-) symmetry-enforced gaplessness by the fusion ring symmetry $\mathrm{FR}(SU(2)_{p-2})$}. By replacing $\mathbf{A}_{\text{ub}}$ and $\mathbf{A'}$, one can apply the same discussions in general.

For the demonstration of the above arguments, it is important that the condition $[\mathcal{Q}_{\alpha'}, \Phi_{a'}]=0 $ results in the relation $\{q_{\alpha',(I')}=q_{\alpha',(a')}\}$, where $I'$ is the identity operator corresponding to the vacuum and $q_{\alpha', (a')}$ is the algebraic generalized quantum dimension (AGQD) of $\alpha'$ defined by the following way, 
\begin{equation}
q_{\alpha',(a')}=\frac{S_{\alpha', a'}}{S_{I',a'}}
\end{equation}
where $S$ is the modular $S$ matrix.
The AGQD reproduces the conventional quantum dimensions in the literature as
\begin{equation}
\text{dim}(\alpha')=q_{\alpha',(I')}
\end{equation}
In short, the fundamental relation can be noted as,
\begin{equation}
\{ [\mathcal{Q}_{\alpha'}, \Phi_{a'}]=0 \} \Rightarrow \{q_{\alpha',(I')}=q_{\alpha',(a')}\},
\end{equation}

In particular, the following quantity is fundamental,
\begin{equation}
q_{|1,2|',(a')}
=2(-1)^{r'+s'}\cos\left(\frac{\pi s'}{p}\right).
\end{equation}
where $|1,2|'$ and $a'=|r',s'|$ is the Kac label of primary fields. At the ring theoretic level, the unbroken symmetry can be expressed as $\mathrm{FR}(SU(2)_{p-2})=\{\mathcal{Q}_{|1,2|'}^{n}\}_{n=0}^{p-2}$, and this simplifies the analysis. This simplification is a benefit of treating the generalized symmetry as a ring or algebra. We stress that this simple operation in linear algebra contains some difficulties when interpreting the symmetry as fusion category symmetry (without $\mathbb{C}-$linearity). 

\subsection{Derivation of the proposals}\label{subsec:derivation}

To show the main proposal, Eq. \eqref{statement_minimal}, we study the necessary condition where the IR theory $H'_{CFT}+H'_{\text{pert}}$ is invariant under the symmetry $\mathrm{FR}(SU(2)_{p-2})$. This relation is nothing but the commutative relation.
\begin{equation}
\{ [\mathcal{Q}_{\alpha'_{\text{ub}}}, \Phi_{a'}]=0 \}_{\alpha'_{\text{ub}}\in \mathrm{FR}(SU(2)_{p-2})}
\end{equation}
when $H'_{\text{pert}}$ is represented by the linear sum of of $\{\Phi_{a'}\}$.
It should be stressed that the direct calculation of the commutator $[\mathcal{Q}_{\alpha'}, \Phi_{a'}]$ is a difficult problem in general, because the commutation relations can generate nonlocal operators, called disorder fields. One can see respective arguments in some class of statistical mechanical models in \cite{Fateev:1985mm,Dotsenko:2003zc,Dotsenko:2003kg,Dotsenko:2003ui}, but the classification of more general disordered fields has not been studied except for a few works\cite{Runkel:2010ym,Konechny:2025rcc}. At the level of abstract algebra,  they will be described by the (extended) Brauer-Picard fusion category $\mathbf{M'}\otimes \mathbf{B'}$, constructed from stacking spherical fusion category ($\mathbf{B'}\sim \mathbf{A'}\boxtimes \overline{\mathbf{A'}}$) to modular tensor category ($\mathbf{M'}\sim \mathbf{A'}$), by applying the techniques in \cite{Fukusumi:2024ejk,Fukusumi:2025ljx}. More intuitively, $\mathbf{B'}$ corresponds to the bulk fields and their nonchiral fusion rings in CFT\cite{Fuchs:1993et,Rida:1999xu,Rida:1999ru,Nivesvivat:2025odb} and $\mathbf{M'}$ is the defects which provide the extension of CFTs. This is an interesting problem, but it is outside of the scope of the present manuscript.

Instead of calculating the commutator directly, the necessary condition has been studied in recent references\cite{Nakayama:2024msv,Gaberdiel:2026sfg,Ambrosino:2026umb,Benedetti:2026drn}. The fundamental points are symmetry action on the quantum states, and the operator-state correspondence. When the commutation relation holds, the following should hold,
\begin{equation}
\mathcal{Q}_{\alpha'}\Phi_{a'}|0'\rangle=q_{\alpha',(I')}\Phi_{a'}|0'\rangle
\end{equation}
where $|0'\rangle$ is the CFT vacuum. We also assume the radial quantization of the CFT and the operator-state correspondence, 
\begin{equation}
|a'\rangle=\Phi_{a'}(t=-\infty)|0\rangle
\end{equation}
where $\Phi_{a'}(t=-\infty)$ corresponds to the elementary many-body excitation corresponding to a generalization of the Lieb-Schultz-Mattis twist operators\cite{Lieb:1961fr,Schultz:1964fv}. In the Lagrangian formalism of QFT, $\phi_{a'}(t=-\infty)$ can be interpreted as a Wilson line operator. By applying the operator-state correspondence, one can obtain the following relation,
\begin{equation}
\mathcal{Q}_{\alpha'}|a'\rangle=q_{\alpha',(a')}|a'\rangle
\end{equation}
Hence, one can derive the desired relation,
\begin{equation}
q_{\alpha',(I')}=q_{\alpha', (a')}
\end{equation}
These arguments apply to a general class of models when the operator-state correspondence and the AGQD are defined.

As a consequence of the massless RG flow $\mathbf{M}(p,p+1)\rightarrow \overline{\mathbf{M}(p-1,p)}$, one can express the unbroken symmetries as follows\cite{Gaiotto:2012np};
\begin{equation}
\mathrm{FR}(SU(2)_{p-2})=\{ \mathcal{Q}_{|1,v'|'}\}_{v'=1}^{p-1}.
\end{equation}
One can check consistency by calculating $\text{dim}(|v',1|)=\text{dim}(|1,v'|')$, implying the preservation of the vacuum in the transition\footnote{However, this is not a sufficient condition for fixing the preserved sectors. However, for our purpose, the detailed data are not necessary.}. Hence, one can detect the candidate of bulk primary fields preserving the fusion ring symmetry by testing the condition $q_{\alpha'_{\text{ub}},(I')}=q_{\alpha'_{\text{ub}}, (a')}$ for each $\{\alpha'_{\text{ub}}=|1,v'|\}_{v'=1}^{p-1}$. However, in this case, the following recursive relation holds,
\begin{equation}
\mathcal{Q}_{|1,2|'}\mathcal{Q}_{|1,v'|'}=\mathcal{Q}_{|1,v'-1|'}+\mathcal{Q}_{|1,v'+1|'}
\end{equation}
This relation implies that an arbitrary object $\mathcal{Q}_{\alpha'_{\text{ub}}}\in \mathrm{FR}(SU(2)_{p-2})$ can be constructed from the linear sum of $\{\mathcal{Q}_{|1,2|'}^{v'}\}_{v'=0}^{p-2}$. For example, one can obtain the following relation,
\begin{equation}
\mathcal{Q}_{|1,3|'}=\mathcal{Q}_{|1,2|'}^{2}-\mathcal{Q}_{I'}
\end{equation} 
Hence, we focus on the case $\alpha'=|1,2|'$, and we denote $|1,2|'$ as $\alpha'_{\text{ub,g}}$ to emphasize the role of this object as a generator of the unbroken symmetry.
It should be kept in mind that the above type of elementary relations is not compatible with the analysis of defects without some modifications, because of the $-$ sign. 

Hence, for the bulk fields $\phi_{a'}$ preserving the unbroken symmetry, the following relation is necessary,
\begin{equation}
q_{\alpha'_{\text{ub.g}},(a')}=q_{\alpha'_{\text{ub,g}},(I')}=2\cos\left(\frac{\pi}{p}\right)
\end{equation}
where $2\cos\left(\pi / p\right)$ is the quantum dimension of $|1,2|'$.
The exact form of AGQD $q_{\alpha'_{\text{ub.g}},(a')}$ can be determined straightforwardly by substituting the established form of the modular $S$ matrix, 
\begin{equation}
\begin{split}
q_{\alpha'_{\text{ub.g}},(a')}&=q_{|1,2|',(a')} \\
&=
(-1)^{r'}
\frac{\sin\left(\frac{2\pi(p-1)s'}{p}\right)}
{\sin\left(\frac{\pi(p-1)s'}{p}\right)} \\
&=
2(-1)^{r'+s'}\cos\left(\frac{\pi s'}{p}\right).
\end{split}
\end{equation} 
where we have labelled $a'=|r',s'|'$

This relation holds, up to modulo the Kac-table identification, if and only if
\begin{equation}
s'=1,\qquad r'\ \text{odd}.
\end{equation}
Thus the relevant candidates are the fields with Kac labels.
\begin{equation}
|r',1|',\qquad r'\ \text{odd}.
\end{equation}

The conformal weight of such a field in $\mathbf{M}(p-1,p)$ is
\begin{equation}
h_{|r',1|'}
=
\frac{\left(pr'-(p-1)\right)^2-1}{4p(p-1)}.
\end{equation}
For the vacuum $r'=1$, this gives $h_{1,1}=0$, and we exclude this trivial case. For every nontrivial solution with $r'$ odd, we have $r\geq 3$. To prove that these fields are irrelevant, i. e. $h_{|r'1|'}>1$ we perform the following straightforward calculation,
\begin{equation}
\begin{split}
&h_{|r',1|'}-1 \\
&=
\frac{\left(pr'-(p-1)\right)^2-1}{4p(p-1)}-1 \\
&=
\frac{(r'+1)(pr'-3p+2)}{4(p-1)}.
\end{split}
\end{equation}
The denominator is positive. Moreover, for every nontrivial odd solution $r\geq 3$,
\begin{equation}
pr'-3p+2
=
p(r'-3)+2
>
0.
\end{equation}
Hence, the following relation holds for all nontrivial solutions,
\begin{equation}
h_{|r',1|}-1>0 
\end{equation}
and this is nothing but $h_{|r',1|}>1$. Therefore, every primary field $\phi_{a'}$ invairant under the unbroken symmetry $\text{FR}(SU(2)_{p-2})$ or satisfying the relation $q_{\alpha_{\text{ub.g}},(I')}=q_{\alpha_{\text{ub.g}},(a')}$ is irrelevant. 

Consequently, the RG flow
\begin{equation}
\mathbf{M}(p,p+1)\to \overline{\mathbf{M}(p-1,p)}
\end{equation}
is stabilized by the unbroken $\mathrm{FR}(SU(2)_{p-2})$ symmetry in the sense that no relevant IR primary field appears under the symmetry. Preservation of the same AGQD as the vacuum for the operator $a'$, i.e., $q_{\alpha_{\text{ub},g},(I')}=q_{\alpha_{\text{ub},g},(a')}$, is fundamental.

\subsection{Emergence in massless renormalization group flow: Difference of defect and symmetry}
In this subsection, we provide phenomenological discussions from the flows that we established and their implications for the emergence\cite{Anderson:1972pca}. The most fundamental point we would like to stress is the distinction of the emergence between $H'_{\text{CFT}}$ and $H'$. We provide a canonical definition of emergent defects ($\mathbf{A'}\setminus \mathbf{A'}_{\text{ub}}$) and emergent symmetries($\mathbf{A'}\setminus \mathbf{A'}_{\text{ex}}$). This is a natural continuation of the corresponding discussions by the authors\cite{Fukusumi:2025clr}.

First, it should be stressed that the unbroken symmetry $\mathbf{A}_{\text{ub}}$ is smaller than the IR theory because of the relation $\mathbf{A'}_{\text{ub}}\subset \mathbf{A'}$. It is tempting to call $\mathbf{A'}\setminus \mathbf{A'}_{\text{ub}}(\subset \mathbf{A'})$ an emergent symmetry, but this terminology requires some caution when focusing on the CFT parts $H_{\text{CFT}}$ and $H'_{\text{CFT}}$ and their fusion ring symmetries $\mathbf{A}$ and $\mathbf{A'}$. It should be stressed that at the level of linear algebra, every IR object should be expressed as an application of $\rho$ to UV objects. Hence, by fixing the channel of the transition (such as primary fields), one can express the action of IR operators as that of UV operators. When interpreting the objects in $\mathbf{A'}\setminus \mathbf{A'}_{\text{ub}}\subset \mathbf{A'}$ as defects, one can observe emergent phenomena. Hence, the distinction between defects and symmetry operators is fundamental when focusing on $H_{\text{CFT}}$ and $H'_{\text{CFT}}$. One can see related discussions in \cite{Fukusumi:2025clr}, and this is in the research direction in \cite{Anderson:1972pca}.

However, there can exist emergent symmetry when focusing on $H'$ rather than $H'_{\text{CFT}}$. In this setting, $H'$ should be interpreted as an (intermediate) UV theory such as microscopic or lattice models, and $H'_{\text{CFT}}$ is an IR theory obtained by the large system size limit because of the irrelevance of $H'_{\text{pert}}$. The difference between $H'$ and $H'_{\text{CFT}}$ comes from the existence of $H'_{\text{pert}}$. $H'_{\text{pert}}$ will break the exact IR symmetry $\mathbf{A'}$ to a smaller symmetry $\mathbf{A'}_{\text{ex}}\subset \mathbf{A'}$ in more realistic settings, but this symmetry is bounded to $\mathbf{A'}_{\text{ex}}\supset \mathbf{A'}_{\text{ub}}$. In this context, what we have shown for the massless flow of minimal models is
\begin{equation}
\text{The theory $H'$ has emergent symmetry $\mathbf{A}_{\text{em}}=\mathbf{A'}\setminus \mathbf{A'}_{\text{ex}}$}
\end{equation}
The emergent symmetry $\mathbf{A}_{\text{em}}$ is bounded to $\mathbf{A}_{\text{em}}\subset\mathbf{A'}\setminus \mathbf{A'}_{\text{ub}}$. The precise form of $\mathbf{A'}_{\text{ex}}$ depends on the respective settings, and this is an interesting research direction. However, we stress that the phenomena or transition of the universality itself is robust at the level of the scaling analysis, and it does not depend on the details of $\mathbf{A'}_{\text{ex}}(\supset \mathbf{A'}_{\text{ub}})$. For more practical calculations, one can implement the UV exact symmetry $\mathbf{A}_{\text{ex}}(\subset \mathbf{A})$ by the mapping $\rho(\mathbf{A}_{\text{ex}})=\mathbf{A'}_{\text{ex}}$, but the calculations will be more involved.

Interestingly, whereas the massless RG flows can be formulated as a sequence of symmetry-enforced gaplessness\cite{Wang:2014lca,Wang:2016gqj,Sodemann:2016mib,Wang:2017txt,Kim:2025tzc}, or cascades of generalized symmetry breakings, the relationship of unbroken symmetry in the hierarchy is not inclusion in general. For example, in $\mathbf{M}(p,p+1)$, the relation $\text{FR}(SU(2)_{p-2}) \subset \text{FR}(SU(2)_{p-1})$ does not hold. In other words, the symmetry breaking of a step of the massless RG flows inevitably generates emergent defects and the corresponding symmetry, and they should play a fundamental role in the next step as unbroken symmetry. Related to this observation, the UV symmetry operator without locality, such as $\mathcal{Q}_{\alpha}+\mathcal{Q}_{\beta}+...$, can be stabilized under the RG flows. This observation will be useful in studying bulk and boundary (or bulk and defect) RG flows\cite{Dorey:2005ak,Green:2007wr,Dorey:2009vg,Fredenhagen:2009tn,Chang:2018iay,Fukusumi:2023vjm}, to establish RG understanding of edge modes in both gapped and gapless systems\cite{Affleck:1987vf,Scaffidi:2017ppg,Verresen:2019igf}. For example, $\mathcal{Q}_{I}+\mathcal{Q}_{\psi}$ and the corresponding defect in the Ising CFT can be stabilized as Majorana edge modes in a topological phase\cite{Kitaev:2001kla,Fukusumi:2020irh} where $\psi$ is the Majorana fermion field satisfying $\psi \times \psi =I$. One can formulate these edge modes even in boundary conformal field theories (BCFTs) \cite{Smith:2021luc,Fukusumi:2021zme,Weizmann}, but they are unstable despite their fundamental importance. In this context, \cite{Graham:2003nc} is the pioneering work providing these edge modes beyond conventional BCFTs\cite{Cardy:1986gw}. These kinds of edge modes or defects are fundamental in realizing qubits in lattice models\cite{Okada:2024qmk}.

\section{Analysis of the simple models}
\label{section_example}
\subsection{Tricritical Ising model to Ising}
First, we study the simplest flow $\mathbf{M}(4,5) \rightarrow \overline{\mathbf{M}(3,4)}$, the flow from the tricritical Ising model to the Ising model. This flow has been realized in the quantum lattice model\cite{Grover:2013rc,Rahmani:2015qpa,OBrien:2017wmx}, and captured attention in the field as the supersymmetric model with the lattice realization. To simplify the notation, we denote objects in a fusion ring $\{ \mathcal{Q}_{\alpha}\}=\mathbf{A}$ as $\alpha$, and apply the same notation to the IR objects.

The fusion ring of the UV theory can be summarized as
\begin{equation}
\mathbf{M}(4,5)=\{ I^{[1]},\psi^{[1]},\sigma^{[1]}\} \otimes \{ I^{[2]}, \tau^{[2]}\}
\end{equation}
where $\{ I^{[1]},\psi^{[1]},\sigma^{[1]}\}$ is the Ising fusion ring satisfying the following algebraic relations,
\begin{align}
\psi^{[1]}\times\psi^{[1]}&=I^{[1]}, \\
\psi^{[1]}\times \sigma^{[1]}&=\sigma^{[1]}, \\
\sigma^{[1]}\times\sigma^{[1]}&=I^{[1]}+\psi^{[1]},
\end{align}
and $\{ I^{[2]}, \tau^{[2]}\}$ is the Fibonacci fusion ring satisfying the following relations,
\begin{equation}
\tau^{[2]}\times \tau^{[2]}= I^{[2]}+\tau^{[2]}.
\end{equation}

The IR theory is the Ising CFT $\{ I',\psi', \sigma' \}$, and the flow can be represented as the reduction,
\begin{equation}
\text{Ising} \otimes \text{Fib}\rightarrow \text{Ising}'
\end{equation}
In this flow, all of the primary operators $\Phi_{\psi'}, \Phi_{\sigma'}$ in the IR theory are charged, and the IR theory can be regarded as a (weak-) symmetry-enforced gapless phase. The spin operator $\Phi_{\sigma'}$ is charged under the $\mathbb{Z}_{2}$ operator $\mathcal{Q}_{\psi'}$. This symmetry operator can be interpreted as a global $\mathbb{Z}_{2}$ spin flip operation. The other operator $\Phi_{\psi'}$ is charged under the Kramer-Wannier dual operatior $\mathcal{Q}_{\sigma'}$. More generally, if the relation $\mathbf{A}_{\text{ub}}=\mathbf{A'}$ holds, we expect all the operators are charged under the unbroken symmetry.

\subsection{Tetracritical Ising model to tricritical Ising model}
Next, we consider the RG flow $\mathbf{M}(5,6)\to \overline{\mathbf{M}(4,5)}$ between unitary minimal models, and identify the primary fields which preserve the AGQDs associated with the unbroken symmetry $\text{FR}(SU(2)_{3})$, identical to the $\mathbb{Z}_{2}\otimes \text{Fib}$. More precisely, we look for labels $a'$ in the IR theory $\mathbf{M}(4,5)$ satisfying the relation
\begin{equation}
q_{\alpha',(I')}=q_{\alpha',(a')},
\end{equation}
The symmetry preserved along this flow is $\mathrm{FR}(SU(2)_3)$, and its generator is represented by $\alpha'_{\text{ub,g}}=|1,2|'$. Hence, it is enough to impose the above condition for $\alpha=|1,2|$. The quantum dimension of this object is
\begin{equation}
\text{dim}(\alpha'_{\text{ub,g}})=\varphi:=\frac{1+\sqrt{5}}{2},
\end{equation}
which, by definition, is equal to $q_{\alpha'_{\text{ub,g}},(I')}$. Therefore, we need to find $a'$ such that
\begin{equation}
q_{\alpha'_{\text{ub,g}},(a')}=\varphi.
\end{equation}
Writing $a'=|r',s'|'$, we have
\begin{equation}
q_{\alpha'_{\text{ub,g}},(a')}
=
\frac{S_{|1,2|',|r,'s'|'}}{S_{|1,1|',|r',s'|'}}
=
2(-1)^{r'}\cos\left(\frac{4\pi s'}{5}\right).
\end{equation}
Since
\begin{equation}
\varphi=2\cos\left(\frac{\pi}{5}\right),
\end{equation}
one finds that the solutions, modulo the usual Kac-table identification, are
\begin{equation}
|r',s'|=|1,1|',|3,1|'.
\end{equation}
The solution $|1,1|'$ is trivial, since it is the vacuum. The only nontrivial solution is therefore $|3,1|'$. Its conformal weight is
\begin{equation}
h_{|3,1|'}=\frac{3}{2}>1,
\end{equation}
so this field is irrelevant.

One can confirm the above arguments more intuitvely by representing the IR theory as $\mathbf{M}(4,5)=\text{Ising'}\otimes \text{Fib'}$ and its unbroken symmetry $\mathbf{A}_{\text{ub}}=\mathrm{FR}(SU(2)_3)=\mathbb{Z}_{2}\otimes \text{Fib}=\{ I'^{[1]},\psi'^{[1]}\}\otimes \{ I'^{[2]},\tau'^{[2]}\}$. By focusing on the structure of the tensor product of unbroken symmetry, one can restrict the symmetric bulk field easily in this case. Because of the $\mathbb{Z}_{2}$ symmetry, one can exclude the $\mathbb{Z}_{2}$ charged fields labelled by $\sigma'^{[1]}\otimes \{ I'^{[2]},\tau'^{[2]}\}$, by applying the arguments in the Ising case. Because of the Fibonacci fusion ring symmetry, one can eliminate the fields labeled by $\{ I'^{[1]},\psi'^{[1]},\sigma'^{[1]}\}\otimes \tau'^{[2]}$. Hence, the nontrivial field invariant under the unbroken $\mathbb{Z}_{2}\otimes \text{Fib}$ symmetry is $\{\psi'^{[1]}\otimes I'^{[2]}\}$, and this $\mathbb{Z}_{2}$ simple current is nothing but $\Phi_{|3,1|'}$ with the chiral conformal dimension $3/2$.

\section{(Half-)integer spin nonsimple current in unbroken symmetry: Coset representation and level-rank duality}
\label{section_Higgs}

In this section, we demonstrate the phenomenology of the unbroken symmetry that stabilizes the IR theories, based on the TQFT representations. We stress that most of the discussions in this section can be explained by a variant of the Higgs mechanism\cite{Higgs:1964pj}, whereas the corresponding mathematical framework is still in development. We kindly note a textbook on category theory\cite{etingof2015tensor}, but we focus our attention on phenomenological arguments on the (half-)integer spin nonsimple current in this section, and do not intend to make rigorous or detailed arguments. The main mathematical structures in the present section are two structures: coset and level-rank duality.

First, we consider the following decomposition of $\mathbf{A}=G$, analogous to a Lie group, by $H=\mathbf{A'}$\footnote{More precisely, we study the $\mathbb{C}$-linear realization of the representation category, but to simplify notation, we do not distinguish the Lie group and representation category.},
\begin{equation}
\mathbf{A}=G=K\otimes H
\label{coset_definition}
\end{equation}
and \emph{formally} denote $K$ as $G/H$\footnote{The coset decomposition involves more complicated processes coming from the identification (or orbifolding) of (non)simple current. We expect this difficulty will be resolved by extending (i.e., inverse of orbifolding) the underlying theory by the (non)simple current or by replacing the tensor product $\otimes$ with the box product $\boxtimes$, but this is a future problem. The latter seems a more standard approach in the field, but the explicit calculations of the coefficients seem more involved. Hence, we explain the strategy of the former direction here. For example, the $SU(2)_{1}\otimes SU(2)_{1}$ model has two primary fields with conformal dimension $1/4$, but the tensor product $\left(\left(SU(2)_{1}\otimes SU(2)_{1}\right)/SU(2)_{2}\right)\otimes SU(2)_{2}=\text{Ising} \otimes SU(2)_{2}$ has only $1$ field with conformal dimension $1/4$ in the bosonic representation. At the level of the characters, the identification makes sense, but there exists a subtlety in identifying quantum states only by using the bosonic representations. In the fermionic representation, $\text{Majorana} \otimes SU(2)_{2,f}$, has four primary fields with conformal dimension $1/4$, and the tensor decomposition of the original model $SU(2)_{1}\otimes SU(2)_{1}$ is realizable.}. Equivalently, the setting can be noted as
\begin{equation}
\mathbf{A}_{\text{DW}}=G/H=G\boxtimes \overline{H}.
\label{domain_wall_Deligne}
\end{equation}
where $G$ and $H$ are anologous to the Wess-Zumino-Witten models\cite{Wess:1971yu,Witten:1983tw,Witten:1983ar}.
In this section, we focus on the chirality of objects and distinguish a chiral object $\alpha$ and an antichiral object $\overline{\alpha}$. However, for $\mathbf{A}$ (and its variants such as $\mathbf{A'}$), we do not assume the chirality of the theories. If it is necessary to introduce the chirality of the theory, we introduce another symbol such as $\mathbf{M}$.

The first expression, Eq. \eqref{coset_definition}, comes from the definition of the coset structure, and it is trivial to some extent. To avoid unnecessary complication, we do not provide a detailed expression, and we note a few pioneering references in the Goddard-Kent-Olive (GKO) coset CFTs \cite{Goddard:1984vk,Goddard:1984hg,Goddard:1986ee,Goddard:1988md,Bowcock:1988vs}. The second expression, Eq. \eqref{domain_wall_Deligne}, is more involved, but it is also known more or less phenomenologically in the literature. This kind of relation has already been applied widely in the studies of Higgs phenomena in the name hidden local symmetry in earlier references \cite{Sakurai:1960ju,Bando:1984ej}(also see a review \cite{Bando:1987br} and references therein), and those of GKO coset conformal field theory\cite{Moore:1989yh,Gannon:1994km,Ishikawa:2002wx}\footnote{There exist some canonical relations between a dual group $H^{\text{c}}$ and antichiral one $\overline{H}$. In this kind of dual group, the level of the WZW model takes negative values, and the central charge also becomes negative. Because of this negative sign, the theory has a connection to the antichiral theory. When focusing on the fusion rules and TQFTs, the distinction between $H^{\text{c}}$ and $\overline{H}$ is not relevant, but in CFTs it seems necessary to take the chiral and antichiral parts more carefully.}. In category theory, $\boxtimes$ in the above is the (relative) Deligne product\cite{davydov2011structure,DavydovMugerNikshychOstrik+2013+135+177}. One can see related arguments in \cite{Frohlich:2003hm,Frohlich:2003hg,Cordova:2023jip,Cordova:2025eim}. 

In this formal (or pseudo) coset representation, one can gap out the $\mathbf{A}_{\text{DW}}=G/H$ part of $G$ by interpreting this part as Higgs modes (or domain wall in TQFT). Hence, the formal coset representation implies the existence of the reduction (or RG flow\cite{Cordova:2025eim,Fukusumi:2025xrj}),
\begin{equation}
\rho: G\rightarrow H
\end{equation}

On the other hand, it has been known that the massless RG has been mapped to the conformal interface problem by applying the folding trick and the technique of the RG domain wall. Hence, one can expect that the following mapping between problems should exist
\begin{equation}
\{\rho : G\rightarrow H\} \Leftrightarrow \{d_{(a_{\text{DW}})}: G\boxtimes \overline{H}\rightarrow \mathbb{C}\}
\end{equation}
where $d_{(a_{\text{DW}})}$ is the AGQD of $G/H=G\boxtimes \overline{H}=\{ \alpha_{\text{DW}}\}$, i.e., $d_{(a_{\text{DW}})}(\alpha_{\text{DW}})=q_{\alpha_{\text{DW}},(a_{\text{DW}})}$.
It should be stressed that because of the folding trick, the orientation of the preserved structure $H$ is reversed. The mapping relation can be made rigorous by introducing the AGQD and doing the following calculation,
\begin{equation}
\begin{split}
d_{(a_{\text{DW}})}\otimes I':(G\boxtimes \overline{H})&\otimes H\rightarrow H \\
d_{a_{\text{DW}}}(\alpha_{\text{DW}}) \otimes \alpha'&= q_{\alpha_{\text{DW}},(a_{\text{DW}})}\alpha' 
\end{split}
\end{equation}
Hence, one can obtain the constructive expression $\rho=d_{(a_{\text{DW}})}\otimes I'$.
Because we have assumed the tensor decomposition, the mapping does not preserve the identity object $I\in \mathbf{A}$ and $I'\in \mathbf{A'}$ in general, and this point is different from a homomorphism or a monoidal functor. Because of the central role of $G\boxtimes \overline{H}\rightarrow \mathbb{C}$, one can interprete this mapping as a variant of profuctors in mathematics.

By applying the formal coset representation, one can obtain the following expression of the coupled theory,
\begin{equation}
 G\boxtimes \overline{H}=\left(\left(G/H\right)\otimes H\right) \boxtimes \overline{H} 
\end{equation}
and it is natural to introduce the identification $H\boxtimes \overline{H}=\mathbb{C}$. Here, we interpret the structure $H\otimes \overline{H}$ and its reduction to $H\boxtimes \overline{H}$ in category theory. In category theory, this structure $H\boxtimes \overline{H}$ is called a spherical fusion category, and this is nothing but a condensable block of theory corresponding to modular invariants. Hence, one can interpret the symbol $\boxtimes$ as the anomaly-free parts of the theory described by the (half-)integer spin nonsimple current, called connected \'{e}tale algebra or condensable algebra. 

We summarize the phenomenology of this kind of Higgs transition: If a chiral theory $\mathbf{M}$ flows to a chiral theory $\mathbf{M'}$ without changing the chirality of objects, $\mathbf{M}\otimes \overline{\mathbf{M'}}$ should contain integer-spin nonsimple currents in $\mathbf{M}\boxtimes \overline{\mathbf{M'}}$. $H\boxtimes \overline{H}$ is the corresponding structure in the previous example. In other words, for the unbroken symmetry, the simple objects satisfy the spin relation $h_{\alpha_{\text{ub}}}-h_{\alpha'_{\text{ub}}}=\mathbb{Z}/2$. We stress that in the massless RG in minimal models, the model satisfies the relation $h_{\alpha_{\text{ub}}}+h_{\alpha'_{\text{ub}}}=\mathbb{Z}/2$, and this is different from the structure from the coset construction. For this purpose, we revisit the structure of unbroken symmetry and their braiding (or spin statistics) relations as has been studied in \cite{Gaiotto:2012np}. 

In the massless RG flow  from $\mathbf{M}(p,p+1)$ to $\mathbf{M}(p-1,p)$ (which should be denoted as $\mathbf{M}(p,p+1)\rightarrow \overline{\mathbf{M}(p-1,p)}$ as we demonstrate below), the unbroken symmetry of the UV theory $\mathbf{M}(p,p+1)$ can be identified as
\begin{equation}
\mathbf{A}_{\text{ub}}=\{ \mathcal{Q}_{|u,1|}\}_{u=1}^{p-1}
\end{equation}
Corresponding to these symmetry operators, the conformal dimensions of the primary fields are
\begin{equation}
h_{|u,1|}=\frac{(u-1)^{2}}{4} +\frac{(u^{2}-1)}{4p}
\end{equation}
At the IR, the symmetry operators can be taken as  
\begin{equation}
\mathbf{A'}_{\text{ub}}=\{ \mathcal{Q}_{|1,u|'}\}_{u=1}^{p-1}
\end{equation}
The corresponding conformal dimensions of the primary operators are
\begin{equation}
h_{|1,u|'}=\frac{(u-1)^{2}}{4} -\frac{(u^{2}-1)}{4p}
\end{equation}
Hence, the following relation holds,
\begin{equation}
h_{|u,1|}+h_{|1,u|'}=\frac{(u-1)^{2}}{2}.
\end{equation}
Consequently, the coupled system $\mathbf{M}(p,p+1)\boxtimes \mathbf{M}(p-1,p)$  provides $\text{FR}(SU(2)_{p-2})\boxtimes \text{FR'}(SU(2)_{p-2})$ as a building block of single-valued wavefunctions of topologically ordered systems \footnote{We have included the information of braiding or conformal spin to the theory, we distinguished the UV and IR fusion ring by the prime symbol '}. This kind of coupling between models has already been proposed several times\cite{Kong:2019cuu,Kaidi:2021gbs,Zhang:2024bye,Seo:2026wmq}. In more recent technical term, this generalization of the Deligne product between different systems are called the relative Deligne product\cite{Etingof:2009yvg,Douglas2013DualizableTC,Douglas2014TheBT,Huston:2022utd,Gannon:2026ttf}. By applying the folding trick, this structure implies the massless flow should be interpreted as the flow changing chirality, and it is more approprate to be denoted as $\mathbf{M}(p,p+1)\rightarrow \overline{\mathbf{M}(p-1,p)}$. 

To resolve this puzzle of chirality, it is necessary to implement a condensable structure in $\mathbf{M}\otimes \mathbf{M'}$, not in $\mathbf{M}\otimes \overline{\mathbf{M'}}$. The condensable structure of this kind of chiral-chiral pair has appeared in the level-rank duality in references\cite{Kuniba:1990im,Kuniba:1990zh,Nakanishi:1990hj,Altschuler:1990th,Naculich:1990hg,Aharony:2016jvv,Hsin:2016blu}, and the massless RG flow in the literature corresponds to a generalization of level-rank duality by applying the folding trick. For example, in unitary minimal models, one can apply the following phenomenological relations,
\begin{equation}
\begin{split}
&\mathbf{M}(p,p+1)\boxtimes \mathbf{M}(p-1,p) \\
&=\frac{SU(2)_{p-2}\otimes SU(2)_{1}}{SU(2)_{p-1}} \boxtimes \frac{SU(2)_{p-3}\otimes SU(2)_{1}}{SU(2)_{p-2}} \\
&= \left((SU(2)_{p-2}\otimes SU(2)_{1}) \boxtimes \overline{SU(2)_{p-1}}\right) \\
&\boxtimes \left( (SU(2)_{p-3}\otimes SU(2)_{1}) \boxtimes \overline{SU(2)_{p-2}}\right) \\
&\sim \left( SU(2)_{p-3}\otimes SU(2)_{1}\otimes SU(2)_{1}\right) \boxtimes \overline{SU(2)_{p-1}} \\
&=\frac{ SU(2)_{p-3}\otimes SU(2)_{1}\otimes SU(2)_{1}} { SU(2)_{p-1}}
\end{split}
\end{equation}
where we have used the (conjectural) Witt equivalence coming from $SU(2)_{p-2}\boxtimes\overline{SU(2)_{p-2}}\sim \mathbb{C}$\cite{davydov2011structure,DavydovMugerNikshychOstrik+2013+135+177}. The structure $ \left(SU(2)_{p-3}\otimes SU(2)_{1}\otimes SU(2)_{1}\right) / SU(2)_{p-1}$ reproduces the coset representation in \cite{Crnkovic:1989ug,Gaiotto:2012np}. From this formal expression, it becomes clearer that the unbroken symmetry $\mathbf{A}_{\text{ub}}=\text{FR}(SU(2)_{p-2})$ of the flow $\mathbf{M}(p,p+1)\rightarrow \overline{\mathbf{M}(p,p-1)}$ comes from the coset representation and the anomaly-free structure of the spherical fusion category, an established example of (half-)integer-spin nonsimple currents. However, only by observing the resultant coset structure, the structure of $\text{FR}(SU(2)_{p-2})$ is absent in the numerator and denominator, and the (half-)integer spin nonsimple current is a hidden structure in this representation. We remark that we have not distinguished the coset representations in CFTs and the formal coset representations coming from the Deligne product, but we expect these should match in some proper formalisms as has been discussed in \cite{Frohlich:2003hm,Frohlich:2003hg}.

Alternatively, one can straightforwardly observe the (half-)integer spin nonsimple current structure of massless RG flows, $\text{FR}(SU(2)_{p-2})$, by the existing level-rank duality. There exists another beautiful representation of the minimal model\cite{Blumenhagen:1994ik,Blumenhagen:1994wg},
\begin{equation}
\mathbf{M}(p,p+1)=\frac{SU(p-1)_{2}}{U(1)_{p-2}\otimes SU(p-2)_{2}}
\end{equation}

Hence, one can obtain the following phenomenological relations,
\begin{equation}
\begin{split}
&\mathbf{M}(p,p+1)\boxtimes \mathbf{M}(p-1,p) \\
&=\frac{SU(p-1)_{2}}{U(1)_{p-2}\otimes SU(p-2)_{2}}\boxtimes \frac{SU(2)_{p-3}\otimes SU(2)_{1}}{SU(2)_{p-2}} \\
&= \left(SU(p-1)_{2}\boxtimes \left(\overline{U(1)_{p-2}}\otimes \overline{SU(p-2)_{2}} \right)\right) \\
&\boxtimes \left(\left(SU(2)_{p-3}\otimes SU(2)_{1}\right) \boxtimes \overline{SU(2)_{p-2}}\right) \\
&\sim \left(SU(p-1)_{2}\otimes SU(2)_{p-3}\otimes SU(2)_{1}\right)  \\
&\boxtimes \left(\overline{SU(2)_{p-2}}\boxtimes \left(\overline{U(1)_{p-2}}\otimes \overline{SU(p-2)_{2}} \right)\right) \\
&=\frac{SU(p-1)_{2}\otimes SU(2)_{p-3}\otimes SU(2)_{1}}{SU(2)_{p-2}\boxtimes \left(U(1)_{p-2}\otimes SU(p-2)_{2} \right) }
\end{split}
\end{equation}
It should be stressed that the part $SU(2)_{p-2}\boxtimes \left(U(1)_{p-2}\otimes SU(p-2)_{2} \right)$ admit the level-rank duality\cite{Kuniba:1990im,Kuniba:1990zh,Nakanishi:1990hj},
\begin{equation} 
SU(2)_{p-2}\boxtimes \left(U(1)_{p-2}\otimes SU(p-2)_{2} \right)\subset GL\left( 2\left(p-2\right)\right)_{1}, 
\end{equation}
and this part can produce a theory only with (half-)integer spin primary fields because of their connection to multicomponent fermionic representations\cite{Nakanishi:1990hj}. In other words, the massless RG flow can be interpreted as a consequence of fermion pair creation (or condensation), whereas the Higgs transition is the boson pair creation coming from the spherical fusion category\cite{Bais:2008ni}. We conjecture that the fusion rule $\text{FR}(SU(2)_{p-2})$ appears from this part. Recently, it has been studied that free fermions can possess non-group-like symmetries (or nonabelian anyonic fusion rules)\cite{Wei:2026fsn}. 

By replacing $SU(2)$ with $SU(N)$, one can apply the same arguments to more general massless flows with $W_{N}-$symmetry\cite{Martins:1991hi,Dunning:2002cu,Ambrosino:2026umb}. To our knowledge, studies on the level-rank duality in this context have still been limited\cite{Cordova:2023jip,Cordova:2025zkz}, and some of the $\boxtimes$ in the above expressions might be replaced with $\otimes$ or vice versa depending on the respective methods. More traditionally, this part produces a local part of the theory, satisfying the fermion or boson spin statistics relations. Hence, this part can be interpreted as a building block of the wavefunction of the TQFT, whereas the ``electron" operators satisfy the nonabelian anyonic fusion rules (see \cite{Kong:2019cuu,Kaidi:2021gbs,Bourgine:2024ycr} for related references). In other words, the above expression clarifies the hidden connection to level-rank duality and the unbroken symmetry in the massless RG flows. Moreover, as we have already observed, the corresponding interface in TQFTs (not in CFTs), $\mathbf{M}(p,p+1)\rightarrow \overline{\mathbf{M}(p-1,p)}$, should change the chirality of anyons, as is known as the Alice ring\cite{SCHWARZ1982141,SCHWARZ1982427}.

The physical process that changes the chirality of a particle has already been studied in particle scattering. For example, the CPT theorem permits the nonpreservation of chiralities\cite{Schwinger:1951xk}. Moreover, in the CPT theorem, the spin statistics for bosonic (or fermionic) particles plays the central role\cite{Pauli:1940zz}, and the (half-)integer spin simple nonsimple current is a CFT and TQFT analogue of this phenomenon.

We note a general framework for the unbroken subalgebraic structure $\mathbf{A}_{\text{ub}}$ in a tensor functor $\mathbf{A}\rightarrow \mathbf{A'}$ at this stage. The first thing to note is that this problem is rather model-dependent. Hence, we assume the coset representation $\mathbf{A}=\prod_{i}G_{i}/(\prod_{j}H_{j})=\prod_{i}G_{i}\boxtimes (\prod_{j}\overline{H_{j}})$, where the coset is the GKO coset. For the UV theory, we assume $\prod_{j}H_{j}\subset \prod_{i}G_{i}$, but one can consider further stacking of the antichiral theories. Hence, the expression itself is robust against this modification. To avoid unnecessary compilcations, we express the IR theory as $\mathbf{A'}=\prod_{i}G'_{i}\boxtimes (\prod_{j}\overline{H'_{j}})$ in both cases, and exclude the constraint $\prod_{j}H'_{j}\subset \prod_{i}G'_{i}$.

The observation is as follows:
\begin{itemize}
\item{The chiral-chiral pairs $\prod_{i}G_{i}\prod_{j}H'_{j}$ or antichiral-antichral pairs $\prod_{j}\overline{H_{j}}\prod_{i}\overline{G'_{i}}$ with level-rank duality corresponds to the unbroken symmetry algebra $\mathbf{A}_{\text{ub}}$.} 
\end{itemize}
and/or,
\begin{itemize}
\item{ The chiral-antichiral or antichiral-chiral pair in $\mathbf{A}\otimes \overline{\mathbf{A'}}$ with (generalized) spherical fusion category corresponds to the unbroken symmetry algebra $\mathbf{A}_{\text{ub}}$.}
\end{itemize}
Compared with the former case, the latter case is more involved, because of the mixing of the chirality. Moreover, in principle, the (half-)integer spin nonsimple current condition does not require the difference between chiral central charge and antichiral central charge to be $0$. For example, one can consider pairing of a chiral CFT and a different antichiral CFT, but with the same fusion rule(see \cite{Harvey:2019qzs}, for example). We address this problem in the forthcoming paper.

\begin{figure}[htbp]
\begin{center}
\includegraphics[width=0.5\textwidth]{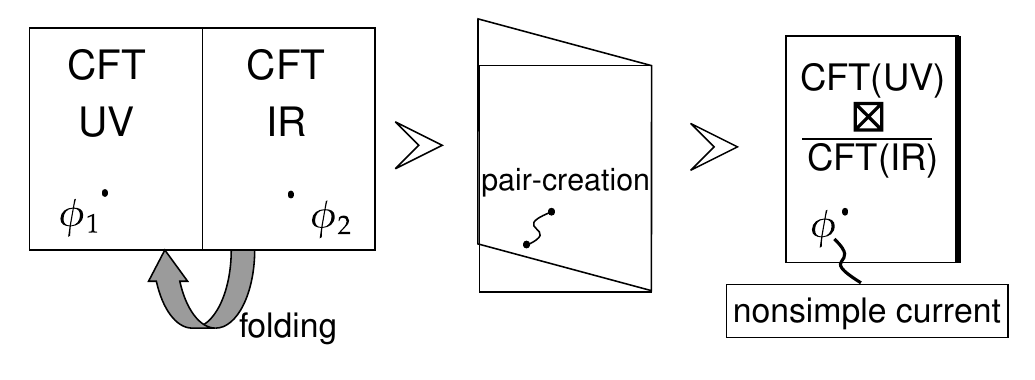}
\caption{This figure shows how the (half-)integer spin nonsimple current emerges after folding. This current corresponds to the unbroken symmetry $\textbf{A}_{\text{ub}}$ through the flow. This is an analogous phenomenon to the BCS pair creation in condensed matter physics.}
\label{fig:folding_nonsimple_current}
\end{center}
\end{figure}

In both cases, the unbroken symmetry $\mathbf{A}_{\text{ub}}\sim \mathbf{A}_{\text{ub}}\boxtimes \overline{\mathbf{A'}_{\text{ub}}}$ will correspond to the (half-)integer spin nonsimple current (see Figure \ref{fig:folding_nonsimple_current}), or nonabelian anyonic analog of the Cooper pair. Hence, they will form single-valued wavefunctions producing  generalized ``electron" operators with anyonic fusion rules isomorphic to $\mathbf{A}_{\text{ub}}$, which can be called (half-)integer spin nonsimple current, but with bosonic or fermionic spin-statistics with (half-)integer conformal spin. The other nonpreserving parts will correspond to ``quasihole operators"\footnote{More precisely, to make the wavefunction single-valued for the insertion of the quasihole operators, it is necessary to attach some additional flux to the theory. However, this procedure can introduce complications involving the mock modularity, and this is out of the scope of the present manuscript.}. Hence, their spin-statistics can be outside of bosonic or fermionic ones. One can see related observations in \cite{Zhang:2024bye,Seo:2026wmq}, and the (half-)integer spin nonsimple current has been studied as a condensable algebra or connected \'{e}tale algebra. However, to our knowledge, their connection to the GKO coset or the level-rank duality has not been studied sufficiently, whereas the older studies \cite{Frohlich:2003hm,Frohlich:2003hg} have close connections to these frameworks. 

We note references on the $\lambda$ deformation, which provides massless RG flows between coset models\cite{LeClair:2001yp,Georgiou:2017jfi,Sfetsos:2017sep,Georgiou:2018gpe}. For further quantum field theoretic understanding of the properties of the corresponding domain walls in CFTs, the research on the reflection or transmission properties of the corresponding conformal interfaces is fundamental\cite{Quella:2006de,Kimura:2014hva,Kimura:2015nka}. As earlier references on conformal interfaces and their application to condensed matter, we note \cite{Affleck:1990by,Kane:1992xse,Kane:1992zza,Wong:1994np,Affleck:1995ge,Fendley_1995,PhysRevLett.77.2604} and related reviews\cite{Saleur:1998hq,affleck2009quantumimpurityproblemscondensed}. Recently, the key structure has been studied as ``phantom" current \cite{Antinucci:2025uvj,Furuta:2025ahl}. However, we remark that this technical term should be distinguished from a more familiar technical term, the phantom field in quantum field theories, which has been widely applied to astrophysics\cite{Caldwell:1999ew} (Moreover, the term ``phantom current" has already been used in laser physics\cite{HOGAN1995570}). In this manuscript, we call it spin-$2$ chiral-chiral nonsimple current. The RG domain wall can be identified as a consequence of a generalized version of the level-rank duality\cite{Fukusumi:2025xrj,Fukusumi:2025fvb}, as we have shown. However, it should be remarked that the corresponding hidden currents have already been studied in \cite{Lashkevich:1993fb}, and the related currents from the coset construction of the CFT have also been studied widely by using the Coulomb gas representations\cite{Dotsenko:1984ad,Dotsenko:1984nm}, known as the Feigin-Fuchs construction\cite{Feigin:1981st} (for a related impurity problem, see \cite{Bernard:2014qia}). In the next section, we clarify the role of the spin-$2$ chiral-chiral nonsimple current in the unfolded model when adding the corresponding perturbations to the UV Hamiltonian.

\subsection{General domain wall phenomena: Algebraic generalized quantum dimensions as domain wall particles}
In this subsection, we summarize the relationship between the folding trick and implications for the domain wall problem when assuming the tensor decomposition of the UV theory. The content is a continuation of the research direction in Appendix B of \cite{Fukusumi:2025clr}, combining the ideas in \cite{Frohlich:2003hg,Frohlich:2003hm,Fukusumi:2025xrj}.
The following observation\cite{Fukusumi:2025xrj}, which is analogous to the coset and level-rank duality, will be relevant.
\begin{equation}
\begin{split}
&\text{If UV theory can be expressed as $\mathbf{A}=\mathbf{A}_{(\text{DW})}\otimes \mathbf{A'}$} \\
&\text{$d_{a_{\text{DW}}}\otimes I' $ produces the tensor functor $\rho$.}
\end{split}
\end{equation}
where $\mathbf{A}_{\text{DW}}$ is a domain wall which provides a decomposition of the UV theory, and $a_{\text{DW}}$ is the anyon at the domain wall (or the domain wall excitation). In category theory $d_{a_{\text{DW}}}$ can be interpreted as a variant of profunctors. Because of the tensor decomposition, the identity $I\in \mathbf{A}$ is not mapped to $I'\in \mathbf{A'}$ in general, and this point contrasts with a monoidal functor or a ring homomorphism. Interestingly, there exists a general method for constructing monoidal functor, by focusing on the idempotents and AGQDs directly \cite{Fukusumi:2025fvb}. The precise relation between the argument in this section and the ring theoretic method (or monoidal functor) is an interesting research direction.

The domain wall should be interpreted as a dual vector, mapping $\mathbf{A}$ to $\mathbf{A'}$, and $a_{\text{DW}}$ characterizes the types of the domain walls. The expression by the Deligne product is,
\begin{equation}
\mathbf{A}_{\text{DW}}=\mathbf{A}\boxtimes \overline{\mathbf{A'}}
\end{equation}
The setting $\mathbf{A}=G$ and $\mathbf{A'}=H$ corresponds to the coset construction of CFT, and $\mathbf{A}=G$ and $\mathbf{A'}=\overline{H}$ corresponds to the level-rank duality, specifically, but one can apply the formalism to more general settings. One can add lower index $(a_{\text{DW}})$, and introduce the modified notation $\mathbf{A}_{\text{DW},(a_{\text{DW}})}=(\mathbf{A}\boxtimes \overline{\mathbf{A'}})_{(a_{\text{DW}})}$, but we drop the lower index to simplify the notation.  By introducing the Deligne product, the decomposition of the UV theory can be reexpressed as,
\begin{equation}
\mathbf{A}=(\mathbf{A}\boxtimes \overline{\mathbf{A'}})\otimes \mathbf{A'}
\end{equation}
In other words, the relative Deligne product \emph{should} provide domain wall or tensor functor $\rho: \mathbf{A}\rightarrow \mathbf{A'}$, when the AGQD $d_{(a_{\text{DW}})}: \mathbf{A}\boxtimes \mathbf{A'}\rightarrow \mathbb{C}$ is well-defined. In this setting, one can express the tensor functor as,
\begin{equation}
\rho=d_{(a_{\text{DW}})}\otimes I'.
\end{equation}

In this manuscript, we focused on the case where the AGQD can be considered as mapping to $\mathbb{C}$ or \emph{measurement} in experimental settings, but one can consider more general AGQD $\mathbf{A}_{\text{DW}}\rightarrow\mathbb{K}$ where $\mathbb{K}$ is a number field or rings (or a set more generally). The structure $\mathbb{K}$ corresponds to charges or domain wall particles (i.e., nonabelian anyons, depending on the respective settings) or \emph{observer} by identifying $\rho$ as a measurement, and this can be regarded as a charged domain wall in physics\cite{Callan:1984sa}. In this case, by modifying the AGQD of the domain wall $d_{(a_{\text{DW}})}:\mathbf{A}\boxtimes \overline{\mathbf{A'}}\rightarrow \mathbb{K}$, one can obtain the homomorphism or mapping
\begin{equation}
d_{(a_{\text{DW}})}\otimes I': \mathbf{A}\rightarrow \mathbb{K}\otimes \mathbf{A'}.
\end{equation}
This generalization is called a profunctor, and its network has been proposed in \cite{Yue:2026qjl}.

For example, by choosing the models with $\mathbb{Z}_{N}$ symmetry and taking $\mathbb{K}$ as the $\mathbb{Z}_{N}$ group ring, the $\mathbb{Z}_{N}$ preserving homomorphism can be interpreted as a charged domain wall. The setting has a close connection to the charged sector (or modular covariant) of the $\mathbb{Z}_{N}$ extended models\cite{Fukusumi:2024ejk,Fukusumi:2025ljx}. For the $\mathbb{Z}_{2}=\{ I_{\text{DW}}, \psi_{\text{DW}}\}$ case, one can observe the UV Neveu-Schwarz (NS) sector $\alpha$ ($\mathbb{Z}_{2}$ charge 0) can contain a tensor product of Ramond (R) sectors $\alpha'$, $\alpha_{\text{DW}}$ ($\mathbb{Z}_{2}$ charge 1) of the IR theory and the domain wall, because of the relation $1+1=0 \ (\text{mod.}2)$. In a profuctor, $\alpha$ can be mapped to $\psi_{\text{DW}}\otimes \alpha'$ even when assuming the $\mathbb{Z}_{2}$ charge preservation, but the mapping $\alpha\rightarrow \alpha'$, from NS to R, is prohibitted in the usual $\mathbb{Z}_{2}$ charge presering tensor functor. In other words, the profunctor defines both gapped and charged domain walls\footnote{It should be remarked that the charge preservation is broken in charged domain walls, and there exists some subtlety in interpreting the phenomena in RGs. However, as a domain wall between TQFTs, it is well-defined.}. By further assuming the existence of the tensor functor, $\mu: \mathbb{K}\otimes \mathbf{A'} \rightarrow \mathbf{A'}$ or $\boxtimes$ structure $\mathbb{K}=\mathbf{A'}\boxtimes \overline{\mathbf{A'}}$, representing  the action (or absorption) of domain wall particle $\mathbb{K}$ to the IR theory $\mathbf{A'}$, one can obtain the tensor functor as
\begin{equation}
\mu\circ (d_{(a_{\text{DW}})}\otimes I'):\mathbf{A}\rightarrow \mathbf{A'}.
\end{equation}

In other words, if the domain wall cannot hold the degree of freedom $\mathbb{K}$, the domain wall particle will flow to IR, and the structure of the tensor functor will appear. The trapping of domain wall particles is an interesting problem, and the setting will be realized in quantum Hall point contacts or wire junctions (see reviews\cite{Affleck:1995ge,Saleur:1998hq} and references therein).

\section{Nontriviality of the changes of the conformal dimensions: Cascade of flows from multiple perturbations}
\label{section_cascade}
\subsection{Integer spin nonsimple current and cascade from the resonance effect}

We mainly focused on the symmetry and scaling property of the IR theory $\overline{\mathbf{M}(p-1, p)}$, by fixing the unbroken symmetry $\mathrm{FR}(SU(2)_{p-2})$. In this section, we discuss those in the UV theory and demonstrate the possibility of the emergence of relevant fields at (intermediate) IR. 

First, we introduce the results on the $\mathrm{FR}(SU(2)_{p-2})$ symmetric field in \cite{Gaiotto:2012np}. In the UV theory $\overline{\mathbf{M}(p-1,p)}$, the set of $\mathrm{FR}(SU(2)_{p-2})$ symmetric bulk fields is 
\begin{equation}
\{ \Phi_{1,s}\}_{s:\text{odd}}.
\end{equation}
The most relevant field in this set is $\Phi_{|1,3|}$, and the other fields are irrelevant. What we have shown in the previous sections is the existence of the obstruction from symmetry which prohibits these irrelevant fields from flowing to relevant fields at the IR $\overline{\mathbf{M}(p-1,p)}$. Hence, the field $\Phi_{|1,3|}$ can be interpreted as a most fundamental operator determining the property of the flow. 

One may expect that a field becomes more irrelevant under the flow because of its connection to thermalization or stabilization\cite{Zamolodchikov:1986gt}. For a field $\Phi_{\alpha}\rightarrow \Phi_{\alpha'}$, one may expect 
\begin{equation}
h_{\alpha}<h_{\alpha'}
\end{equation} 
through an RG flow or corresponding application of the RG domain wall. Hence, the prohibition of the existence of relevant fields in $H'_{\text{pert}}$ from the irrelevant fields in $H_{\text{pert}}$ may seem trivial from this naive observation. However, this observation is not true in general. Even in the RG flow we studied, there exist cases where a field becomes more relevant through the RG flow. Moreover, there can exist cases where irrelevant operators at UV become relevant at IR. The same phenomena have been studied in supersymmetric models\cite{Seiberg:1994pq,Kutasov:1995np,Kutasov:1995ve,Intriligator:1995ff,Kutasov:1995ss,Leigh:1996ds} and the walking in the weak-first order transitions\cite{Gorbenko:2018ncu}, and we interpret the effect as a resonance effect from the coexistence of relevant and irrelevant terms. This resonance effect can result in the cascade of phase transitions $H\rightarrow H' \rightarrow H''$ as we demonstrate below. The fields labelling the unbroken symmetry $\mathbf{A}_{\text{ub}}$ are a fundamental structure producing the resonance effect and cascade of phase transitions. By introducing only these irrelevant terms, the UV Hamiltonian does not flow, but the system flows nontrivially with the coexistence of the relevant perturbation. For the flow $\mathbf{M}(p,p+1)\rightarrow \overline{\mathbf{M}(p-1,p)}$ triggered by $\Phi_{|1,3|}$, the field $\Phi_{|3,1|}$ is the corresponding irrelevant perturbation. 

As we have already introduced in Sec. \ref{section_Higgs}, the UV unbroken symmetry can be written as 
\begin{equation}
\mathbf{A}_{\text{ub}}=\{ \mathcal{Q}_{|u,1|}\}_{u=1}^{p-1}
\end{equation}
The corresponding operators $\{ \Phi_{|u,1|}\}_{u=1}^{p-1}$ have the chiral conformal dimensions,
\begin{equation}
h_{|u,1|}=\frac{(u-1)^{2}}{4} +\frac{(u^{2}-1)}{4p}.
\end{equation}
In IR, the symmetry operator can be written as
\begin{equation}
\mathbf{A'}_{\text{ub}}=\{ \mathcal{Q}_{|1,u|'}\}_{u=1}^{p-1}
\end{equation}
and the conformal dimensions of the corresponding fields $\{ \Phi_{|1,u|'}\}_{u=1}^{p-1}$ are,
\begin{equation}
h_{|1,u|'}=\frac{(u-1)^{2}}{4} -\frac{(u^{2}-1)}{4p}.
\end{equation}
Hence, this exacly satisfies $h_{\alpha}>h_{\alpha'}$, if the integer spin nonsimple current structure ($h_{\alpha}+h_{\alpha'}=\mathbb{Z}/2$) holds. At the level of the operator, the $\Phi_{|1,3|}$ perturbation maps $\Phi_{|u,1|}\rightarrow \Phi_{|1,u|'}$, and this makes the operator more relevant as a consequence of the flow.

Even in the massless RG flow, the RG can make a UV-irrelevant field relevant through the flow, and this point has not been emphasized in references to our knowledge. The setting with the parameter $u=3$ corresponds to an example, because of the relation 
\begin{equation}
h_{|3,1|}>1>h_{|1,3|'} 
\end{equation}
Interestingly, the pair $\Phi_{|3,1|}\Phi_{|1,3|'}$ plays a central role in constructing the spin-$2$ chiral-chiral nonsimple current\cite{Furuta:2025ahl}. Moreover, the field $\Phi_{|1,3|'}$ triggers the RG flow $\overline{\mathbf{M}(p-1,p)}\rightarrow \mathbf{M}(p-2,p-1)$. In other words, only by introducing the irrelevant perturbation $\Phi_{|3,1|}$ together with $\Phi_{|1,3|}$ at UV, one cannot exclude the possibility of the flow $\mathbf{M}(p,p+1)\rightarrow \mathbf{M}(p-2,p-1)$. In other words, 
\begin{equation}
\begin{split}
&\text{\emph{Universal properties of RGs with multiple perturbations}} \\ 
&\text{\emph{cannot be determined only by the UV relevant fields.}}  
\end{split}
\end{equation}
More specifically, we conjecture 
\begin{equation}
\text{$\Phi_{|1,3|}+\Phi_{|3,1|}$ triggers $\mathbf{M}(p,p+1)\rightarrow \mathbf{M}(p-2,p-1)$}
\label{cascade_example}
\end{equation}
For analytical and numerical tests, it seems necessary to fix the strength of the coupling constants of these two terms, and we leave this as an open problem. The Hamiltonian truncation method will be fundamental in this research direction. There also exists the alternative scenario where the system flows to another fixed point (possibly complex CFTs) at the intermediate stage of the flow $\mathbf{M}(p,p+1)\rightarrow \overline{\mathbf{M}(p-1,p)}$\cite{Gukov:2015qea,Gorbenko:2018ncu}. We expect some interpolation between the cascade and this intermediate change of the flow may occur depending on the respective settings, and do not provide conclusive arguments in the present manuscript.

When focusing on the integrablity of the flow, $\mathbf{M}(p,p+1)\rightarrow \overline{\mathbf{M}(p-1,p)}\rightarrow \mathbf{M}(p-2,p-1)$, the proposed flow has very interesting property. The first flow, $\mathbf{M}(p,p+1)\rightarrow \overline{\mathbf{M}(p-1,p)}$, is nonintegrable because of the coexistence of the irrelevant perturbation $\Phi_{|3,1|}$. However, it recovers the integrability in the flow $\overline{\mathbf{M}(p-1,p)}\rightarrow \mathbf{M}(p-2,p-1)$. In other words,
\begin{equation}
\begin{split}
&\text{The intermidiate flow $\overline{\mathbf{M}(p-1,p)}\rightarrow \mathbf{M}(p-2,p-1)$}\\
&\text{has emergent integrable structures.}
\end{split}
\end{equation}
When applying this observation to the corresponding lattice model, it will signal separation (or fragmentation) of integrable and nonintegrable sectors (see a review\cite{Moudgalya:2021xlu} and references therein). Fortunately, this phenomenon has already been studied as partially solvable models\cite{Matsui_2024}. 

In a general massless flow $\mathbf{A}\rightarrow \mathbf{A'}$ triggered by a relevant perturbation $\Phi_{a_{\text{rel}}}$, one can apply the discussions above.  The spin-$2$ chiral-chiral integer spin nonsimple current $J\overline{J}=\Phi_{\alpha_{\text{c}}}\Phi_{\alpha'_{\text{c}}}$ constructed from the unbroken symmetry $ \mathcal{Q}_{\alpha_{\text{c}}}\in\mathbf{A}$ and $ \mathcal{Q}_{\alpha'_{\text{c}}}\in\mathbf{A'}$, can trigger such cascades, when assuming the UV operator is irrelevant, i.e. $h_{\alpha_{\text{c}}}>1$. One can confirm the relevance of the IR operator $\Phi_{\alpha'_{\text{c}}}$, by the primary relation of the definition of the spin-$2$ nonsimple current current,
\begin{equation}
h_{\alpha_{\text{c}}}+h_{\alpha'_{\text{c}}}=2,
\end{equation}
Hence, combined with the UV relation $h_{\alpha_{\text{c}}}>1$, the relation $h_{\alpha'_{\text{c}}}<1$ holds, and the perturbation $\Phi_{\alpha'_{\text{c}}}$ is relevant in $\mathbf{A'}$. This implies that the addition of the irrelevant perturbation $\Phi_{a_{\text{c}}}$ at UV causes a flow to another model $\mathbf{A''}$. Hence, the generalization of the phenomena can be derived straightforwardly as  
\begin{equation}
\begin{split}
&\text{$\Phi_{a_{\text{rel}}}+\Phi_{\alpha_{\text{c}}}$ with $h_{\alpha_{\text{c}}}>1$} \\ 
&\text{triggers the cascade $\mathbf{A}\rightarrow \mathbf{A'}\rightarrow \mathbf{A''}$.}
\end{split}
\label{cascade_general}
\end{equation}
In the above arguments, we assumed $\Phi_{\alpha_{\text{c}}}$ flows to $\Phi_{\alpha'_{\text{c}}}$ under the relevant perturbation $\Phi_{a_{\text{rel}}}$, and this is a consequence of (half-)integer spin nonsimple current conditions. For example, the monotonic decrease of the conformal dimensions in the nonunitary model has been noticed in \cite{Kikuchi:2025sso}, and one can apply the same arguments in the models\cite{Nakayama:2024msv}. 

More generally, one can consider the cascade from the unbroken symmetry and (half-)integer spin nonsimple current condition, but $h_{\alpha_{\text{c}}}>1$ does not imply the condition $h_{\alpha'_{\text{c}}}<1$ straightforwardly. In this sense, the spin-$2$ chiral-chiral nonsimple current is exceptional in the sense that it does not require detailed data of conformal dimensions. From the scaling analysis, the following modification holds,
\begin{equation}
\begin{split}
&\text{$\Phi_{a_{\text{rel}}}+\Phi_{\alpha_{\text{c}}}$ with $h_{\alpha_{\text{c}}}>1$ and $h_{\alpha'_{\text{c}}}<1$} \\ 
&\text{triggers the cascade $\mathbf{A}\rightarrow \mathbf{A'}\rightarrow \mathbf{A''}$.}
\end{split}
\label{cascade_higher_spin}
\end{equation}
This applies to more general chiral-chiral or chiral-antichiral (half-)integer spin nonsimple current. We stress that the above statement is exact at the level of scaling analysis (1) by interpreting the effect of UV-relevant perturbation $\Phi_{a_{\text{rel}}}$ as an RG domain wall $\rho$, i.e., 
\begin{equation}
\left(H_{\text{CFT}}+\Phi_{a_{\text{rel}}}\sim \rho\right): \Phi_{\alpha_{\text{c}}}\rightarrow \Phi_{\alpha'_{\text{c}}},
\end{equation} 
and (2) by applying the perturbative formalism to the consequent $H'$ with the relevant perturbation $\Phi'_{\alpha_{\text{c}}}$, i.e.,
\begin{equation}
\text{$\Phi'_{\alpha_{\text{c}}}$ with $h_{\alpha'_{\text{c}}}<1$ triggers $H'\rightarrow H''$}
\end{equation}
This combination of the RG domain wall and perturbative analysis can be interpreted as a modern version of the discussion on the dangerously irrelevant perturbations, as can be seen in \cite{Seiberg:1994pq,Kutasov:1995np,Kutasov:1995ve,Intriligator:1995ff,Kutasov:1995ss,Leigh:1996ds}. 

For further concrete understanding, we note arguments focusing on the perturbative methods. Let us assume the $H_{\text{pert}}$ is decomposed into the summation of the relevant term $H_{\text{pert,rel}}$ and irrelevant term $H_{\text{pert,irr}}$. The introduction of an irrelevant field $H_{\text{pert,irr}}$ will break the UV fusion ring irrelevantly in $H_{\text{CFT}}+H_{\text{pert,irr}}$. One may expect this small breaking of symmetry does not affect the unbroken symmetry structure from the relevant fields $H_{\text{pert,rel}}$, but this is not true. Even when assuming the system is first flowed by the relevant perturbations $H_{\text{pert,rel}}$ and arrives at the expression $H'=H'_{\text{CFT}}+H'_{\text{pert}}$, the breaking of symmetry (or ordering) can be enhanced through the flow. This is because $H'$ can further flow to another CFT because of the effect of $H_{\text{pert,irr}}$, $H''=H''_{\text{CFT}}+H''_{\text{pert}}$.

Conventionally, one can summarize the flows as
\begin{align}
H_{\text{CFT}}+H_{\text{pert,irr}}&\rightarrow H_{\text{CFT}},\\
H_{\text{CFT}}+H_{\text{pert,rel}}&\rightarrow H'_{\text{CFT}}+H'_{\text{pert,irr}} \rightarrow H'_{\text{CFT}},\\
H\rightarrow H'_{\text{CFT}}+H'_{\text{pert}}&\rightarrow H''_{\text{CFT}}+H''_{\text{pert,irr}}\rightarrow H''_{\text{CFT}}
\end{align}
The emergent symmetry will appear corresponding to the vanishing of irrelevant parts of the above processes, and the determination of these emergent symmetries requires further analysis.
Hence, one can state
\begin{equation}
\begin{split}
&\text{The UV descriptions, $H_{\text{CFT}}+H_{\text{pert,irr}}$, $H_{\text{CFT}}+H_{\text{pert,rel}}$,} \\
&\text{ and $H$,  can result in different IR theories.}
\end{split}
\end{equation}

This phenomenon seems unconventional, but it does not contradict the $c$-theorem\cite{Zamolodchikov:1986gt} or the law of thermodynamics because the emergent relevant field triggers the RG flow. When studying the realization of CFTs and their RG flows in the lattice models, this effect from the coexistence of relevant and irrelevant fields is worth further study. The most fundamental phenomenon in the above discussions is,
\begin{equation}
\begin{split}
&\text{Effects from the irrelevant perturbations $H_{\text{pert,irr}}$ can be} \\
&\text{\emph{enhanced} by the relevant perturbations $H_{\text{pert,rel}}$.}
\end{split}
\end{equation}
Extraction of unnecessary irrelevant fields can be fundamental in realizing a particular RG flow $H_{\text{CFT}}\rightarrow H'_{\text{CFT}}$, because the introduction of a relevant field can cause the undesired cascade of the flow $H_{\text{CFT}}\rightarrow H''_{\text{CFT}}$. Our organization by generalized symmetry will be more important, because we have prohibited such a cascade for the simplest flows. Here, for further numerical or combinatorial studies, we note several pioneering works\cite{Andrews:1984af,Huse:1984mn,Kuniba:1990ci,Chui:2001kw,Chui:2002bp} and recent ones\cite{Belletete:2018eua,Belletete:2020gst,Aasen:2020jwb,Pearce:2025aqh} (for references in tensor-network method, see the references in the reviews \cite{Cirac:2020obd,Vancraeynest-DeCuiper:2025msv}).

Studies on the effect of multiple perturbations have still been limited. We also note \cite{Gukov:2015qea,Gukov:2016tnp,Konechny:2023xvo} as studies containing a general discussion, and as those on the double sine-Gordon (or related) model\cite{Delfino:1997ya,Toth:2004bi}. The nontrivial phenomena from the coexistence of marginal perturbation and irrelevant perturbation have been studied in \cite{Fukusumi:2021qwa}.

\subsection{Unstoppable cascade from perturbative results: Nontrivial result in the minimal models}
In the previous subsection, we focused on the transformation law of the operator $\Phi_{|3,1|}\rightarrow \Phi_{|1,3|'}$. This law itself is evident from both the integer spin nonsimple current condition and the perturbative calculations only by admitting the flow $\mathbf{M}(p,p+1)\rightarrow \overline{\mathbf{M}(p-1,p)}$. In this section, we comment on the more nontrivial effect when assuming the perturbative calculation on the transformation law of the relevant operator $\Phi_{|1,3|}$, which triggers the flow.  

Based on the perturbative calculation and RG domain wall in \cite{Zamolodchikov:1987ti,Gaiotto:2012np}, one can see that $\Phi_{|1,3|}$ flows to $\Phi_{|3,1|'}$. Based on this perturbative results and the transformation law $\Phi_{|3,1|}\rightarrow \Phi_{|1,3|'}$, following transformation in the flow $\mathbf{M}(p,p+1)\rightarrow \overline{\mathbf{M}(p-1,p)}$ appears,
\begin{equation} 
\Phi_{|1,3|}+\Phi_{|3,1|}\rightarrow \Phi_{|1,3|'}+\Phi_{|3,1|'}
\end{equation}
In other words, when admitting this perturbative result, the cascade of RG flows never stops at any value of $p$ of the UV model $\mathbf{M}(p,p+1)$. Finally, at the Ising point $\mathbf{M}(3,4)$, the flow becomes massive, and the system flows to the $\mathbb{Z}_{2}$ symmetric disordered phase by the energy operator $\Phi_{\epsilon}$ with chiral conformal dimension $1/2$ (and stress tensors $T\overline{T}$). In short,
\begin{equation}
\begin{split}
&\text{$\Phi_{|1,3|}+\Phi_{|3,1|}$ can trigger the cascade} \\
&\text{$\mathbf{M}(p,p+1)$ to $\mathbb{Z}_{2}$ disordered phase.}
\end{split}
\end{equation}

\begin{figure}[htbp]
\begin{center}
\includegraphics[width=0.5\textwidth]{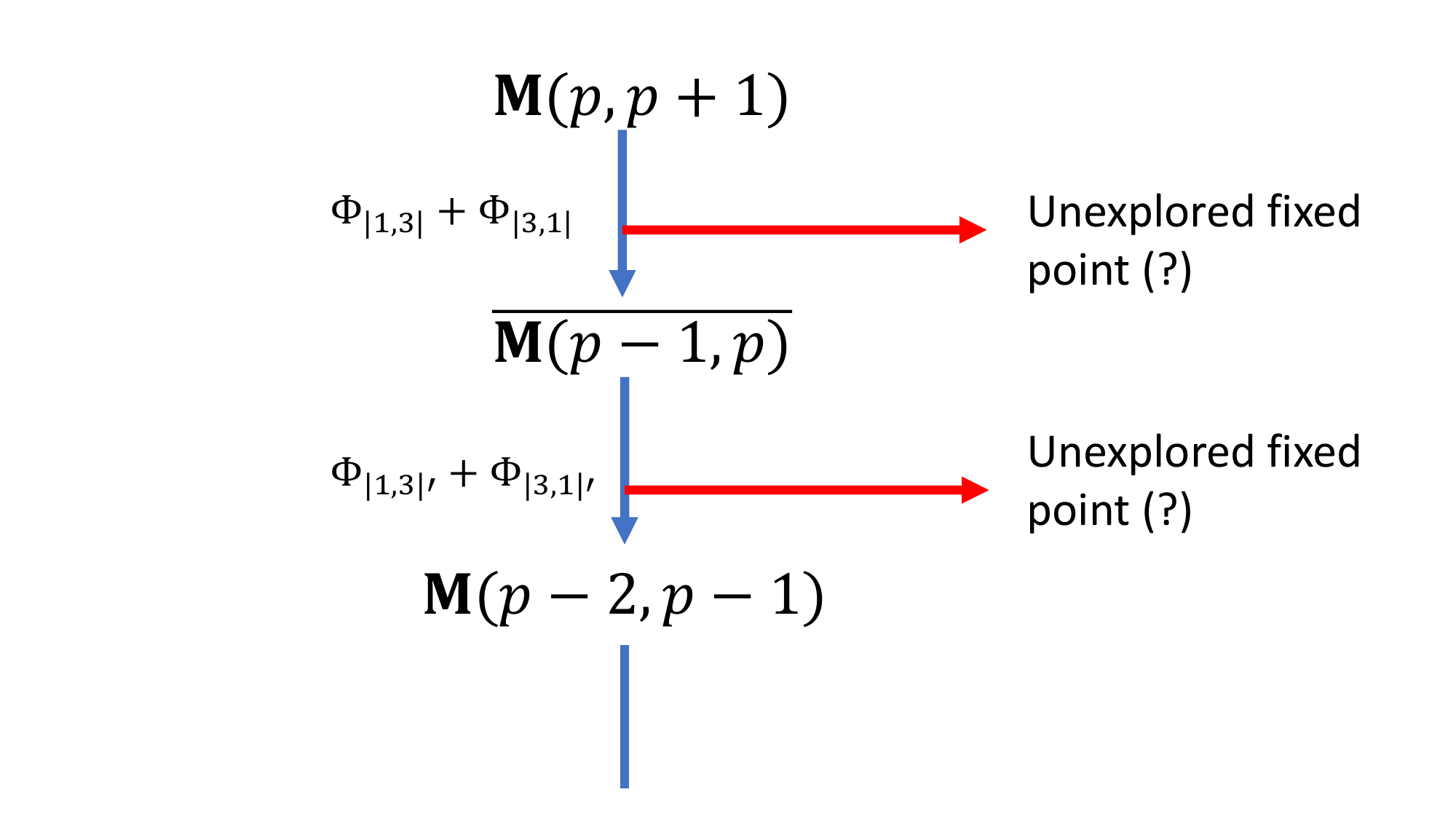}
\caption{Possible scenario for the resonance effect between relevant perturbations and irrelevant perturbations. If one believes the RG domain wall and perturbative analysis, the cascade will appear. On the other hand, a variant of the Morse theory in \cite{Gukov:2015qea} may indicate a new series of fixed points. Their relation to complex fixed points or weak-first-order transitions is an interesting problem\cite{Gorbenko:2018ncu}. The existence of the first-order transitions in the lattice models has been pointed out in \cite{Huse:1984mn} with a remark on the $Q$-state Potts models with $Q>4$. More or less, both cascades and the first-order transition seem unconventional when applying CFTs, and further clarification by other methods is necessary.}
\label{fig:cascade}
\end{center}
\end{figure}

This phenomenon is interesting in the sense that the role of the relevant perturbation and the irrelevant perturbation exchanges each step of the flow. One can straightforwardly generalize this argument. First, let us assume $H_{\text{CFT}}^{(i)}\rightarrow H_{\text{CFT}}^{(i+1)}$ triggered by a relevant perturbation $\Phi_{a_{\text{rel}}}^{(i)}$ for $i=0$ to $i=N_{\text{f}}$. At each step of the flow, we assume the following transformation law,
\begin{equation}
\begin{split}
\Phi_{a_{\text{rel}}}^{(i)}&\rightarrow \Phi_{a_{\text{irr}}}^{(i+1)} \\
\Phi_{a_{\text{irr}}}^{(i)}&\rightarrow \Phi_{a_{\text{rel}}}^{(i+1)} \\
\end{split}
\end{equation}
The integer spin current condition provides a way to construct such exchange of the role of relevant and irrelevant perturbations.
This condition exchanges the role of relevant and irrelevant operators recursively until $i=N_{\text{f}}$. Hence 
\begin{equation}
\begin{split}
&\text{  $\Phi_{a_{\text{rel}}}^{(0)}+\Phi_{a_{\text{irr}}}^{(0)}$ can trigger the cascade} \\
&\text{$H^{(0)}_{\text{CFT}}$ to $H_{\text{CFT}}^{(N_{\text{f}})}$.}
\end{split}
\end{equation}
Numerical or analytical studies on the respective settings are worth further study. 

We also note that the appearance of such resonance effects has been pointed out to be rare in \cite{Gukov:2015qea}. Interestingly, it has been argued that a different RG flow will appear when an irrelevant perturbation becomes marginal at the continuous RG trajectory by the relevant perturbation. Hence, we do not make a conclusive statement. We only remark that there may exist an interpolation between such an emergent marginal perturbation in \cite{Gukov:2015qea} and the cascade in the present manuscript (Figure \ref{fig:cascade}). 

\section{Conclusion}
\label{section_conclusion}
In this manuscript, we have demonstrated that the RG flow between unitary minimal CFTs can be stabilized by assuming a particular class of unbroken fusion ring symmetry $\text{FR}(SU(2)_{p-2})$. In more recent terminology in theoretical physics, one can state the massless RG flows as a sequence of (weak-)symmetry-enforced gaplessness\cite{Wang:2014lca,Wang:2016gqj,Sodemann:2016mib,Wang:2017txt,Kim:2025tzc} by the fusion ring symmetry $\text{FR}(SU(2)_{p-2})$. Related to this point, there exists a related phenomenon called strong-symmetry enforced gaplessness\cite{Kim:2025tzc,Ando:2026ffy}, and the strong version of our arguments is an interesting research direction.

Moreover, the phenomenological context of this unbroken fusion ring symmetry has also been explained by combining the coset representation of CFTs and the folding trick. We clarified the structure as a (half-)integer spin nonsimple current in the level-rank duality. For future study, more explicit studies of the corresponding (half-)integer spin nonsimple current will be important. This structure is closely related to connected \'{e}tale algebra or condensable algebra, in the literature on TQFTs, but its relation to the coset construction and the level-rank duality in CFTs has not been studied sufficiently to our knowledge. For this purpose, the coupling or pairing of dual coset models will be fundamental\cite{Bowcock:1988vs,Altschuler:1988mg,Bouwknegt:1992wg,Blumenhagen:1994ik,Blumenhagen:1994wg}. We expect that one will obtain a series of chiral topological orders with electron operators satisfying the nonabelian anyonic fusion rules, by stacking two chiral CFTs related by massless RGs and dual coset representations. We emphasize that the rigorous categorical arguments of the phenomenology in Sec. \ref{section_Higgs} are still in development. In particular, the relative Deligne product of two different theories and its implications for profunctors are not straightforward.

Finally, we remark on the consequence from Sec. \ref{section_cascade}. The main point we clarified at the level of symmetry and scaling analysis is the following observation on the dangerously irrelevant perturbations\footnote{We kindly remind again that we studied the dangerously irrelevant perturbation in quantum field theory, but it is different from that in statistical physics.};
\begin{equation}
\begin{split}
&\text{Irrelevant perturbations can change the IR universalities} \\
&\text{under the existence of relevant perturbations.}
\end{split}
\end{equation}
Moreover, we have clarified the fundamental role of spin-2 chiral-chiral nonsimple current in the folded model. When considering lattice models, this effect of cascade of flows is inevitable without restricting the RG by symmetries. In other words, the unbroken symmetry can be interpreted as a \emph{stopper} prohibiting the flow of cascades. One can observe the analogy to this resonance mechanism and domino mechanism of the cascade of phase transitions, more generally. The symmetry provides a constraint on the resultant quantum phase even in the domino effects.

\section{Acknowledgement}
Yoshiki Fukusumi thanks Shuma Nakashiba and Taishi Kawamoto for related collaborations and helpful comments. He also thanks Takamasa Ando for the discussion on the symmetry-enforced gaplessness, Justin Kaidi for the discussion on integer spin nonsimple current in string theories, and Yunqin Zheng for reminding us of the $W-$symmetric models and nontrivial structures in fermionic models. He thanks the support from NCTS.

\appendix

\section{Short review on topological symmetry in minimal models}
\label{section_review}
\subsection{Modular property and topological symmetry}
In this section, we introduce some basic aspects of minimal conformal field theories and their topological symmetries. Related arguments can be seen in the corresponding review parts of the works by the authors and collaborators\cite{Fukusumi:2025clr,Fukusumi:2025xrj}. We note a few textbooks and reviews for references\cite{Ginsparg:1988ui,DiFrancesco:1997nk,Recknagel:2013uja,Northe:2024tnm}.

First, let us introduce the partition function of the diagonal $A$-series model as follows:
\begin{equation}
Z(\tau,\overline{\tau})=\sum_{\alpha\in \mathbf{A}} |\chi_{\alpha}(\tau)|^{2}
\end{equation}
where $\tau$ ($\overline{\tau}$) is the (anti)holomorphic modular parameter, $\chi$ is the (anti)chiral character, $\alpha\in \mathbf{A}$ is the label of primary fields, and $H_{CFT}$ is the Hamiltonian of CFT. We also assume the chiral character can be written as 
\begin{equation}
\chi_{\alpha}(\tau)=\sum_{M} \langle \alpha,M|e^{2 \pi i\tau (L_{0}-c/24)}|\alpha, M\rangle
\end{equation}
where $L_{0}$ is a generator of the Virasoro algebra, $M$ is the label of descendant fields, and $c$ is the central charge. We kindly note the relation of the Virasoro generators $\{ L_{m}\}_{m\in \mathbb{Z}}$,
\begin{equation}
[L_{m},L_{n}]= (m-n)L_{m+n}+(m^{3}-m)\delta_{m+n, 0}\frac{c}{12}
\end{equation}
where $\delta$ is the Kronecker delta. We also assume the same structure for the antichiral parts. In this setting one can express the CFT Hamiltonian as $H_{CFT}=L_{0}+\overline{L_{0}}-c/12$ and express the partition function as,
\begin{equation}
Z(\tau)=\text{Tr} (e^{2\pi i H_{CFT}})
\end{equation}
where we have introduced a condition $\overline{\tau}=-\tau$, which often applies to lattice models. In the main text, we mainly focus on the bulk primary field $\Phi_{\alpha}$ corresponding to the modular invariant. The chiral antichiral decomposition of the bulk field can be written as $\Phi_{\alpha}=\phi_{\alpha}\boxtimes \overline{\phi_{\alpha}}$, where $\phi$ ($\overline{\phi}$) is the (anti-) chiral field. The chiral and bulk fields satisfy the same fusion rule in a diagonal model\cite{Moore:1988ss,Moore:1988qv,Moore:1989vd,Fuchs:2002cm}. Hence, we express the fusion ring as $\mathbf{A}=\{\phi_{\alpha}\}=\{ \Phi_{\alpha}\}$, and often denote $\mathbf{A}=\{ \alpha\}$ to focus on the algebraic structure.

The modular transformations $T$ and $S$ act on the modular parameter and chiral characters as follows,
\begin{equation}
\begin{split}
T&: \tau\rightarrow \tau+1 \\
T&(\chi)(\tau)= e^{2 \pi i \tau \left(h_{\alpha}-c/24\right)} \chi_{\alpha} (\tau)
\end{split}
\end{equation}
\begin{equation}
\begin{split}
S&: \tau\rightarrow -1/\tau \\
S&(\chi)(\tau)= \sum_{\beta}S_{\alpha,\beta} \chi_{\alpha} (\tau)
\end{split}
\end{equation}
where $S$ is called the modular $S$ matrix and plays a fundamental role in defining the topological symmetry operators.

Then, we introduce the topological symmetry operators as a linear sum of projections\cite{Petkova:2000ip},
\begin{equation}
\mathcal{Q}_{\alpha}=\sum_{a,M, \overline{M}} \frac{S_{\alpha,a}}{S_{I,a}} |a,M\rangle \langle a,M| \otimes |a,\overline{M}\rangle \langle a,\overline{M}|
\end{equation}
Because a topological symmetry operator $\mathcal{Q}_{\alpha}$ is represented as a linear sum of projections, it commutes with the Hamiltonian, 
\begin{equation}
 [\mathcal{Q}_{\alpha},H_{\text{CFT}}]=0.
\end{equation}
Hence, $\{\mathcal{Q}_{\alpha}\}$ is a particular class of conserved charges in a CFT, and its preservation or breaking by perturbations can lead to a nontrivial quantum phase diagram.
Related to the introduction of the topological symmetry operator, we note the algebraic generalized quantum dimension (AGQD)\cite{Fukusumi:2025xrj,Fukusumi:2025fir},
\begin{equation}
q_{\alpha,(a)}=\frac{S_{\alpha,a}}{S_{I,a}}
\end{equation}

The topological symmetry operators satisfy the following (commutative) fusion rules $\mathbf{A}$, and they are called a fusion ring in abstract algebra. 
\begin{equation}
\mathcal{Q}_{\alpha}\times \mathcal{Q}_{\beta}=\sum_{\gamma}N^{\gamma}_{\alpha,\beta}\mathcal{Q}_{\gamma}
\end{equation}
where $N$ is the fusion coefficient, and it can be determined by the modular $S$ matrices through the following Verlinde formula\cite{Verlinde:1988sn} 
\begin{equation}
N^{\gamma}_{\alpha,\beta}=\sum_{\delta}\frac{S_{I,\alpha}S_{I,\beta}\overline{S_{\gamma,\delta}}}{S_{I,\delta}}
\end{equation}
By definition, the AGQD satisfies the relations $q_{\alpha,(a)}q_{\beta,(a)}=\sum_{\gamma}N_{\alpha,\beta}^{\gamma}q_{\gamma,(a)}$, and this implies the AGQD can be regarded as a quantity couting degrees of freedom and charges of anyons.

By treating this set of conserved charges as a generalization of symmetry, we call it fusion ring symmetry. This expression satisfies the following remarkable relation\cite{Petkova:2000ip},
\begin{equation}
\text{Tr}\left( e^{2\pi i \tau (L_{0}+\overline{L_{0}}-c/12)} \mathcal{Q}_{\alpha} \right)=\sum_{\beta, \gamma} N_{\alpha, \beta}^{\gamma} \chi_{\gamma}(-1/\tau) \overline{\chi_{\beta}} (-1/\overline{\tau})
\end{equation}
In other words, at the level of the partition function, the modular $S$ transformation relates the insertion of the topological symmetry operator $\mathcal{Q}_{\alpha}$ to the addition of the chiral particle $\phi_{\alpha}$. This (formal) addition of the chiral particle can be regarded as the insertion of a topological defect $\mathcal{D}_{\alpha}$ into the system, and the topological symmetry operators form a \emph{particular} set corresponding to the defects. This relation is quite nontrivial in that a coefficient before a symmetry operator $\mathcal{Q}_{\alpha}$ can be a $\mathbb{C}$ number, but that of a topological defect $\mathcal{D}_{\alpha}$ is constrained to a nonnegative integer. More mathematically, generalized symmetry for a fixed or particular model is a paradigm relating NIM-rep to $\mathbb{C}$-linear algebra. This relationship has provided many fruitful research directions connecting physics of conserved charges, anyons, and defects. However, we stress that when treating RG flow and studying the quantum phase transition between models, the conserved charge should be carefully distinguished from topological defects.

The projection $e_{(a)}=\sum_{M,\overline{M}}|a,M\rangle \langle a,M| \otimes |a,\overline{M}\rangle \langle a,\overline{M}|$ is called idempotent satisfying the relation $e_{(a)}e_{(b)}=\delta_{a,b}e_{(a)}$. The idempotent plays a fundamental role in studying the low-energy interpolation properties of symmetries in the massless renormalization group and its analysis by ring homomorphism\cite{Fukusumi:2025fvb} (or monoidal tensor functor classifying fusion rules), but we do not focus on this aspect in the present manuscript. Rather, we focus on the symmetry action of these symmetry operators on primary states,
\begin{equation}
 \mathcal{Q}_{\alpha}|a\rangle=q_{\alpha,(a)}|a\rangle.
\end{equation}
As we demonstrate in the main text, this relation is important when restricting the form of the perturbations under fusion ring symmetries.

\subsection{Elementary data of minimal conformal field theory}

The minimal model is a landmark conformal field theory established in \cite{Belavin:1984vu}. In this subsection, we provide the elementary data of the model that we used in the main text. The model can be characterized by two integer $p,q$ and denoted as $\mathbf{M}(p,q)$, and the central charge of the model is,
\begin{equation}
c=1-6\frac{(q-p)^{2}}{pq}
\end{equation}
The primary fields can be labelled by the Kac indices $|r,s|$ and the chiral conformal dimension of the bulk field $\Phi_{|r,s|}$ can be written as,
\begin{equation}
h_{|r,s|}=\frac{(qr-ps)^{2}-(p-q)^{2}}{4pq}
\end{equation}
Corresponding to the fields or states, the modular matrix can be written as 
\begin{equation}
\begin{split}
&S_{|r_{1},s_{1}|,|r_{2},s_{2}|} \\
&=-\sqrt{\frac{8}{pq}}(-1)^{r_{1}s_{2}+r_{2}s_{1}} \\
&\times \text{sin}\left( \frac{q}{p}r_{1}r_{2}\right) \text{sin}\left( \frac{p}{q}s_{1}s_{2}\right)
\end{split}
\end{equation}

\section{Tanaka-Nakayama flow}

In this section, we study the Tanaka-Nakayama RG flows between non-unitary minimal models\cite{Nakayama:2024msv} and their stability under the fusion ring symmetry $\text{FR}(SU(2)_{q-2})$. We write the flow as
\begin{equation}
\mathbf{M}(kq+I,q)\to \overline{\mathbf{M}(kq-I,q)},
\end{equation}
and set
\begin{equation}
P:=kq-I.
\end{equation}
We assume the following bounds,
\begin{equation}
q\geq 3,\qquad k\geq 1,\qquad 1\leq I\leq q-1,\qquad \gcd(I,q)=1.
\end{equation}
The Kac labels of the IR theory are
\begin{equation}
1\leq r'\leq P-1,\qquad 1\leq s'\leq q-1,
\end{equation}
with the usual identification
\begin{equation}
|r',s'|'\sim|P-r',q-s|' .
\end{equation}
With this convention, the preserved symmetry is generated by the line
\begin{equation}
\alpha'_{\text{ub,g}}=|1,2|',
\end{equation}
which generates the fusion ring $\mathrm{FR}(SU(2)_{q-2})$.

Throughout this section, both the AGQD and the conventional quantum dimension are normalized by the ordinary vacuum
\begin{equation}
I'=|1,1|'.
\end{equation}
For a primary field $a'=|r',s'|'$, the AGQD is defined as
\begin{equation}
q_{|1,2|',(a')}
:=
\frac{S_{|1,2|',a'}}{S_{|1,1|',a'}}.
\end{equation}
Using the modular $S$ matrix of $\mathbf{M}(P,q)$, this becomes
\begin{align}
q_{|1,2|',(|r',s'|')}
&=
(-1)^{r'}
\frac{\sin\left(\frac{2\pi P s'}{q}\right)}
{\sin\left(\frac{\pi P s'}{q}\right)}
\nonumber\\
&=
2(-1)^{r'}\cos\left(\frac{\pi P s'}{q}\right)
\nonumber\\
&=
2(-1)^{r'+ks'}\cos\left(\frac{\pi I s'}{q}\right).
\end{align}
The conventional quantum dimension of the preserved generator is therefore
\begin{equation}
\dim(|1,2|')
:=
q_{|1,2|',(|1,1|')}
=
\frac{S_{|1,2|',|1,1|'}}{S_{|1,1|',|1,1|'}},
\end{equation}
and is given by
\begin{equation}
\dim(|1,2|')
=
2(-1)^{k+1}\cos\left(\frac{\pi I}{q}\right).
\label{eq:TN_vacuum_qdim}
\end{equation}
Unlike the unitary case, this AGQD is not necessarily positive.

We now solve the equality condition
\begin{equation}
q_{|1,2|',(|r',s'|')}
=
\dim(|1,2|').
\label{eq:TN_vacuum_equality}
\end{equation}
The AGQD can be written as
\begin{equation}
q_{|1,2|',(|r',s'|')}
=
2\cos\left(\frac{\pi(Ps'+qr')}{q}\right),
\end{equation}
whereas
\begin{equation}
\dim(|1,2|')
=
2\cos\left(\frac{\pi(P+q)}{q}\right).
\end{equation}
Hence Eq.~\eqref{eq:TN_vacuum_equality} is equivalent to
\begin{equation}
Ps'+qr'\equiv\pm(P+q)\pmod{2q}.
\label{eq:TN_vacuum_congruence}
\end{equation}
Reducing this relation modulo $q$ gives
\begin{equation}
s'=1
\qquad\text{or}\qquad
s'=q-1.
\end{equation}
For $s'=1$, Eq.~\eqref{eq:TN_vacuum_congruence} requires
\begin{equation}
r'\equiv1\pmod2.
\end{equation}
The solutions with $s'=q-1$ are related to these by the Kac-label identification. Therefore, a complete set of inequivalent solutions is
\begin{equation}
\mathcal{S}_{\mathrm{vac}}
=
\left\{
|r',1|'
\ \middle|\ 
1\leq r'\leq P-1,
\quad
r'\equiv1\pmod2
\right\}.
\label{eq:TN_vacuum_general_solution}
\end{equation}

The conformal weight of an IR primary is
\begin{equation}
h_{|r',s'|'}
=
\frac{(qr'-Ps')^2-(P-q)^2}{4Pq},
\end{equation}
and its bulk scaling dimension is
\begin{equation}
\Delta_{|r',s'|'}
=
\frac{(qr'-Ps')^2-(P-q)^2}{2Pq}.
\end{equation}
Consequently, the primary field is relevant if and only if
\begin{equation}
|qr'-Ps'|<P+q.
\label{eq:TN_vacuum_relevance_general}
\end{equation}
For the complete family in Eq.~\eqref{eq:TN_vacuum_general_solution}, this reduces to
\begin{equation}
|qr'-P|<P+q,
\qquad
r'\equiv1\pmod2.
\label{eq:TN_vacuum_relevance_solution}
\end{equation}
Thus the existence of a non-vacuum relevant primary preserving the same AGQD as the ordinary vacuum can be decided directly from Eqs.~\eqref{eq:TN_vacuum_general_solution} and \eqref{eq:TN_vacuum_relevance_solution}.

\subsection{Examples}

We provide two explicit examples in which Eq.~\eqref{eq:TN_vacuum_equality} can be solved completely. In the first example, every non-vacuum solution is irrelevant. In the second example, the solution set contains a non-vacuum relevant primary.

\subsubsection{An example without non-vacuum relevant solutions}

Let us first take
\begin{equation}
q=7,\qquad k=1,\qquad I=2,
\end{equation}
so that $P=5$ and the flow is
\begin{equation}
\mathbf{M}(9,7)\to\overline{\mathbf{M}(5,7)}.
\end{equation}
The AGQD and the conventional quantum dimension are
\begin{align}
q_{|1,2|',(|r',s'|')}
&=
2(-1)^{r'+s'}\cos\left(\frac{2\pi s'}{7}\right),\\
\dim(|1,2|')
&=
2\cos\left(\frac{2\pi}{7}\right).
\end{align}
Equation~\eqref{eq:TN_vacuum_general_solution} gives the complete set of inequivalent solutions,
\begin{equation}
\mathcal{S}_{\mathrm{vac}}
=
\left\{|1,1|',|3,1|'\right\}.
\end{equation}
The non-vacuum solution has
\begin{equation}
\Delta_{|3,1|'}
=
\frac{(7\cdot3-5)^2-(5-7)^2}{2\cdot5\cdot7}
=
\frac{18}{5}>2.
\end{equation}
Therefore, every non-vacuum primary satisfying Eq.~\eqref{eq:TN_vacuum_equality} is irrelevant in this example.

\subsubsection{An example containing a non-vacuum relevant solution}

We next take
\begin{equation}
q=5,\qquad k=2,\qquad I=2,
\end{equation}
so that $P=8$ and the flow is
\begin{equation}
\mathbf{M}(12,5)\to\overline{\mathbf{M}(8,5)}.
\end{equation}
The AGQD and the conventional quantum dimension are
\begin{align}
q_{|1,2|',(|r',s'|')}
&=
2(-1)^{r'}\cos\left(\frac{2\pi s'}{5}\right),\\
\dim(|1,2|')
&=
-2\cos\left(\frac{2\pi}{5}\right)
=
\frac{1-\sqrt{5}}{2}.
\end{align}
Equation~\eqref{eq:TN_vacuum_general_solution} gives the complete set of inequivalent solutions,
\begin{equation}
\mathcal{S}_{\mathrm{vac}}
=
\left\{|1,1|',|3,1|',|5,1|',|7,1|'\right\}.
\end{equation}
Their bulk scaling dimensions are
\begin{equation}
\begin{array}{c|cccc}
|r',s'|' & |1,1|' & |3,1|' & |5,1|' & |7,1|' \\
\hline
\Delta_{|r',s'|'} & 0 & \frac{1}{2} & \frac{7}{2} & 9
\end{array}
\end{equation}
Thus $|3,1|'$ is a non-vacuum relevant primary preserving the same AGQD as the ordinary vacuum, whereas $|5,1|'$ and $|7,1|'$ are irrelevant.

\clearpage

\appendix

\bibliographystyle{ytphys}
\bibliography{weak-sym}

\providecommand{\href}[2]{#2}\begingroup\raggedright\begin{thebibliography}{100}

\bibitem{Cobanera:2009as}
E.~Cobanera, G.~Ortiz, and Z.~Nussinov, {\slshape {Unified approach to Quantum
  and Classical Dualities},}
  \href{http://dx.doi.org/10.1103/PhysRevLett.104.020402}{{\em Phys. Rev.
  Lett.} {\bfseries 104} (2010) 020402},
  \href{http://arxiv.org/abs/0907.0733}{{
  arXiv:0907.0733~[cond-mat.stat-mech]}}.

\bibitem{Cobanera:2011wn}
E.~Cobanera, G.~Ortiz, and Z.~Nussinov, {\slshape {The Bond-Algebraic Approach
  to Dualities},} \href{http://dx.doi.org/10.1080/00018732.2011.619814}{{\em
  Adv. Phys.} {\bfseries 60} (2011) 679--798},
  \href{http://arxiv.org/abs/1103.2776}{{
  arXiv:1103.2776~[cond-mat.stat-mech]}}.

\bibitem{Cobanera:2012dc}
E.~Cobanera, G.~Ortiz, and Z.~Nussinov, {\slshape {Holographic symmetries and
  generalized order parameters for topological matter},}
  \href{http://dx.doi.org/10.1103/PhysRevB.87.041105}{{\em Phys. Rev. B}
  {\bfseries 87} (2013) 041105}, \href{http://arxiv.org/abs/1211.0564}{{
  arXiv:1211.0564~[cond-mat.stat-mech]}}.

\bibitem{Gaiotto:2014kfa}
D.~Gaiotto, A.~Kapustin, N.~Seiberg, and B.~Willett, {\slshape {Generalized
  Global Symmetries},} \href{http://dx.doi.org/10.1007/JHEP02(2015)172}{{\em
  JHEP} {\bfseries 02} (2015) 172}, \href{http://arxiv.org/abs/1412.5148}{{
  arXiv:1412.5148~[hep-th]}}.

\bibitem{McGreevy:2022oyu}
J.~McGreevy, {\slshape {Generalized Symmetries in Condensed Matter},}
  \href{http://dx.doi.org/10.1146/annurev-conmatphys-040721-021029}{{\em Ann.
  Rev. Condensed Matter Phys.} {\bfseries 14} (2023) 57--82},
  \href{http://arxiv.org/abs/2204.03045}{{
  arXiv:2204.03045~[cond-mat.str-el]}}.

\bibitem{Cordova:2022ruw}
C.~Cordova, T.~T. Dumitrescu, K.~Intriligator, and S.-H. Shao, {\slshape
  {Snowmass White Paper: Generalized Symmetries in Quantum Field Theory and
  Beyond},} in {\em {Snowmass 2021}}.
\newblock 5, 2022.
\newblock \href{http://arxiv.org/abs/2205.09545}{{ arXiv:2205.09545~[hep-th]}}.

\bibitem{Bhardwaj:2023kri}
L.~Bhardwaj, L.~E. Bottini, L.~Fraser-Taliente, L.~Gladden, D.~S.~W. Gould,
  A.~Platschorre, and H.~Tillim, {\slshape {Lectures on generalized
  symmetries},} \href{http://dx.doi.org/10.1016/j.physrep.2023.11.002}{{\em
  Phys. Rept.} {\bfseries 1051} (2024) 1--87},
  \href{http://arxiv.org/abs/2307.07547}{{ arXiv:2307.07547~[hep-th]}}.

\bibitem{Petkova:2000ip}
V.~B. Petkova and J.~B. Zuber, {\slshape {Generalized Twisted Partition
  Functions},} \href{http://dx.doi.org/10.1016/S0370-2693(01)00276-3}{{\em
  Phys. Lett. B} {\bfseries 504} (2001) 157--164},
  \href{http://arxiv.org/abs/hep-th/0011021}{{ arXiv:hep-th/0011021}}.

\bibitem{Cardy:1989ir}
J.~L. Cardy, {\slshape {Boundary Conditions, Fusion Rules and the Verlinde
  Formula},} \href{http://dx.doi.org/10.1016/0550-3213(89)90521-X}{{\em Nucl.
  Phys. B} {\bfseries 324} (1989) 581--596}.

\bibitem{Behrend:1999bn}
R.~E. Behrend, P.~A. Pearce, V.~B. Petkova, and J.-B. Zuber, {\slshape
  {Boundary conditions in rational conformal field theories},}
  \href{http://dx.doi.org/10.1016/S0550-3213(99)00592-1}{{\em Nucl. Phys. B}
  {\bfseries 570} (2000) 525--589},
  \href{http://arxiv.org/abs/hep-th/9908036}{{ arXiv:hep-th/9908036}}.

\bibitem{Leinaas:1977fm}
J.~M. Leinaas and J.~Myrheim, {\slshape {On the theory of identical
  particles},} \href{http://dx.doi.org/10.1007/BF02727953}{{\em Nuovo Cim. B}
  {\bfseries 37} (1977) 1--23}.

\bibitem{Goldin:1979ki}
G.~A. Goldin, R.~Menikoff, and D.~H. Sharp, {\slshape {Particle Statistics From
  Induced Representations of a Local Current Group},}
  \href{http://dx.doi.org/10.1063/1.524510}{{\em J. Math. Phys.} {\bfseries 21}
  (1980) 650}.

\bibitem{article}
A.~Ocneanu, {\slshape Paths on coxeter diagrams: From platonic solids and
  singularities to minimal models and subfactors,} {\em Lectures on Operator
  Theory} (01, 2000) .

\bibitem{Bockenhauer:1999wt}
J.~Bockenhauer, D.~E. Evans, and Y.~Kawahigashi, {\slshape {Chiral structure of
  modular invariants for subfactors},}
  \href{http://dx.doi.org/10.1007/s002200050798}{{\em Commun. Math. Phys.}
  {\bfseries 210} (2000) 733--784}, \href{http://arxiv.org/abs/math/9907149}{{
  arXiv:math/9907149}}.

\bibitem{Kawahigashi:2021hds}
Y.~Kawahigashi, {\slshape {Two-dimensional topological order and operator
  algebras},} \href{http://dx.doi.org/10.1142/S0217979221300036}{{\em Int. J.
  Mod. Phys. B} {\bfseries 35} (2021) 2130003},
  \href{http://arxiv.org/abs/2102.10953}{{ arXiv:2102.10953~[math-ph]}}.

\bibitem{Zamolodchikov:1987ti}
A.~B. Zamolodchikov, {\slshape {Renormalization Group and Perturbation Theory
  Near Fixed Points in Two-Dimensional Field Theory},} {\em Sov. J. Nucl.
  Phys.} {\bfseries 46} (1987) 1090.

\bibitem{Zamolodchikov:1987jf}
A.~B. Zamolodchikov, {\slshape {Higher Order Integrals of Motion in
  Two-Dimensional Models of the Field Theory with a Broken Conformal
  Symmetry},} {\em JETP Lett.} {\bfseries 46} (1987) 160--164.

\bibitem{Zamolodchikov:1989hfa}
A.~B. Zamolodchikov, {\slshape {Integrable field theory from conformal field
  theory},} {\em Adv. Stud. Pure Math.} {\bfseries 19} (1989) 641--674.

\bibitem{Kitaev:2011dxc}
A.~Kitaev and L.~Kong, {\slshape {Models for Gapped Boundaries and Domain
  Walls},} \href{http://dx.doi.org/10.1007/s00220-012-1500-5}{{\em Commun.
  Math. Phys.} {\bfseries 313} (2012) 351--373},
  \href{http://arxiv.org/abs/1104.5047}{{ arXiv:1104.5047~[cond-mat.str-el]}}.

\bibitem{Lan:2014uaa}
T.~Lan, J.~C. Wang, and X.-G. Wen, {\slshape {Gapped Domain Walls, Gapped
  Boundaries and Topological Degeneracy},}
  \href{http://dx.doi.org/10.1103/PhysRevLett.114.076402}{{\em Phys. Rev.
  Lett.} {\bfseries 114} (2015) 076402},
  \href{http://arxiv.org/abs/1408.6514}{{ arXiv:1408.6514~[cond-mat.str-el]}}.

\bibitem{Laughlin:1983fy}
R.~B. Laughlin, {\slshape {Anomalous quantum Hall effect: An Incompressible
  quantum fluid with fractionallycharged excitations},}
  \href{http://dx.doi.org/10.1103/PhysRevLett.50.1395}{{\em Phys. Rev. Lett.}
  {\bfseries 50} (1983) 1395}.

\bibitem{Witten:1988hf}
E.~Witten, {\slshape {Quantum Field Theory and the Jones Polynomial},}
  \href{http://dx.doi.org/10.1007/BF01217730}{{\em Commun. Math. Phys.}
  {\bfseries 121} (1989) 351--399}.

\bibitem{Moore:1991ks}
G.~W. Moore and N.~Read, {\slshape {Nonabelions in the fractional quantum Hall
  effect},} \href{http://dx.doi.org/10.1016/0550-3213(91)90407-O}{{\em Nucl.
  Phys. B} {\bfseries 360} (1991) 362--396}.

\bibitem{Hung:2015hfa}
L.-Y. Hung and Y.~Wan, {\slshape {Generalized ADE classification of topological
  boundaries and anyon condensation},}
  \href{http://dx.doi.org/10.1007/JHEP07(2015)120}{{\em JHEP} {\bfseries 07}
  (2015) 120}, \href{http://arxiv.org/abs/1502.02026}{{
  arXiv:1502.02026~[cond-mat.str-el]}}.

\bibitem{Wan:2016php}
Y.~Wan and C.~Wang, {\slshape {Fermion Condensation and Gapped Domain Walls in
  Topological Orders},} \href{http://dx.doi.org/10.1007/JHEP03(2017)172}{{\em
  JHEP} {\bfseries 03} (2017) 172}, \href{http://arxiv.org/abs/1607.01388}{{
  arXiv:1607.01388~[cond-mat.str-el]}}.

\bibitem{Zeev:2022cnv}
R.~B. Zeev, B.~Ergun, E.~Milan, and S.~S. Razamat, {\slshape {Categorical
  structure of the set of all CFTs},}
  \href{http://dx.doi.org/10.1103/PhysRevD.110.025019}{{\em Phys. Rev. D}
  {\bfseries 110} (2024) 025019}, \href{http://arxiv.org/abs/2212.11022}{{
  arXiv:2212.11022~[hep-th]}}.

\bibitem{Fukusumi_2022_c}
Y.~Fukusumi, {\slshape {Composing parafermions: a construction of $Z_{N}$
  fractional quantum Hall systems and a modern understanding of confinement and
  duality},} \href{http://arxiv.org/abs/2212.12999}{{
  arXiv:2212.12999~[cond-mat.str-el]}}.

\bibitem{Fukusumi:2024ejk}
Y.~Fukusumi, {\slshape {Gauging or extending bulk and boundary conformal field
  theories: Application to bulk and domain wall problem in topological matter
  and their descriptions by mock modular covariant},}
  \href{http://dx.doi.org/10.1103/f1bp-fzq5}{{\em Phys. Rev. B} {\bfseries 112}
  (2025) 075144}, \href{http://arxiv.org/abs/2412.19577}{{
  arXiv:2412.19577~[hep-th]}}.

\bibitem{Antinucci:2025fjp}
A.~Antinucci, C.~Copetti, Y.~Gai, and S.~Schafer-Nameki, {\slshape {Categorical
  Anomaly Matching},} \href{http://arxiv.org/abs/2508.00982}{{
  arXiv:2508.00982~[hep-th]}}.

\bibitem{Kong:2013aya}
L.~Kong, {\slshape {Anyon condensation and tensor categories},}
  \href{http://dx.doi.org/10.1016/j.nuclphysb.2014.07.003}{{\em Nucl. Phys. B}
  {\bfseries 886} (2014) 436--482}, \href{http://arxiv.org/abs/1307.8244}{{
  arXiv:1307.8244~[cond-mat.str-el]}}.

\bibitem{Note1}
$\protect \mathbb {C}$ is the field of complex number. In this manuscript, we
  also use $\protect \mathbb {Z}$ as the set of integers.

\bibitem{Zhao:2023wtg}
Y.~Zhao, H.~Wang, Y.~Hu, and Y.~Wan, {\slshape {Symmetry fractionalized
  (irrationalized) fusion rules and two domain-wall Verlinde formulae},}
  \href{http://dx.doi.org/10.1007/JHEP04(2024)115}{{\em JHEP} {\bfseries 04}
  (2024) 115}, \href{http://arxiv.org/abs/2304.08475}{{
  arXiv:2304.08475~[cond-mat.str-el]}}.

\bibitem{Fukusumi:2025clr}
Y.~Fukusumi and Y.~Furuta, {\slshape {Homomorphism, substructure, and ideal:
  Elementary but rigorous aspects of renormalization group or hierarchical
  structure of topological orders},}
  \href{http://dx.doi.org/10.1103/1t7d-sjcq}{{\em Phys. Rev. B} {\bfseries 113}
  (2026) 155103}, \href{http://arxiv.org/abs/2506.23155}{{
  arXiv:2506.23155~[hep-th]}}.

\bibitem{Fukusumi:2025ljx}
Y.~Fukusumi and S.~Yahagi, {\slshape {Extending fusion rules with finite
  subgroups: A general construction of $Z_{N}$ extended conformal field
  theories and their orbifoldings},}
  \href{http://dx.doi.org/10.21468/SciPostPhys.20.5.136}{{\em SciPost Phys.}
  {\bfseries 20} (2026) 136}, \href{http://arxiv.org/abs/2508.08639}{{
  arXiv:2508.08639~[hep-th]}}.

\bibitem{Fukusumi:2025xrj}
Y.~Fukusumi and T.~Kawamoto, {\slshape {Generalizing quantum dimensions:
  Symmetry-based classification of local pseudo-Hermitian systems and the
  corresponding domain walls},} \href{http://arxiv.org/abs/2511.11059}{{
  arXiv:2511.11059~[hep-th]}}.

\bibitem{Fukusumi:2025fir}
Y.~Fukusumi and T.~Kawamoto, {\slshape {Generalizing fusion rules by shuffle:
  Symmetry-based classifications of nonlocal systems constructed from
  similarity transformations},} \href{http://arxiv.org/abs/2512.02139}{{
  arXiv:2512.02139~[hep-th]}}.

\bibitem{Fukusumi:2025fvb}
Y.~Fukusumi, {\slshape {Classifying fusion rules of anyons or SymTFTs: A
  general algebraic formula for domain wall problems and quantum phase
  transitions},} \href{http://arxiv.org/abs/2512.21687}{{
  arXiv:2512.21687~[hep-th]}}.

\bibitem{Brunner:2007ur}
I.~Brunner and D.~Roggenkamp, {\slshape {Defects and bulk perturbations of
  boundary Landau-Ginzburg orbifolds},}
  \href{http://dx.doi.org/10.1088/1126-6708/2008/04/001}{{\em JHEP} {\bfseries
  04} (2008) 001}, \href{http://arxiv.org/abs/0712.0188}{{
  arXiv:0712.0188~[hep-th]}}.

\bibitem{Gaiotto:2012np}
D.~Gaiotto, {\slshape {Domain Walls for Two-Dimensional Renormalization Group
  Flows},} \href{http://dx.doi.org/10.1007/JHEP12(2012)103}{{\em JHEP}
  {\bfseries 12} (2012) 103}, \href{http://arxiv.org/abs/1201.0767}{{
  arXiv:1201.0767~[hep-th]}}.

\bibitem{Stanishkov:2016pvi}
M.~Stanishkov, {\slshape {RG domain wall for the general $ \widehat{su}(2) $
  coset models},} \href{http://dx.doi.org/10.1007/JHEP08(2016)096}{{\em JHEP}
  {\bfseries 08} (2016) 096}, \href{http://arxiv.org/abs/1606.03605}{{
  arXiv:1606.03605~[hep-th]}}.

\bibitem{Stanishkov:2016rgv}
M.~Stanishkov, {\slshape {Second order RG flow in general $
  \widehat{\mathrm{su}}(2) $ coset models},}
  \href{http://dx.doi.org/10.1007/JHEP09(2016)040}{{\em JHEP} {\bfseries 09}
  (2016) 040}, \href{http://arxiv.org/abs/1606.04328}{{
  arXiv:1606.04328~[hep-th]}}.

\bibitem{Poghosyan:2022mfw}
H.~Poghosyan and R.~Poghossian, {\slshape {RG flow between W$_{3}$ minimal
  models by perturbation and domain wall approaches},}
  \href{http://dx.doi.org/10.1007/JHEP08(2022)307}{{\em JHEP} {\bfseries 08}
  (2022) 307}, \href{http://arxiv.org/abs/2205.05091}{{
  arXiv:2205.05091~[hep-th]}}.

\bibitem{Poghosyan:2022ecv}
H.~Poghosyan and R.~Poghossian, {\slshape {RG flows between $W_3$ minimal
  models},} \href{http://dx.doi.org/10.22323/1.412.0039}{{\em PoS} {\bfseries
  Regio2021} (2022) 039}.

\bibitem{Higgs:1964pj}
P.~W. Higgs, {\slshape {Broken Symmetries and the Masses of Gauge Bosons},}
  \href{http://dx.doi.org/10.1103/PhysRevLett.13.508}{{\em Phys. Rev. Lett.}
  {\bfseries 13} (1964) 508--509}.

\bibitem{Klos:2019axh}
F.~Klos and D.~Roggenkamp, {\slshape {Realizing IR theories by projections in
  the UV},} \href{http://dx.doi.org/10.1007/JHEP01(2020)097}{{\em JHEP}
  {\bfseries 01} (2020) 097}, \href{http://arxiv.org/abs/1907.12339}{{
  arXiv:1907.12339~[hep-th]}}.

\bibitem{Klos:2020upw}
F.~Klos and D.~Roggenkamp, {\slshape {Complementary projection defects and
  decomposition},} \href{http://dx.doi.org/10.1007/JHEP03(2021)195}{{\em JHEP}
  {\bfseries 03} (2021) 195}, \href{http://arxiv.org/abs/2006.08961}{{
  arXiv:2006.08961~[hep-th]}}.

\bibitem{Frohlich:2003hg}
J.~Frohlich, J.~Fuchs, I.~Runkel, and C.~Schweigert, {\slshape {Algebras in
  tensor categories and coset conformal field theories},}
  \href{http://dx.doi.org/10.1002/prop.200310162}{{\em Fortsch. Phys.}
  {\bfseries 52} (2004) 672--677}, \href{http://arxiv.org/abs/hep-th/0309269}{{
  arXiv:hep-th/0309269}}.

\bibitem{Frohlich:2003hm}
J.~Frohlich, J.~Fuchs, I.~Runkel, and C.~Schweigert, {\slshape {Correspondences
  of ribbon categories},}
  \href{http://dx.doi.org/10.1016/j.aim.2005.04.007}{{\em Adv. Math.}
  {\bfseries 199} (2006) 192--329}, \href{http://arxiv.org/abs/math/0309465}{{
  arXiv:math/0309465}}.

\bibitem{Cordova:2025eim}
C.~Cordova, D.~Garc\'\i{}a-Sep\'ulveda, and K.~Ohmori, {\slshape {Higgsing
  Transitions from Topological Field Theory \& Non-Invertible Symmetry in
  Chern-Simons Matter Theories},} \href{http://arxiv.org/abs/2504.03614}{{
  arXiv:2504.03614~[hep-th]}}.

\bibitem{Wong:1994np}
E.~Wong and I.~Affleck, {\slshape {Tunneling in quantum wires: A Boundary
  conformal field theory approach},}
  \href{http://dx.doi.org/10.1016/0550-3213(94)90479-0}{{\em Nucl. Phys. B}
  {\bfseries 417} (1994) 403--438},
  \href{http://arxiv.org/abs/cond-mat/9311040}{{ arXiv:cond-mat/9311040}}.

\bibitem{Bais:2008ni}
F.~A. Bais and J.~K. Slingerland, {\slshape {Condensate induced transitions
  between topologically ordered phases},}
  \href{http://dx.doi.org/10.1103/PhysRevB.79.045316}{{\em Phys. Rev. B}
  {\bfseries 79} (2009) 045316}, \href{http://arxiv.org/abs/0808.0627}{{
  arXiv:0808.0627~[cond-mat.mes-hall]}}.

\bibitem{Cheng:2026qax}
M.~Cheng and N.~Seiberg, {\slshape {Proliferation transitions from a
  topological phase in $2+1$ dimensions},}
  \href{http://arxiv.org/abs/2603.00245}{{
  arXiv:2603.00245~[cond-mat.str-el]}}.

\bibitem{Vafa:1986wx}
C.~Vafa, {\slshape {Modular Invariance and Discrete Torsion on Orbifolds},}
  \href{http://dx.doi.org/10.1016/0550-3213(86)90379-2}{{\em Nucl. Phys. B}
  {\bfseries 273} (1986) 592--606}.

\bibitem{Dijkgraaf:1989hb}
R.~Dijkgraaf, C.~Vafa, E.~P. Verlinde, and H.~L. Verlinde, {\slshape {The
  Operator Algebra of Orbifold Models},}
  \href{http://dx.doi.org/10.1007/BF01238812}{{\em Commun. Math. Phys.}
  {\bfseries 123} (1989) 485}.

\bibitem{Schellekens:1989uf}
A.~N. Schellekens and S.~Yankielowicz, {\slshape {Field Identification Fixed
  Points in the Coset Construction},}
  \href{http://dx.doi.org/10.1016/0550-3213(90)90657-Y}{{\em Nucl. Phys. B}
  {\bfseries 334} (1990) 67--102}.

\bibitem{Schellekens:1990ys}
A.~N. Schellekens, {\slshape {Fusion rule automorphisms from integer spin
  simple currents},} \href{http://dx.doi.org/10.1016/0370-2693(90)90065-E}{{\em
  Phys. Lett. B} {\bfseries 244} (1990) 255--260}.

\bibitem{Gato-Rivera:1990lxi}
B.~Gato-Rivera and A.~N. Schellekens, {\slshape {Complete classification of
  simple current automorphisms},}
  \href{http://dx.doi.org/10.1016/0550-3213(91)90346-Y}{{\em Nucl. Phys. B}
  {\bfseries 353} (1991) 519--537}.

\bibitem{Kreuzer:1993tf}
M.~Kreuzer and A.~N. Schellekens, {\slshape {Simple currents versus orbifolds
  with discrete torsion: A Complete classification},}
  \href{http://dx.doi.org/10.1016/0550-3213(94)90055-8}{{\em Nucl. Phys. B}
  {\bfseries 411} (1994) 97--121}, \href{http://arxiv.org/abs/hep-th/9306145}{{
  arXiv:hep-th/9306145}}.

\bibitem{Fuchs:2000gv}
J.~Fuchs, C.~Schweigert, and J.~Walcher, {\slshape {Projections in string
  theory and boundary states for Gepner models},}
  \href{http://dx.doi.org/10.1016/S0550-3213(00)00487-9}{{\em Nucl. Phys. B}
  {\bfseries 588} (2000) 110--148},
  \href{http://arxiv.org/abs/hep-th/0003298}{{ arXiv:hep-th/0003298}}.

\bibitem{Cappelli:2010jv}
A.~Cappelli and G.~Viola, {\slshape {Partition Functions of Non-Abelian Quantum
  Hall States},} \href{http://dx.doi.org/10.1088/1751-8113/44/7/075401}{{\em J.
  Phys. A} {\bfseries 44} (2011) 075401},
  \href{http://arxiv.org/abs/1007.1732}{{
  arXiv:1007.1732~[cond-mat.mes-hall]}}.

\bibitem{Schoutens:2015uia}
K.~Schoutens and X.-G. Wen, {\slshape {Simple-current algebra constructions of
  2+1-dimensional topological orders},}
  \href{http://dx.doi.org/10.1103/PhysRevB.93.045109}{{\em Phys. Rev. B}
  {\bfseries 93} (2016) 045109}, \href{http://arxiv.org/abs/1508.01111}{{
  arXiv:1508.01111~[cond-mat.str-el]}}.

\bibitem{Hansson_2017}
T.~Hansson, M.~Hermanns, S.~Simon, and S.~Viefers, {\slshape Quantum hall
  physics: Hierarchies and conformal field theory techniques,}
  \href{http://dx.doi.org/10.1103/RevModPhys.89.025005}{{\em Reviews of Modern
  Physics} {\bfseries 89} (May, 2017) 025005}.

\bibitem{DavydovMugerNikshychOstrik+2013+135+177}
A.~Davydov, M.~Muger, D.~Nikshych, and V.~Ostrik, {\slshape The witt group of
  non-degenerate braided fusion categories,}
  \href{https://doi.org/10.1515/crelle.2012.014}{{\em Journal für die reine
  und angewandte Mathematik (Crelles Journal)} {\bfseries 2013} (2013)
  135--177}.

\bibitem{Bardeen:1957mv}
J.~Bardeen, L.~N. Cooper, and J.~R. Schrieffer, {\slshape {Theory of
  superconductivity},} \href{http://dx.doi.org/10.1103/PhysRev.108.1175}{{\em
  Phys. Rev.} {\bfseries 108} (1957) 1175--1204}.

\bibitem{PhysRevB.84.125434}
F.~J. Burnell, S.~H. Simon, and J.~K. Slingerland, {\slshape Condensation of
  achiral simple currents in topological lattice models: Hamiltonian study of
  topological symmetry breaking,}
  \href{https://link.aps.org/doi/10.1103/PhysRevB.84.125434}{{\em Phys. Rev. B}
  {\bfseries 84} (Sep, 2011) 125434}.

\bibitem{Kaidi:2021gbs}
J.~Kaidi, Z.~Komargodski, K.~Ohmori, S.~Seifnashri, and S.-H. Shao, {\slshape
  {Higher central charges and topological boundaries in 2+1-dimensional
  TQFTs},} \href{http://dx.doi.org/10.21468/SciPostPhys.13.3.067}{{\em SciPost
  Phys.} {\bfseries 13} (2022) 067}, \href{http://arxiv.org/abs/2107.13091}{{
  arXiv:2107.13091~[hep-th]}}.

\bibitem{Huston:2022utd}
P.~Huston, F.~Burnell, C.~Jones, and D.~Penneys, {\slshape {Composing
  topological domain walls and anyon mobility},}
  \href{http://dx.doi.org/10.21468/SciPostPhys.15.3.076}{{\em SciPost Phys.}
  {\bfseries 15} (2023) 076}, \href{http://arxiv.org/abs/2208.14018}{{
  arXiv:2208.14018~[cond-mat.str-el]}}.

\bibitem{Fukusumi:2022xxe}
Y.~Fukusumi and B.~Yang, {\slshape {Fermionic fractional quantum Hall states: A
  modern approach to systems with bulk-edge correspondence},}
  \href{http://dx.doi.org/10.1103/PhysRevB.108.085123}{{\em Phys. Rev. B}
  {\bfseries 108} (2023) 085123}, \href{http://arxiv.org/abs/2212.12993}{{
  arXiv:2212.12993~[cond-mat.str-el]}}.

\bibitem{Zhang:2024bye}
C.~Zhang, A.~Vishwanath, and X.-G. Wen, {\slshape {Hierarchy construction for
  non-Abelian fractional quantum Hall states via anyon condensation},}
  \href{http://dx.doi.org/10.1103/jndb-435f}{{\em Phys. Rev. B} {\bfseries 112}
  (2025) 125116}, \href{http://arxiv.org/abs/2406.12068}{{
  arXiv:2406.12068~[cond-mat.str-el]}}.

\bibitem{Kikuchi:2024ibt}
K.~Kikuchi, K.-S. Kam, and F.-H. Huang, {\slshape {Anyon condensation in
  mixed-state topological order},} \href{http://arxiv.org/abs/2406.14320}{{
  arXiv:2406.14320~[hep-th]}}.

\bibitem{Kikuchi:2024cjd}
K.~Kikuchi, {\slshape {Rational RG flow, extension, and Witt class},}
  \href{http://arxiv.org/abs/2412.08935}{{ arXiv:2412.08935~[hep-th]}}.

\bibitem{davydov2011structure}
A.~Davydov, D.~Nikshych, and V.~Ostrik, {\slshape On the structure of the witt
  group of braided fusion categories,} 2011.

\bibitem{Polyakov:1974gs}
A.~M. Polyakov, {\slshape {Nonhamiltonian approach to conformal quantum field
  theory},} {\em Zh. Eksp. Teor. Fiz.} {\bfseries 66} (1974) 23--42.

\bibitem{Note2}
However, we stress that the exact and detailed construction of this relative
  tensor product structure between theories has not been studied sufficiently.
  Moreover the resultant model can be outside of the conventional models such
  as modular tensor categories (MTCs) because of the field identification
  problems even when taking $\protect \mathbf {A}$ and $\protect \mathbf {A'}$
  as MTCs. Inversely, to obtain the resultant bimodules as a MTC, it can be
  necessary to take $\protect \mathbf {A}$ and $\protect \mathbf {A'}$ outside
  of MTCs. For example, in the case with $\protect \mathbf {A}_{\protect \text
  {ub}}=\protect \mathbb {Z}_{2}$, the underlying models $\protect \mathbf {A}$
  and $\protect \mathbf {A'}$ can be fermionic models outside of MTCs. This
  modification seems to correspond to a variant of Karoubi envelop in
  mathematics and parton construction in condensed matter. We study this aspect
  elsewhere.

\bibitem{Cardy:1986ie}
J.~L. Cardy, {\slshape {Operator Content of Two-Dimensional Conformally
  Invariant Theories},}
  \href{http://dx.doi.org/10.1016/0550-3213(86)90552-3}{{\em Nucl. Phys. B}
  {\bfseries 270} (1986) 186--204}.

\bibitem{Yurov:1989yu}
V.~P. Yurov and A.~B. Zamolodchikov, {\slshape {TRUNCATED CONFORMAL SPACE
  APPROACH TO SCALING LEE-YANG MODEL},}
  \href{http://dx.doi.org/10.1142/S0217751X9000218X}{{\em Int. J. Mod. Phys. A}
  {\bfseries 5} (1990) 3221--3246}.

\bibitem{Yurov:1991my}
V.~P. Yurov and A.~B. Zamolodchikov, {\slshape {Truncated fermionic space
  approach to the critical 2-D Ising model with magnetic field},}
  \href{http://dx.doi.org/10.1142/S0217751X91002161}{{\em Int. J. Mod. Phys. A}
  {\bfseries 6} (1991) 4557--4578}.

\bibitem{Hogervorst:2014rta}
M.~Hogervorst, S.~Rychkov, and B.~C. van Rees, {\slshape {Truncated conformal
  space approach in d dimensions: A cheap alternative to lattice field
  theory?},} \href{http://dx.doi.org/10.1103/PhysRevD.91.025005}{{\em Phys.
  Rev. D} {\bfseries 91} (2015) 025005},
  \href{http://arxiv.org/abs/1409.1581}{{ arXiv:1409.1581~[hep-th]}}.

\bibitem{James_2018}
A.~J.~A. James, R.~M. Konik, P.~Lecheminant, N.~J. Robinson, and A.~M. Tsvelik,
  {\slshape Non-perturbative methodologies for low-dimensional
  strongly-correlated systems: From non-abelian bosonization to truncated
  spectrum methods,} \href{https://doi.org/10.1088%2F1361-6633%2Faa91ea}{{\em
  Reports on Progress in Physics} {\bfseries 81} (Feb, 2018) 046002}.

\bibitem{Grimm:1990ch}
U.~Grimm, {\slshape {The Quantum Ising Chain With a Generalized Defect},}
  \href{http://dx.doi.org/10.1016/0550-3213(90)90462-M}{{\em Nucl. Phys. B}
  {\bfseries 340} (1990) 633--658},
  \href{http://arxiv.org/abs/hep-th/0310089}{{ arXiv:hep-th/0310089}}.

\bibitem{Grimm:2001dr}
U.~Grimm, {\slshape {Spectrum of a duality twisted Ising quantum chain},}
  \href{http://dx.doi.org/10.1088/0305-4470/35/3/101}{{\em J. Phys. A}
  {\bfseries 35} (2002) L25--L30}, \href{http://arxiv.org/abs/hep-th/0111157}{{
  arXiv:hep-th/0111157}}.

\bibitem{Belletete:2018eua}
J.~Bellet\^ete, A.~M. Gainutdinov, J.~L. Jacobsen, H.~Saleur, and T.~S.
  Tavares, {\slshape {Topological Defects in Lattice Models and Affine
  Temperley\textendash{}Lieb Algebra},}
  \href{http://dx.doi.org/10.1007/s00220-022-04618-0}{{\em Commun. Math. Phys.}
  {\bfseries 400} (2023) 1203--1254}, \href{http://arxiv.org/abs/1811.02551}{{
  arXiv:1811.02551~[hep-th]}}.

\bibitem{Belletete:2020gst}
J.~Bellet\^ete, A.~M. Gainutdinov, J.~L. Jacobsen, H.~Saleur, and T.~S.
  Tavares, {\slshape {Topological defects in periodic RSOS models and anyonic
  chains},} \href{http://arxiv.org/abs/2003.11293}{{
  arXiv:2003.11293~[math-ph]}}.

\bibitem{Giokas:2011ix}
P.~Giokas and G.~Watts, {\slshape {The renormalisation group for the truncated
  conformal space approach on the cylinder},}
  \href{http://arxiv.org/abs/1106.2448}{{ arXiv:1106.2448~[hep-th]}}.

\bibitem{Poghosyan:2013qta}
A.~Poghosyan and H.~Poghosyan, {\slshape {Mixing with descendant fields in
  perturbed minimal CFT models},}
  \href{http://dx.doi.org/10.1007/JHEP10(2013)131}{{\em JHEP} {\bfseries 10}
  (2013) 131}, \href{http://arxiv.org/abs/1305.6066}{{
  arXiv:1305.6066~[hep-th]}}.

\bibitem{Poghossian:2013fda}
R.~Poghossian, {\slshape {Two Dimensional Renormalization Group Flows in Next
  to Leading Order},} \href{http://dx.doi.org/10.1007/JHEP01(2014)167}{{\em
  JHEP} {\bfseries 01} (2014) 167}, \href{http://arxiv.org/abs/1303.3015}{{
  arXiv:1303.3015~[hep-th]}}.

\bibitem{Belavin:1984vu}
A.~A. Belavin, A.~M. Polyakov, and A.~B. Zamolodchikov, {\slshape {Infinite
  Conformal Symmetry in Two-Dimensional Quantum Field Theory},}
  \href{http://dx.doi.org/10.1016/0550-3213(84)90052-X}{{\em Nucl. Phys. B}
  {\bfseries 241} (1984) 333--380}.

\bibitem{Wang:2014lca}
C.~Wang and T.~Senthil, {\slshape {Interacting fermionic topological
  insulators/superconductors in three dimensions},}
  \href{http://dx.doi.org/10.1103/PhysRevB.89.195124}{{\em Phys. Rev. B}
  {\bfseries 89} (2014) 195124}, \href{http://arxiv.org/abs/1401.1142}{{
  arXiv:1401.1142~[cond-mat.str-el]}}. [Erratum: Phys.Rev.B 91, 239902 (2015)].

\bibitem{Wang:2016gqj}
C.~Wang and T.~Senthil, {\slshape {Composite fermi liquids in the lowest Landau
  level},} \href{http://dx.doi.org/10.1103/PhysRevB.94.245107}{{\em Phys. Rev.
  B} {\bfseries 94} (2016) 245107}, \href{http://arxiv.org/abs/1604.06807}{{
  arXiv:1604.06807~[cond-mat.str-el]}}.

\bibitem{Sodemann:2016mib}
I.~Sodemann, I.~Kimchi, C.~Wang, and T.~Senthil, {\slshape {Composite fermion
  duality for half-filled multicomponent Landau Levels},}
  \href{http://dx.doi.org/10.1103/PhysRevB.95.085135}{{\em Phys. Rev. B}
  {\bfseries 95} (2017) 085135}, \href{http://arxiv.org/abs/1609.08616}{{
  arXiv:1609.08616~[cond-mat.str-el]}}.

\bibitem{Wang:2017txt}
C.~Wang, A.~Nahum, M.~A. Metlitski, C.~Xu, and T.~Senthil, {\slshape
  {Deconfined quantum critical points: symmetries and dualities},}
  \href{http://dx.doi.org/10.1103/PhysRevX.7.031051}{{\em Phys. Rev. X}
  {\bfseries 7} (2017) 031051}, \href{http://arxiv.org/abs/1703.02426}{{
  arXiv:1703.02426~[cond-mat.str-el]}}.

\bibitem{Kim:2025tzc}
M.~L. Kim, S.~D. Pace, and S.-H. Shao, {\slshape {Symmetry-Enforced Fermi
  Surfaces},} \href{http://dx.doi.org/10.1103/wpdt-4t1c}{{\em Phys. Rev. Lett.}
  {\bfseries 136} (2026) 176502}, \href{http://arxiv.org/abs/2512.04150}{{
  arXiv:2512.04150~[cond-mat.str-el]}}.

\bibitem{Goddard:1984vk}
P.~Goddard, A.~Kent, and D.~I. Olive, {\slshape {Virasoro Algebras and Coset
  Space Models},} \href{http://dx.doi.org/10.1016/0370-2693(85)91145-1}{{\em
  Phys. Lett. B} {\bfseries 152} (1985) 88--92}.

\bibitem{Goddard:1984hg}
P.~Goddard and D.~I. Olive, {\slshape {Kac-Moody Algebras, Conformal Symmetry
  and Critical Exponents},}
  \href{http://dx.doi.org/10.1016/0550-3213(85)90344-X}{{\em Nucl. Phys. B}
  {\bfseries 257} (1985) 226--252}.

\bibitem{Goddard:1986ee}
P.~Goddard, A.~Kent, and D.~I. Olive, {\slshape {Unitary Representations of the
  Virasoro and Supervirasoro Algebras},}
  \href{http://dx.doi.org/10.1007/BF01464283}{{\em Commun. Math. Phys.}
  {\bfseries 103} (1986) 105--119}.

\bibitem{Goddard:1988md}
P.~Goddard and A.~Schwimmer, {\slshape {Unitary Construction of Extended
  Conformal Algebras},}
  \href{http://dx.doi.org/10.1016/0370-2693(88)91263-4}{{\em Phys. Lett. B}
  {\bfseries 206} (1988) 62--70}.

\bibitem{Kuniba:1990im}
A.~Kuniba and T.~Nakanishi, {\slshape {LEVEL RANK DUALITY IN FUSION RSOS
  MODELS},} in {\em {International Colloquium on Modern Quantum Field Theory}}.
\newblock 1, 1990.

\bibitem{Kuniba:1990zh}
A.~Kuniba, T.~Nakanishi, and J.~Suzuki, {\slshape {Ferromagnetizations and
  antiferromagnetizations in RSOS models},}
  \href{http://dx.doi.org/10.1016/0550-3213(91)90385-B}{{\em Nucl. Phys. B}
  {\bfseries 356} (1991) 750--774}.

\bibitem{Nakanishi:1990hj}
T.~Nakanishi and A.~Tsuchiya, {\slshape {Level rank duality of WZW models in
  conformal field theory},} \href{http://dx.doi.org/10.1007/BF02101097}{{\em
  Commun. Math. Phys.} {\bfseries 144} (1992) 351--372}.

\bibitem{Altschuler:1990th}
D.~Altschuler, M.~Bauer, and H.~Saleur, {\slshape {Level rank duality in
  nonunitary coset theories},} {\em J. Phys. A} {\bfseries 23} (1990)
  L789--L794.

\bibitem{Naculich:1990hg}
S.~G. Naculich and H.~J. Schnitzer, {\slshape {Duality Between SU($N$)-k and
  SU(k)-$N$ {WZW} Models},}
  \href{http://dx.doi.org/10.1016/0550-3213(90)90380-V}{{\em Nucl. Phys. B}
  {\bfseries 347} (1990) 687--742}.

\bibitem{Aharony:2016jvv}
O.~Aharony, F.~Benini, P.-S. Hsin, and N.~Seiberg, {\slshape
  {Chern-Simons-matter dualities with $SO$ and $USp$ gauge groups},}
  \href{http://dx.doi.org/10.1007/JHEP02(2017)072}{{\em JHEP} {\bfseries 02}
  (2017) 072}, \href{http://arxiv.org/abs/1611.07874}{{
  arXiv:1611.07874~[cond-mat.str-el]}}.

\bibitem{Hsin:2016blu}
P.-S. Hsin and N.~Seiberg, {\slshape {Level/rank Duality and
  Chern-Simons-Matter Theories},}
  \href{http://dx.doi.org/10.1007/JHEP09(2016)095}{{\em JHEP} {\bfseries 09}
  (2016) 095}, \href{http://arxiv.org/abs/1607.07457}{{
  arXiv:1607.07457~[hep-th]}}.

\bibitem{Blumenhagen:1994ik}
R.~Blumenhagen, W.~Eholzer, A.~Honecker, K.~Hornfeck, and R.~Hubel, {\slshape
  {Unifying W algebras},}
  \href{http://dx.doi.org/10.1016/0370-2693(94)90857-5}{{\em Phys. Lett. B}
  {\bfseries 332} (1994) 51--60}, \href{http://arxiv.org/abs/hep-th/9404113}{{
  arXiv:hep-th/9404113}}.

\bibitem{Blumenhagen:1994wg}
R.~Blumenhagen, W.~Eholzer, A.~Honecker, K.~Hornfeck, and R.~Hubel, {\slshape
  {Coset realization of unifying W algebras},}
  \href{http://dx.doi.org/10.1142/S0217751X95001157}{{\em Int. J. Mod. Phys. A}
  {\bfseries 10} (1995) 2367--2430},
  \href{http://arxiv.org/abs/hep-th/9406203}{{ arXiv:hep-th/9406203}}.

\bibitem{Crnkovic:1989ug}
C.~Crnkovic, R.~Paunov, G.~M. Sotkov, and M.~Stanishkov, {\slshape {Fusions of
  Conformal Models},}
  \href{http://dx.doi.org/10.1016/0550-3213(90)90445-J}{{\em Nucl. Phys. B}
  {\bfseries 336} (1990) 637--690}.

\bibitem{Martins:1991hi}
M.~J. Martins, {\slshape {The Thermodynamic Bethe ansatz for deformed W A(N)-1
  conformal field theories},}
  \href{http://dx.doi.org/10.1016/0370-2693(92)90750-X}{{\em Phys. Lett. B}
  {\bfseries 277} (1992) 301--305},
  \href{http://arxiv.org/abs/hep-th/9201032}{{ arXiv:hep-th/9201032}}.

\bibitem{Dunning:2002cu}
C.~Dunning, {\slshape {Massless flows between minimal W models},}
  \href{http://dx.doi.org/10.1016/S0370-2693(02)01938-X}{{\em Phys. Lett. B}
  {\bfseries 537} (2002) 297--305},
  \href{http://arxiv.org/abs/hep-th/0204090}{{ arXiv:hep-th/0204090}}.

\bibitem{Note3}
We thank Yunqin Zheng for the corresponding discussion.

\bibitem{Ambrosino:2026umb}
F.~Ambrosino and T.~Proch{\'a}zka, {\slshape {RG flows of minimal $\mathcal
  W$-algebra CFTs via non-invertible symmetries},}
  \href{http://arxiv.org/abs/2601.18667}{{ arXiv:2601.18667~[hep-th]}}.

\bibitem{Antinucci:2025uvj}
A.~Antinucci, C.~Copetti, G.~Galati, and G.~Rizi, {\slshape {Defect Conformal
  Manifolds from Phantom (Non-Invertible) Symmetries},}
  \href{http://arxiv.org/abs/2505.09668}{{ arXiv:2505.09668~[hep-th]}}.

\bibitem{Furuta:2025ahl}
Y.~Furuta, Y.~Kusuki, and T.~Onagi, {\slshape {Transmission coefficients from
  phantom currents},} \href{http://dx.doi.org/10.1103/fdfb-tkz4}{{\em Phys.
  Rev. D} {\bfseries 113} (2026) 045008},
  \href{http://arxiv.org/abs/2511.00356}{{ arXiv:2511.00356~[hep-th]}}.

\bibitem{Zhang:2026gqp}
J.-R. Zhang, J.-H. Jin, T.-K. Chen, and J.~Chen, {\slshape {Defect Conformal
  Manifolds along RG Domain Walls between $\mathbb Z_N$-Parafermions and
  Minimal Models},} \href{http://arxiv.org/abs/2605.24978}{{
  arXiv:2605.24978~[hep-th]}}.

\bibitem{Gukov:2015qea}
S.~Gukov, {\slshape {Counting RG flows},}
  \href{http://dx.doi.org/10.1007/JHEP01(2016)020}{{\em JHEP} {\bfseries 01}
  (2016) 020}, \href{http://arxiv.org/abs/1503.01474}{{
  arXiv:1503.01474~[hep-th]}}.

\bibitem{Seiberg:1994pq}
N.~Seiberg, {\slshape {Electric - magnetic duality in supersymmetric nonAbelian
  gauge theories},} \href{http://dx.doi.org/10.1016/0550-3213(94)00023-8}{{\em
  Nucl. Phys. B} {\bfseries 435} (1995) 129--146},
  \href{http://arxiv.org/abs/hep-th/9411149}{{ arXiv:hep-th/9411149}}.

\bibitem{Kutasov:1995np}
D.~Kutasov and A.~Schwimmer, {\slshape {On duality in supersymmetric Yang-Mills
  theory},} \href{http://dx.doi.org/10.1016/0370-2693(95)00676-C}{{\em Phys.
  Lett. B} {\bfseries 354} (1995) 315--321},
  \href{http://arxiv.org/abs/hep-th/9505004}{{ arXiv:hep-th/9505004}}.

\bibitem{Kutasov:1995ve}
D.~Kutasov, {\slshape {A Comment on duality in N=1 supersymmetric nonAbelian
  gauge theories},} \href{http://dx.doi.org/10.1016/0370-2693(95)00392-X}{{\em
  Phys. Lett. B} {\bfseries 351} (1995) 230--234},
  \href{http://arxiv.org/abs/hep-th/9503086}{{ arXiv:hep-th/9503086}}.

\bibitem{Intriligator:1995ff}
K.~A. Intriligator, {\slshape {New RG fixed points and duality in
  supersymmetric SP(N(c)) and SO(N(c)) gauge theories},}
  \href{http://dx.doi.org/10.1016/0550-3213(95)00296-5}{{\em Nucl. Phys. B}
  {\bfseries 448} (1995) 187--198},
  \href{http://arxiv.org/abs/hep-th/9505051}{{ arXiv:hep-th/9505051}}.

\bibitem{Kutasov:1995ss}
D.~Kutasov, A.~Schwimmer, and N.~Seiberg, {\slshape {Chiral rings, singularity
  theory and electric - magnetic duality},}
  \href{http://dx.doi.org/10.1016/0550-3213(95)00599-4}{{\em Nucl. Phys. B}
  {\bfseries 459} (1996) 455--496},
  \href{http://arxiv.org/abs/hep-th/9510222}{{ arXiv:hep-th/9510222}}.

\bibitem{Leigh:1996ds}
R.~G. Leigh and M.~J. Strassler, {\slshape {Accidental symmetries and N=1
  duality in supersymmetric gauge theory},}
  \href{http://dx.doi.org/10.1016/S0550-3213(97)00204-6}{{\em Nucl. Phys. B}
  {\bfseries 496} (1997) 132--148},
  \href{http://arxiv.org/abs/hep-th/9611020}{{ arXiv:hep-th/9611020}}.

\bibitem{Gorbenko:2018ncu}
V.~Gorbenko, S.~Rychkov, and B.~Zan, {\slshape {Walking, Weak first-order
  transitions, and Complex CFTs},}
  \href{http://dx.doi.org/10.1007/JHEP10(2018)108}{{\em JHEP} {\bfseries 10}
  (2018) 108}, \href{http://arxiv.org/abs/1807.11512}{{
  arXiv:1807.11512~[hep-th]}}.

\bibitem{Gukov:2016tnp}
S.~Gukov, {\slshape {RG Flows and Bifurcations},}
  \href{http://dx.doi.org/10.1016/j.nuclphysb.2017.03.025}{{\em Nucl. Phys. B}
  {\bfseries 919} (2017) 583--638}, \href{http://arxiv.org/abs/1608.06638}{{
  arXiv:1608.06638~[hep-th]}}.

\bibitem{Amit:1982az}
D.~J. Amit and L.~Peliti, {\slshape {ON DANGEROUS IRRELEVANT OPERATORS},}
  \href{http://dx.doi.org/10.1016/0003-4916(82)90159-2}{{\em Annals Phys.}
  {\bfseries 140} (1982) 207}.

\bibitem{Nakayama:2024msv}
Y.~Nakayama and T.~Tanaka, {\slshape {Infinitely many new renormalization group
  flows between Virasoro minimal models from non-invertible symmetries},}
  \href{http://dx.doi.org/10.1007/JHEP11(2024)137}{{\em JHEP} {\bfseries 11}
  (2024) 137}, \href{http://arxiv.org/abs/2407.21353}{{
  arXiv:2407.21353~[hep-th]}}.

\bibitem{Note4}
This incompatibility with $\protect \mathbb {C}$-linear structure in category
  theories is fundamental in studying boundary conditions and defects, but this
  is sometimes inconvenient in studying \protect \emph {operators} and their
  \protect \emph {algebra}.

\bibitem{Cardy:1996xt}
J.~L. Cardy, {\em {Scaling and renormalization in statistical physics}}.
\newblock 1996.

\bibitem{Buican:2017rxc}
M.~Buican and A.~Gromov, {\slshape {Anyonic Chains, Topological Defects, and
  Conformal Field Theory},}
  \href{http://dx.doi.org/10.1007/s00220-017-2995-6}{{\em Commun. Math. Phys.}
  {\bfseries 356} (2017) 1017--1056}, \href{http://arxiv.org/abs/1701.02800}{{
  arXiv:1701.02800~[hep-th]}}.

\bibitem{Gaberdiel:2026sfg}
M.~R. Gaberdiel and L.~Merkens, {\slshape {Defects in $ \mathcal{N}=1 $ minimal
  models and RG flows},} \href{http://dx.doi.org/10.1007/JHEP05(2026)018}{{\em
  JHEP} {\bfseries 05} (2026) 018}, \href{http://arxiv.org/abs/2601.03879}{{
  arXiv:2601.03879~[hep-th]}}.

\bibitem{Benedetti:2026drn}
V.~Benedetti, P.~Fendley, and J.~M. Magan, {\slshape {Non-invertible symmetries
  and selection rules for RG flows of coset models},}
  \href{http://arxiv.org/abs/2603.09591}{{ arXiv:2603.09591~[hep-th]}}.

\bibitem{Note5}
After constructing a series of homomorphisms or monoidal functors, we noticed
  that the connection between homomorphisms and the massless RGs is not
  straightforward. There exist many homomorphisms by fixing a UV and IR theory,
  and these homomorphisms should be interpreted as interpolation (or breaking
  process) of symmetries from the mixing by relevant perturbations. In TQFTs,
  they provide classifications of fusion rules, but there exist other
  classifications by the profunctor, as we study in the present manuscript.

\bibitem{DiFrancesco:1997nk}
P.~Di~Francesco, P.~Mathieu, and D.~Senechal,
  \href{http://dx.doi.org/10.1007/978-1-4612-2256-9}{{\em {Conformal Field
  Theory}}}.
\newblock Graduate Texts in Contemporary Physics. Springer-Verlag, New York,
  1997.

\bibitem{Ginsparg:1988ui}
P.~H. Ginsparg, {\slshape {APPLIED CONFORMAL FIELD THEORY},} in {\em {Les
  Houches Summer School in Theoretical Physics: Fields, Strings, Critical
  Phenomena}}.
\newblock 9, 1988.
\newblock \href{http://arxiv.org/abs/hep-th/9108028}{{ arXiv:hep-th/9108028}}.

\bibitem{Ribault:2016sla}
S.~Ribault, {\slshape {Minimal lectures on two-dimensional conformal field
  theory},} \href{http://dx.doi.org/10.21468/SciPostPhysLectNotes.1}{{\em
  SciPost Phys. Lect. Notes} {\bfseries 1} (2018) 1},
  \href{http://arxiv.org/abs/1609.09523}{{ arXiv:1609.09523~[hep-th]}}.

\bibitem{Graham:2003nc}
K.~Graham and G.~M.~T. Watts, {\slshape {Defect lines and boundary flows},}
  \href{http://dx.doi.org/10.1088/1126-6708/2004/04/019}{{\em JHEP} {\bfseries
  04} (2004) 019}, \href{http://arxiv.org/abs/hep-th/0306167}{{
  arXiv:hep-th/0306167}}.

\bibitem{Recknagel:2013uja}
A.~Recknagel and V.~Schomerus,
  \href{http://dx.doi.org/10.1017/CBO9780511806476}{{\em {Boundary Conformal
  Field Theory and the Worldsheet Approach to D-Branes}}}.
\newblock Cambridge Monographs on Mathematical Physics. Cambridge University
  Press, 11, 2013.

\bibitem{Northe:2024tnm}
C.~Northe, {\slshape {Young Researchers School 2024 Maynooth: Lectures on CFT,
  BCFT and DCFT},} \href{http://arxiv.org/abs/2411.03381}{{
  arXiv:2411.03381~[hep-th]}}.

\bibitem{Lieb:1961fr}
E.~H. Lieb, T.~Schultz, and D.~Mattis, {\slshape {Two soluble models of an
  antiferromagnetic chain},}
  \href{http://dx.doi.org/10.1016/0003-4916(61)90115-4}{{\em Annals Phys.}
  {\bfseries 16} (1961) 407--466}.

\bibitem{Schultz:1964fv}
T.~D. Schultz, D.~C. Mattis, and E.~H. Lieb, {\slshape {Two-Dimensional Ising
  Model as a Soluble Problem of Many Fermions},}
  \href{http://dx.doi.org/10.1103/RevModPhys.36.856}{{\em Rev. Mod. Phys.}
  {\bfseries 36} (1964) 856--871}.

\bibitem{Furuya:2015coa}
S.~C. Furuya and M.~Oshikawa, {\slshape {Symmetry Protection of Critical Phases
  and a Global Anomaly in $1+1$ Dimensions},}
  \href{http://dx.doi.org/10.1103/PhysRevLett.118.021601}{{\em Phys. Rev.
  Lett.} {\bfseries 118} (2017) 021601},
  \href{http://arxiv.org/abs/1503.07292}{{
  arXiv:1503.07292~[cond-mat.stat-mech]}}.

\bibitem{Lecheminant:2015iga}
P.~Lecheminant, {\slshape {Massless renormalization group flow in SU(N)$_k$
  perturbed conformal field theory},}
  \href{http://dx.doi.org/10.1016/j.nuclphysb.2015.11.004}{{\em Nucl. Phys. B}
  {\bfseries 901} (2015) 510--525}, \href{http://arxiv.org/abs/1509.01680}{{
  arXiv:1509.01680~[cond-mat.str-el]}}.

\bibitem{Numasawa:2017crf}
T.~Numasawa and S.~Yamaguch, {\slshape {Mixed Global Anomalies and Boundary
  Conformal Field Theories},}
  \href{http://dx.doi.org/10.1007/JHEP11(2018)202}{{\em JHEP} {\bfseries 11}
  (2018) 202}, \href{http://arxiv.org/abs/1712.09361}{{
  arXiv:1712.09361~[hep-th]}}.

\bibitem{Yao:2018kel}
Y.~Yao, C.-T. Hsieh, and M.~Oshikawa, {\slshape {Anomaly matching and
  symmetry-protected critical phases in $SU(N)$ spin systems in 1+1
  dimensions},} \href{http://dx.doi.org/10.1103/PhysRevLett.123.180201}{{\em
  Phys. Rev. Lett.} {\bfseries 123} (2019) 180201},
  \href{http://arxiv.org/abs/1805.06885}{{
  arXiv:1805.06885~[cond-mat.str-el]}}.

\bibitem{Tanizaki:2018xto}
Y.~Tanizaki and T.~Sulejmanpasic, {\slshape {Anomaly and global inconsistency
  matching: $\theta$-angles, $SU(3)/U(1)^2$ nonlinear sigma model, $SU(3)$
  chains and its generalizations},}
  \href{http://dx.doi.org/10.1103/PhysRevB.98.115126}{{\em Phys. Rev. B}
  {\bfseries 98} (2018) 115126}, \href{http://arxiv.org/abs/1805.11423}{{
  arXiv:1805.11423~[cond-mat.str-el]}}.

\bibitem{Wess:1971yu}
J.~Wess and B.~Zumino, {\slshape {Consequences of anomalous Ward identities},}
  \href{http://dx.doi.org/10.1016/0370-2693(71)90582-X}{{\em Phys. Lett. B}
  {\bfseries 37} (1971) 95--97}.

\bibitem{Haldane:1981zza}
F.~D.~M. Haldane, {\slshape {Luttinger liquid theory of one-dimensional quantum
  fluids. I. Properties of the Luttinger model and their extension to the
  general 1D interacting spinless Fermi gas},}
  \href{http://dx.doi.org/10.1088/0022-3719/14/19/010}{{\em J. Phys. C}
  {\bfseries 14} (1981) 2585--2609}.

\bibitem{Haldane:1983ru}
F.~D.~M. Haldane, {\slshape {Nonlinear field theory of large spin Heisenberg
  antiferromagnets. Semiclassically quantized solitons of the one-dimensional
  easy Axis Neel state},}
  \href{http://dx.doi.org/10.1103/PhysRevLett.50.1153}{{\em Phys. Rev. Lett.}
  {\bfseries 50} (1983) 1153--1156}.

\bibitem{Wamer:2019oge}
K.~Wamer, M.~Lajk\'o, F.~Mila, and I.~Affleck, {\slshape {Generalization of the
  Haldane conjecture to SU($n$) chains},}
  \href{http://dx.doi.org/10.1016/j.nuclphysb.2020.114932}{{\em Nucl. Phys. B}
  {\bfseries 952} (2020) 114932}, \href{http://arxiv.org/abs/1910.08196}{{
  arXiv:1910.08196~[cond-mat.str-el]}}.

\bibitem{MOshikawa_1992}
M.~Oshikawa, {\slshape Hidden $z_{2}\times z_{2}$ symmetry in quantum spin
  chains with arbitrary integer spin,}
  \href{https://dx.doi.org/10.1088/0953-8984/4/36/019}{{\em Journal of Physics:
  Condensed Matter} {\bfseries 4} (Sep, 1992) 7469}.

\bibitem{Pollmann:2009ryx}
F.~Pollmann, A.~M. Turner, E.~Berg, and M.~Oshikawa, {\slshape {Entanglement
  spectrum of a topological phase in one dimension},}
  \href{http://dx.doi.org/10.1103/PhysRevB.81.064439}{{\em Phys. Rev. B}
  {\bfseries 81} (2010) 064439}, \href{http://arxiv.org/abs/0910.1811}{{
  arXiv:0910.1811~[cond-mat.str-el]}}.

\bibitem{Pollmann:2009mhk}
F.~Pollmann, E.~Berg, A.~M. Turner, and M.~Oshikawa, {\slshape {Symmetry
  protection of topological phases in one-dimensional quantum spin systems},}
  \href{http://dx.doi.org/10.1103/PhysRevB.85.075125}{{\em Phys. Rev. B}
  {\bfseries 85} (2012) 075125}, \href{http://arxiv.org/abs/0909.4059}{{
  arXiv:0909.4059~[cond-mat.str-el]}}.

\bibitem{Scaffidi:2017ppg}
T.~Scaffidi, D.~E. Parker, and R.~Vasseur, {\slshape {Gapless Symmetry
  Protected Topological Order},}
  \href{http://dx.doi.org/10.1103/PhysRevX.7.041048}{{\em Phys. Rev. X}
  {\bfseries 7} (2017) 041048}, \href{http://arxiv.org/abs/1705.01557}{{
  arXiv:1705.01557~[cond-mat.str-el]}}.

\bibitem{Note6}
We drop space-time coordinates and spatial integrals to simplify the notations.

\bibitem{Moore:1988ss}
G.~W. Moore and N.~Seiberg, {\slshape {Naturality in Conformal Field Theory},}
  \href{http://dx.doi.org/10.1016/0550-3213(89)90511-7}{{\em Nucl. Phys. B}
  {\bfseries 313} (1989) 16--40}.

\bibitem{Moore:1988qv}
G.~W. Moore and N.~Seiberg, {\slshape {Classical and Quantum Conformal Field
  Theory},} \href{http://dx.doi.org/10.1007/BF01238857}{{\em Commun. Math.
  Phys.} {\bfseries 123} (1989) 177}.

\bibitem{Moore:1989vd}
G.~W. Moore and N.~Seiberg, {\slshape {Lectures on RCFT},} in {\em {Strings
  '89, Proceedings of the Trieste Spring School on Superstrings.}}
\newblock World Scientific, 1990.
\newblock \url{{http://www.physics.rutgers.edu/~gmoore/LecturesRCFT.pdf}}.

\bibitem{Fuchs:2002cm}
J.~Fuchs, I.~Runkel, and C.~Schweigert, {\slshape {TFT construction of RCFT
  correlators 1. Partition functions},}
  \href{http://dx.doi.org/10.1016/S0550-3213(02)00744-7}{{\em Nucl. Phys. B}
  {\bfseries 646} (2002) 353--497},
  \href{http://arxiv.org/abs/hep-th/0204148}{{ arXiv:hep-th/0204148}}.

\bibitem{Fateev:1985mm}
V.~A. Fateev and A.~B. Zamolodchikov, {\slshape {Parafermionic Currents in the
  Two-Dimensional Conformal Quantum Field Theory and Selfdual Critical Points
  in Z(n) Invariant Statistical Systems},}
  \href{http://www.jetp.ac.ru/cgi-bin/e/index/e/62/2/p215?a=list}{{\em Sov.
  Phys. JETP} {\bfseries 62} (1985) 215--225}.

\bibitem{Dotsenko:2003zc}
V.~S. Dotsenko, J.~L. Jacobsen, and S.~Raoul, {\slshape {Parafermionic theory
  with the symmetry Z(N), for N odd},}
  \href{http://dx.doi.org/10.1016/S0550-3213(03)00391-2}{{\em Nucl. Phys. B}
  {\bfseries 664} (2003) 477--511},
  \href{http://arxiv.org/abs/hep-th/0303126}{{ arXiv:hep-th/0303126}}.

\bibitem{Dotsenko:2003kg}
V.~S. Dotsenko, J.~L. Jacobsen, and R.~Santachiara, {\slshape {Conformal field
  theories with Z(N) and Lie algebra symmetries},}
  \href{http://dx.doi.org/10.1016/j.physletb.2004.01.033}{{\em Phys. Lett. B}
  {\bfseries 584} (2004) 186--191},
  \href{http://arxiv.org/abs/hep-th/0310102}{{ arXiv:hep-th/0310102}}.

\bibitem{Dotsenko:2003ui}
V.~S. Dotsenko, J.~L. Jacobsen, and R.~Santachiara, {\slshape {Parafermionic
  theory with the symmetry Z(N), for N even},}
  \href{http://dx.doi.org/10.1016/j.nuclphysb.2003.11.019}{{\em Nucl. Phys. B}
  {\bfseries 679} (2004) 464--494},
  \href{http://arxiv.org/abs/hep-th/0310131}{{ arXiv:hep-th/0310131}}.

\bibitem{Runkel:2010ym}
I.~Runkel, {\slshape {Non-local conserved charges from defects in perturbed
  conformal field theory},}
  \href{http://dx.doi.org/10.1088/1751-8113/43/36/365206}{{\em J. Phys. A}
  {\bfseries 43} (2010) 365206}, \href{http://arxiv.org/abs/1004.1909}{{
  arXiv:1004.1909~[hep-th]}}.

\bibitem{Konechny:2025rcc}
A.~Konechny and V.~Vergioglou, {\slshape {On local fields invariant under the
  action of topological defects},}
  \href{http://dx.doi.org/10.1007/JHEP09(2025)114}{{\em JHEP} {\bfseries 09}
  (2025) 114}, \href{http://arxiv.org/abs/2505.04316}{{
  arXiv:2505.04316~[hep-th]}}.

\bibitem{Fuchs:1993et}
J.~Fuchs, {\slshape {Fusion rules in conformal field theory},}
  \href{http://dx.doi.org/10.1002/prop.2190420102}{{\em Fortsch. Phys.}
  {\bfseries 42} (1994) 1--48}, \href{http://arxiv.org/abs/hep-th/9306162}{{
  arXiv:hep-th/9306162}}.

\bibitem{Rida:1999xu}
A.~Rida and T.~Sami, {\slshape {The nonchiral fusion rules in rational
  conformal field theories},}
  \href{http://dx.doi.org/10.1023/A:1014599117130}{{\em Lett. Math. Phys.}
  {\bfseries 58} (2001) 239--248}, \href{http://arxiv.org/abs/hep-th/9907137}{{
  arXiv:hep-th/9907137}}.

\bibitem{Rida:1999ru}
A.~Rida and T.~Sami, {\slshape {Nonchiral fusion rules, structure constants of
  D(m) minimal models},} \href{http://arxiv.org/abs/hep-th/9910070}{{
  arXiv:hep-th/9910070}}.

\bibitem{Nivesvivat:2025odb}
R.~Nivesvivat and S.~Ribault, {\slshape {Fusion rules and structure constants
  of E-series minimal models},}
  \href{http://dx.doi.org/10.21468/SciPostPhys.18.5.163}{{\em SciPost Phys.}
  {\bfseries 18} (2025) 163}, \href{http://arxiv.org/abs/2502.14295}{{
  arXiv:2502.14295~[hep-th]}}.

\bibitem{Note7}
However, this is not a sufficient condition for fixing the preserved sectors.
  However, for our purpose, the detailed data are not necessary.

\bibitem{Anderson:1972pca}
P.~W. Anderson, {\slshape {More Is Different},}
  \href{http://dx.doi.org/10.1126/science.177.4047.393}{{\em Science}
  {\bfseries 177} (1972) 393--396}.

\bibitem{Dorey:2005ak}
P.~Dorey, A.~Lishman, C.~Rim, and R.~Tateo, {\slshape {Reflection factors and
  exact g-functions for purely elastic scattering theories},}
  \href{http://dx.doi.org/10.1016/j.nuclphysb.2006.02.043}{{\em Nucl. Phys. B}
  {\bfseries 744} (2006) 239--276},
  \href{http://arxiv.org/abs/hep-th/0512337}{{ arXiv:hep-th/0512337}}.

\bibitem{Green:2007wr}
D.~R. Green, M.~Mulligan, and D.~Starr, {\slshape {Boundary Entropy Can
  Increase Under Bulk RG Flow},}
  \href{http://dx.doi.org/10.1016/j.nuclphysb.2008.01.010}{{\em Nucl. Phys. B}
  {\bfseries 798} (2008) 491--504}, \href{http://arxiv.org/abs/0710.4348}{{
  arXiv:0710.4348~[hep-th]}}.

\bibitem{Dorey:2009vg}
P.~Dorey, C.~Rim, and R.~Tateo, {\slshape {Exact g-function flow between
  conformal field theories},}
  \href{http://dx.doi.org/10.1016/j.nuclphysb.2010.03.010}{{\em Nucl. Phys. B}
  {\bfseries 834} (2010) 485--501}, \href{http://arxiv.org/abs/0911.4969}{{
  arXiv:0911.4969~[hep-th]}}.

\bibitem{Fredenhagen:2009tn}
S.~Fredenhagen, M.~R. Gaberdiel, and C.~Schmidt-Colinet, {\slshape {Bulk flows
  in Virasoro minimal models with boundaries},}
  \href{http://dx.doi.org/10.1088/1751-8113/42/49/495403}{{\em J. Phys. A}
  {\bfseries 42} (2009) 495403}, \href{http://arxiv.org/abs/0907.2560}{{
  arXiv:0907.2560~[hep-th]}}.

\bibitem{Chang:2018iay}
C.-M. Chang, Y.-H. Lin, S.-H. Shao, Y.~Wang, and X.~Yin, {\slshape {Topological
  Defect Lines and Renormalization Group Flows in Two Dimensions},}
  \href{http://dx.doi.org/10.1007/JHEP01(2019)026}{{\em JHEP} {\bfseries 01}
  (2019) 026}, \href{http://arxiv.org/abs/1802.04445}{{
  arXiv:1802.04445~[hep-th]}}.

\bibitem{Fukusumi:2023vjm}
Y.~Fukusumi, {\slshape {Protected edge modes based on the bulk and boundary
  renormalization group: A relationship between duality and generalized
  symmetry},} \href{http://arxiv.org/abs/2312.12887}{{
  arXiv:2312.12887~[hep-th]}}.

\bibitem{Affleck:1987vf}
I.~Affleck, T.~Kennedy, E.~H. Lieb, and H.~Tasaki, {\slshape {Rigorous Results
  on Valence Bond Ground States in Antiferromagnets},}
  \href{http://dx.doi.org/10.1103/PhysRevLett.59.799}{{\em Phys. Rev. Lett.}
  {\bfseries 59} (1987) 799}.

\bibitem{Verresen:2019igf}
R.~Verresen, R.~Thorngren, N.~G. Jones, and F.~Pollmann, {\slshape {Gapless
  Topological Phases and Symmetry-Enriched Quantum Criticality},}
  \href{http://dx.doi.org/10.1103/PhysRevX.11.041059}{{\em Phys. Rev. X}
  {\bfseries 11} (2021) 041059}, \href{http://arxiv.org/abs/1905.06969}{{
  arXiv:1905.06969~[cond-mat.str-el]}}.

\bibitem{Kitaev:2001kla}
A.~Y. Kitaev, {\slshape {Unpaired Majorana Fermions in Quantum Wires},}
  \href{http://dx.doi.org/10.1070/1063-7869/44/10S/$S^2$9}{{\em Phys. Usp.}
  {\bfseries 44} (2001) 131--136},
  \href{http://arxiv.org/abs/cond-mat/0010440}{{ arXiv:cond-mat/0010440}}.

\bibitem{Fukusumi:2020irh}
Y.~Fukusumi and S.~Iino, {\slshape {Open spin chain realization of a
  topological defect in a one-dimensional Ising model: Boundary and bulk
  symmetry},} \href{http://dx.doi.org/10.1103/PhysRevB.104.125418}{{\em Phys.
  Rev. B} {\bfseries 104} (2021) 125418},
  \href{http://arxiv.org/abs/2004.04415}{{ arXiv:2004.04415~[hep-th]}}.

\bibitem{Smith:2021luc}
P.~Boyle~Smith, {\slshape {Boundary States and Anomalous Symmetries of
  Fermionic Minimal Models},} \href{http://arxiv.org/abs/2102.02203}{{
  arXiv:2102.02203~[hep-th]}}.

\bibitem{Fukusumi:2021zme}
Y.~Fukusumi, Y.~Tachikawa, and Y.~Zheng, {\slshape {Fermionization and boundary
  states in 1+1 dimensions},}
  \href{http://dx.doi.org/10.21468/SciPostPhys.11.4.082}{{\em SciPost Phys.}
  {\bfseries 11} (2021) 082}, \href{http://arxiv.org/abs/2103.00746}{{
  arXiv:2103.00746~[hep-th]}}.

\bibitem{Weizmann}
H.~Ebisu and M.~Watanabe, {\slshape {Fermionization of conformal boundary
  states},} \href{http://dx.doi.org/10.1103/PhysRevB.104.195124}{{\em Phys.
  Rev. B} {\bfseries 104} (2021) 195124},
  \href{http://arxiv.org/abs/2103.01101}{{ arXiv:2103.01101~[hep-th]}}.

\bibitem{Cardy:1986gw}
J.~L. Cardy, {\slshape {Effect of Boundary Conditions on the Operator Content
  of Two-Dimensional Conformally Invariant Theories},}
  \href{http://dx.doi.org/10.1016/0550-3213(86)90596-1}{{\em Nucl. Phys. B}
  {\bfseries 275} (1986) 200--218}.

\bibitem{Okada:2024qmk}
M.~Okada and Y.~Tachikawa, {\slshape {Non-invertible symmetries act locally by
  quantum operations},} \href{http://arxiv.org/abs/2403.20062}{{
  arXiv:2403.20062~[hep-th]}}.

\bibitem{Grover:2013rc}
T.~Grover, D.~N. Sheng, and A.~Vishwanath, {\slshape {Emergent Space-Time
  Supersymmetry at the Boundary of a Topological Phase},}
  \href{http://dx.doi.org/10.1126/science.1248253}{{\em Science} {\bfseries
  344} (2014) 280--283}, \href{http://arxiv.org/abs/1301.7449}{{
  arXiv:1301.7449~[cond-mat.str-el]}}.

\bibitem{Rahmani:2015qpa}
A.~Rahmani, X.~Zhu, M.~Franz, and I.~Affleck, {\slshape {Emergent Supersymmetry
  from Strongly Interacting Majorana Zero Modes},}
  \href{http://dx.doi.org/10.1103/PhysRevLett.115.166401}{{\em Phys. Rev.
  Lett.} {\bfseries 115} (2015) 166401},
  \href{http://arxiv.org/abs/1504.05192}{{
  arXiv:1504.05192~[cond-mat.str-el]}}. [Erratum: Phys.Rev.Lett. 116, 109901
  (2016)].

\bibitem{OBrien:2017wmx}
E.~O'Brien and P.~Fendley, {\slshape {Lattice supersymmetry and order-disorder
  coexistence in the tricritical Ising model},}
  \href{http://dx.doi.org/10.1103/PhysRevLett.120.206403}{{\em Phys. Rev.
  Lett.} {\bfseries 120} (2018) 206403},
  \href{http://arxiv.org/abs/1712.06662}{{
  arXiv:1712.06662~[cond-mat.stat-mech]}}.

\bibitem{etingof2015tensor}
P.~Etingof, S.~Gelaki, D.~Nikshych, and V.~Ostrik, {\em Tensor categories},
  vol.~205.
\newblock American Mathematical Soc., 2015.

\bibitem{Note8}
More precisely, we study the $\protect \mathbb {C}$-linear realization of the
  representation category, but to simplify notation, we do not distinguish the
  Lie group and representation category.

\bibitem{Note9}
The coset decomposition involves more complicated processes coming from the
  identification (or orbifolding) of (non)simple current. We expect this
  difficulty will be resolved by extending (i.e., inverse of orbifolding) the
  underlying theory by the (non)simple current or by replacing the tensor
  product $\otimes $ with the box product $\boxtimes $, but this is a future
  problem. The latter seems a more standard approach in the field, but the
  explicit calculations of the coefficients seem more involved. Hence, we
  explain the strategy of the former direction here. For example, the
  $SU(2)_{1}\otimes SU(2)_{1}$ model has two primary fields with conformal
  dimension $1/4$, but the tensor product $\left (\left (SU(2)_{1}\otimes
  SU(2)_{1}\right )/SU(2)_{2}\right )\otimes SU(2)_{2}=\protect \text {Ising}
  \otimes SU(2)_{2}$ has only $1$ field with conformal dimension $1/4$ in the
  bosonic representation. At the level of the characters, the identification
  makes sense, but there exists a subtlety in identifying quantum states only
  by using the bosonic representations. In the fermionic representation,
  $\protect \text {Majorana} \otimes SU(2)_{2,f}$, has four primary fields with
  conformal dimension $1/4$, and the tensor decomposition of the original model
  $SU(2)_{1}\otimes SU(2)_{1}$ is realizable.

\bibitem{Witten:1983tw}
E.~Witten, {\slshape {Global Aspects of Current Algebra},}
  \href{http://dx.doi.org/10.1016/0550-3213(83)90063-9}{{\em Nucl. Phys. B}
  {\bfseries 223} (1983) 422--432}.

\bibitem{Witten:1983ar}
E.~Witten, {\slshape {Nonabelian Bosonization in Two-Dimensions},}
  \href{http://dx.doi.org/10.1007/BF01215276}{{\em Commun. Math. Phys.}
  {\bfseries 92} (1984) 455--472}.

\bibitem{Bowcock:1988vs}
P.~Bowcock and P.~Goddard, {\slshape {Coset Constructions and Extended
  Conformal Algebras},}
  \href{http://dx.doi.org/10.1016/0550-3213(88)90122-8}{{\em Nucl. Phys. B}
  {\bfseries 305} (1988) 685}.

\bibitem{Sakurai:1960ju}
J.~J. Sakurai, {\slshape {Theory of strong interactions},}
  \href{http://dx.doi.org/10.1016/0003-4916(60)90126-3}{{\em Annals Phys.}
  {\bfseries 11} (1960) 1--48}.

\bibitem{Bando:1984ej}
M.~Bando, T.~Kugo, S.~Uehara, K.~Yamawaki, and T.~Yanagida, {\slshape {Is rho
  Meson a Dynamical Gauge Boson of Hidden Local Symmetry?},}
  \href{http://dx.doi.org/10.1103/PhysRevLett.54.1215}{{\em Phys. Rev. Lett.}
  {\bfseries 54} (1985) 1215}.

\bibitem{Bando:1987br}
M.~Bando, T.~Kugo, and K.~Yamawaki, {\slshape {Nonlinear Realization and Hidden
  Local Symmetries},}
  \href{http://dx.doi.org/10.1016/0370-1573(88)90019-1}{{\em Phys. Rept.}
  {\bfseries 164} (1988) 217--314}.

\bibitem{Moore:1989yh}
G.~W. Moore and N.~Seiberg, {\slshape {Taming the Conformal Zoo},}
  \href{http://dx.doi.org/10.1016/0370-2693(89)90897-6}{{\em Phys. Lett. B}
  {\bfseries 220} (1989) 422--430}.

\bibitem{Gannon:1994km}
T.~Gannon and M.~A. Walton, {\slshape {On the classification of diagonal coset
  modular invariants},} \href{http://dx.doi.org/10.1007/BF02100186}{{\em
  Commun. Math. Phys.} {\bfseries 173} (1995) 175--198},
  \href{http://arxiv.org/abs/hep-th/9407055}{{ arXiv:hep-th/9407055}}.

\bibitem{Ishikawa:2002wx}
H.~Ishikawa and T.~Tani, {\slshape {Novel construction of boundary states in
  coset conformal field theories},}
  \href{http://dx.doi.org/10.1016/S0550-3213(02)01011-8}{{\em Nucl. Phys. B}
  {\bfseries 649} (2003) 205--242},
  \href{http://arxiv.org/abs/hep-th/0207177}{{ arXiv:hep-th/0207177}}.

\bibitem{Note10}
There exist some canonical relations between a dual group $H^{\protect \text
  {c}}$ and antichiral one $\protect \overline {H}$. In this kind of dual
  group, the level of the WZW model takes negative values, and the central
  charge also becomes negative. Because of this negative sign, the theory has a
  connection to the antichiral theory. When focusing on the fusion rules and
  TQFTs, the distinction between $H^{\protect \text {c}}$ and $\protect
  \overline {H}$ is not relevant, but in CFTs it seems necessary to take the
  chiral and antichiral parts more carefully.

\bibitem{Cordova:2023jip}
C.~Cordova and D.~Garc\'\i{}a-Sep\'ulveda, {\slshape {Non-Invertible Anyon
  Condensation and Level-Rank Dualities},}
  \href{http://arxiv.org/abs/2312.16317}{{ arXiv:2312.16317~[hep-th]}}.

\bibitem{Note11}
We have included the information of braiding or conformal spin to the theory,
  we distinguished the UV and IR fusion ring by the prime symbol '.

\bibitem{Kong:2019cuu}
L.~Kong and H.~Zheng, {\slshape {A mathematical theory of gapless edges of 2d
  topological orders. Part II},}
  \href{http://dx.doi.org/10.1016/j.nuclphysb.2021.115384}{{\em Nucl. Phys. B}
  {\bfseries 966} (2021) 115384}, \href{http://arxiv.org/abs/1912.01760}{{
  arXiv:1912.01760~[cond-mat.str-el]}}.

\bibitem{Seo:2026wmq}
D.~Seo, T.~Lee, and G.~Y. Cho, {\slshape {A Unified Categorical Description of
  Quantum Hall Hierarchy and Anyon Superconductivity},}
  \href{http://arxiv.org/abs/2602.03848}{{
  arXiv:2602.03848~[cond-mat.str-el]}}.

\bibitem{Etingof:2009yvg}
P.~Etingof, D.~Nikshych, V.~Ostrik, and w.~a. a. b.~E. Meir, {\slshape {Fusion
  categories and homotopy theory},} \href{http://arxiv.org/abs/0909.3140}{{
  arXiv:0909.3140~[math.QA]}}.

\bibitem{Douglas2013DualizableTC}
C.~L. Douglas, C.~J. Schommer-Pries, and N.~Snyder, {\slshape Dualizable tensor
  categories,} \href{https://api.semanticscholar.org/CorpusID:117517305}{{\em
  Memoirs of the American Mathematical Society} (2013) }.

\bibitem{Douglas2014TheBT}
C.~L. Douglas, C.~J. Schommer-Pries, and N.~Snyder, {\slshape The balanced
  tensor product of module categories,}
  \href{https://api.semanticscholar.org/CorpusID:54602948}{{\em Kyoto Journal
  of Mathematics} (2014) }.

\bibitem{Gannon:2026ttf}
T.~Gannon and B.~C. Rayhaun, {\slshape {Hypergroup Symmetry in Relative Quantum
  Field Theories and Chiral Algebras},}
  \href{http://arxiv.org/abs/2606.05279}{{ arXiv:2606.05279~[hep-th]}}.

\bibitem{Wei:2026fsn}
P.~Wei and Y.~Zheng, {\slshape {Non-invertible Symmetries in Weyl Fermions, and
  Applications to Fermion-Boundary Scattering Problem},}
  \href{http://arxiv.org/abs/2605.19363}{{ arXiv:2605.19363~[hep-th]}}.

\bibitem{Cordova:2025zkz}
C.~Cordova, D.~Garc{\'\i}a-Sep{\'u}lveda, and J.~A. Harvey, {\slshape
  {Generalized Level-Rank Duality, Holomorphic Conformal Field Theory, and
  Non-Invertible Anyon Condensation},} \href{http://arxiv.org/abs/2512.24419}{{
  arXiv:2512.24419~[hep-th]}}.

\bibitem{Bourgine:2024ycr}
J.-E. Bourgine and Y.~Matsuo, {\slshape {Calogero model for the non-Abelian
  quantum Hall effect},}
  \href{http://dx.doi.org/10.1103/PhysRevB.109.155158}{{\em Phys. Rev. B}
  {\bfseries 109} (2024) 155158}, \href{http://arxiv.org/abs/2401.03087}{{
  arXiv:2401.03087~[hep-th]}}.

\bibitem{SCHWARZ1982141}
A.~Schwarz, {\slshape Field theories with no local conservation of the electric
  charge,}
  \href{https://www.sciencedirect.com/science/article/pii/0550321382901900}{{\em
  Nuclear Physics B} {\bfseries 208} (1982) 141--158}.

\bibitem{SCHWARZ1982427}
A.~Schwarz and Y.~Tyupkin, {\slshape Grand unification and mirror particles,}
  \href{https://www.sciencedirect.com/science/article/pii/0550321382902656}{{\em
  Nuclear Physics B} {\bfseries 209} (1982) 427--432}.

\bibitem{Schwinger:1951xk}
J.~S. Schwinger, {\slshape {The Theory of quantized fields. 1.},}
  \href{http://dx.doi.org/10.1103/PhysRev.82.914}{{\em Phys. Rev.} {\bfseries
  82} (1951) 914--927}.

\bibitem{Pauli:1940zz}
W.~Pauli, {\slshape {The Connection Between Spin and Statistics},}
  \href{http://dx.doi.org/10.1103/PhysRev.58.716}{{\em Phys. Rev.} {\bfseries
  58} (1940) 716--722}.

\bibitem{Harvey:2019qzs}
J.~A. Harvey, Y.~Hu, and Y.~Wu, {\slshape {Galois Symmetry Induced by Hecke
  Relations in Rational Conformal Field Theory and Associated Modular Tensor
  Categories},} \href{http://dx.doi.org/10.1088/1751-8121/ab8e03}{{\em J. Phys.
  A} {\bfseries 53} (2020) 334003}, \href{http://arxiv.org/abs/1912.11955}{{
  arXiv:1912.11955~[hep-th]}}.

\bibitem{Note12}
More precisely, to make the wavefunction single-valued for the insertion of the
  quasihole operators, it is necessary to attach some additional flux to the
  theory. However, this procedure can introduce complications involving the
  mock modularity, and this is out of the scope of the present manuscript.

\bibitem{LeClair:2001yp}
A.~LeClair, {\slshape {Chiral stabilization of the renormalization group for
  flavor and color anisotropic current interactions},}
  \href{http://dx.doi.org/10.1016/S0370-2693(01)01089-9}{{\em Phys. Lett. B}
  {\bfseries 519} (2001) 183--187},
  \href{http://arxiv.org/abs/hep-th/0105092}{{ arXiv:hep-th/0105092}}.

\bibitem{Georgiou:2017jfi}
G.~Georgiou and K.~Sfetsos, {\slshape {Integrable flows between exact CFTs},}
  \href{http://dx.doi.org/10.1007/JHEP11(2017)078}{{\em JHEP} {\bfseries 11}
  (2017) 078}, \href{http://arxiv.org/abs/1707.05149}{{
  arXiv:1707.05149~[hep-th]}}.

\bibitem{Sfetsos:2017sep}
K.~Sfetsos and K.~Siampos, {\slshape {Integrable deformations of the $G_{k_1}
  \times G_{k_2}/G_{k_1+k_2}$ coset CFTs},}
  \href{http://dx.doi.org/10.1016/j.nuclphysb.2017.12.011}{{\em Nucl. Phys. B}
  {\bfseries 927} (2018) 124--139}, \href{http://arxiv.org/abs/1710.02515}{{
  arXiv:1710.02515~[hep-th]}}.

\bibitem{Georgiou:2018gpe}
G.~Georgiou and K.~Sfetsos, {\slshape {The most general $\lambda$-deformation
  of CFTs and integrability},}
  \href{http://dx.doi.org/10.1007/JHEP03(2019)094}{{\em JHEP} {\bfseries 03}
  (2019) 094}, \href{http://arxiv.org/abs/1812.04033}{{
  arXiv:1812.04033~[hep-th]}}.

\bibitem{Quella:2006de}
T.~Quella, I.~Runkel, and G.~M.~T. Watts, {\slshape {Reflection and
  transmission for conformal defects},}
  \href{http://dx.doi.org/10.1088/1126-6708/2007/04/095}{{\em JHEP} {\bfseries
  04} (2007) 095}, \href{http://arxiv.org/abs/hep-th/0611296}{{
  arXiv:hep-th/0611296}}.

\bibitem{Kimura:2014hva}
T.~Kimura and M.~Murata, {\slshape {Current Reflection and Transmission at
  Conformal Defects: Applying BCFT to Transport Process},}
  \href{http://dx.doi.org/10.1016/j.nuclphysb.2014.05.026}{{\em Nucl. Phys. B}
  {\bfseries 885} (2014) 266--279}, \href{http://arxiv.org/abs/1402.6705}{{
  arXiv:1402.6705~[hep-th]}}.

\bibitem{Kimura:2015nka}
T.~Kimura and M.~Murata, {\slshape {Transport Process in Multi-Junctions of
  Quantum Systems},} \href{http://dx.doi.org/10.1007/JHEP07(2015)072}{{\em
  JHEP} {\bfseries 07} (2015) 072}, \href{http://arxiv.org/abs/1505.05275}{{
  arXiv:1505.05275~[hep-th]}}.

\bibitem{Affleck:1990by}
I.~Affleck and A.~W.~W. Ludwig, {\slshape {The Kondo effect, conformal field
  theory and fusion rules},}
  \href{http://dx.doi.org/10.1016/0550-3213(91)90109-B}{{\em Nucl. Phys. B}
  {\bfseries 352} (1991) 849--862}.

\bibitem{Kane:1992xse}
C.~L. Kane and M.~P.~A. Fisher, {\slshape {Transport in a one-channel Luttinger
  liquid},} \href{http://dx.doi.org/10.1103/PhysRevLett.68.1220}{{\em Phys.
  Rev. Lett.} {\bfseries 68} (1992) 1220}.

\bibitem{Kane:1992zza}
C.~L. Kane and M.~P.~A. Fisher, {\slshape {Transmission through barriers and
  resonant tunneling in an interacting one-dimensional electron gas},}
  \href{http://dx.doi.org/10.1103/PhysRevB.46.15233}{{\em Phys. Rev. B}
  {\bfseries 46} (1992) 15233--15262}.

\bibitem{Affleck:1995ge}
I.~Affleck, {\slshape {Conformal Field Theory Approach to the Kondo Effect},}
  {\em Acta Phys. Polon. B} {\bfseries 26} (1995) 1869--1932,
  \href{http://arxiv.org/abs/cond-mat/9512099}{{ arXiv:cond-mat/9512099}}.

\bibitem{Fendley_1995}
P.~Fendley, A.~W.~W. Ludwig, and H.~Saleur, {\slshape Exact nonequilibrium
  transport through point contacts in quantum wires and fractional quantum hall
  devices,} \href{https://doi.org/10.1103%2Fphysrevb.52.8934}{{\em Physical
  Review B} {\bfseries 52} (Sep, 1995) 8934--8950}.

\bibitem{PhysRevLett.77.2604}
M.~Oshikawa and I.~Affleck, {\slshape Defect lines in the ising model and
  boundary states on orbifolds,}
  \href{https://link.aps.org/doi/10.1103/PhysRevLett.77.2604}{{\em Phys. Rev.
  Lett.} {\bfseries 77} (Sep, 1996) 2604--2607}.

\bibitem{Saleur:1998hq}
H.~Saleur, {\slshape {Lectures on nonperturbative field theory and quantum
  impurity problems},} \href{http://arxiv.org/abs/cond-mat/9812110}{{
  arXiv:cond-mat/9812110}}.

\bibitem{affleck2009quantumimpurityproblemscondensed}
I.~Affleck, {\slshape Quantum impurity problems in condensed matter physics,}
  2009.
\newblock \url{https://arxiv.org/abs/0809.3474}.

\bibitem{Caldwell:1999ew}
R.~R. Caldwell, {\slshape {A Phantom menace?},}
  \href{http://dx.doi.org/10.1016/S0370-2693(02)02589-3}{{\em Phys. Lett. B}
  {\bfseries 545} (2002) 23--29},
  \href{http://arxiv.org/abs/astro-ph/9908168}{{ arXiv:astro-ph/9908168}}.

\bibitem{HOGAN1995570}
G.~Hogan and C.~Webb, {\slshape Pre-ionization and discharge breakdown in the
  copper vapour laser: the phantom current,}
  \href{https://www.sciencedirect.com/science/article/pii/003040189500143V}{{\em
  Optics Communications} {\bfseries 117} (1995) 570--579}.

\bibitem{Lashkevich:1993fb}
M.~Y. Lashkevich, {\slshape {Coset construction of minimal models},}
  \href{http://dx.doi.org/10.1142/S0217751X93002241}{{\em Int. J. Mod. Phys. A}
  {\bfseries 8} (1993) 5673--5700},
  \href{http://arxiv.org/abs/hep-th/9304116}{{ arXiv:hep-th/9304116}}.

\bibitem{Dotsenko:1984ad}
V.~S. Dotsenko and V.~A. Fateev, {\slshape {Four Point Correlation Functions
  and the Operator Algebra in the Two-Dimensional Conformal Invariant Theories
  with the Central Charge c \ensuremath{<} 1},}
  \href{http://dx.doi.org/10.1016/S0550-3213(85)80004-3}{{\em Nucl. Phys. B}
  {\bfseries 251} (1985) 691--734}.

\bibitem{Dotsenko:1984nm}
V.~S. Dotsenko and V.~A. Fateev, {\slshape {Conformal Algebra and Multipoint
  Correlation Functions in Two-Dimensional Statistical Models},}
  \href{http://dx.doi.org/10.1016/0550-3213(84)90269-4}{{\em Nucl. Phys. B}
  {\bfseries 240} (1984) 312}.

\bibitem{Feigin:1981st}
B.~L. Feigin and D.~B. Fuks, {\slshape {Invariant skew symmetric differential
  operators on the line and verma modules over the Virasoro algebra},}
  \href{http://dx.doi.org/10.1007/BF01081626}{{\em Funct. Anal. Appl.}
  {\bfseries 16} (1982) 114--126}.

\bibitem{Bernard:2014qia}
D.~Bernard, B.~Doyon, and J.~Viti, {\slshape {Non-Equilibrium Conformal Field
  Theories with Impurities},}
  \href{http://dx.doi.org/10.1088/1751-8113/48/5/05FT01}{{\em J. Phys. A}
  {\bfseries 48} (2015) 05FT01}, \href{http://arxiv.org/abs/1411.0470}{{
  arXiv:1411.0470~[math-ph]}}.

\bibitem{Callan:1984sa}
C.~G. Callan, Jr. and J.~A. Harvey, {\slshape {Anomalies and Fermion Zero Modes
  on Strings and Domain Walls},}
  \href{http://dx.doi.org/10.1016/0550-3213(85)90489-4}{{\em Nucl. Phys. B}
  {\bfseries 250} (1985) 427--436}.

\bibitem{Yue:2026qjl}
G.~Yue, A.~Bai, L.~Wu, and T.~Lan, {\slshape {Pro-Tensor Network},}
  \href{http://arxiv.org/abs/2605.06661}{{
  arXiv:2605.06661~[cond-mat.str-el]}}.

\bibitem{Note13}
It should be remarked that the charge preservation is broken in charged domain
  walls, and there exists some subtlety in interpreting the phenomena in RGs.
  However, as a domain wall between TQFTs, it is well-defined.

\bibitem{Zamolodchikov:1986gt}
A.~B. Zamolodchikov, {\slshape {Irreversibility of the Flux of the
  Renormalization Group in a 2D Field Theory},} {\em JETP Lett.} {\bfseries 43}
  (1986) 730--732.

\bibitem{Moudgalya:2021xlu}
S.~Moudgalya, B.~A. Bernevig, and N.~Regnault, {\slshape {Quantum many-body
  scars and Hilbert space fragmentation: a review of exact results},}
  \href{http://dx.doi.org/10.1088/1361-6633/ac73a0}{{\em Rept. Prog. Phys.}
  {\bfseries 85} (2022) 086501}, \href{http://arxiv.org/abs/2109.00548}{{
  arXiv:2109.00548~[cond-mat.str-el]}}.

\bibitem{Matsui_2024}
C.~Matsui, {\slshape Exactly solvable subspaces of nonintegrable spin chains
  with boundaries and quasiparticle interactions,}
  \href{http://dx.doi.org/10.1103/PhysRevB.109.104307}{{\em Physical Review B}
  {\bfseries 109} (Mar., 2024) }.

\bibitem{Kikuchi:2025sso}
K.~Kikuchi, {\slshape {Monotonicities of Tanaka-Nakayama flows},}
  \href{http://arxiv.org/abs/2503.04587}{{ arXiv:2503.04587~[hep-th]}}.

\bibitem{Andrews:1984af}
G.~E. Andrews, R.~J. Baxter, and P.~J. Forrester, {\slshape {Eight vertex SOS
  model and generalized Rogers-Ramanujan type identities},}
  \href{http://dx.doi.org/10.1007/BF01014383}{{\em J. Statist. Phys.}
  {\bfseries 35} (1984) 193--266}.

\bibitem{Huse:1984mn}
D.~A. Huse, {\slshape {Exact exponents for infinitely many new multicritical
  points},} \href{http://dx.doi.org/10.1103/PhysRevB.30.3908}{{\em Phys. Rev.
  B} {\bfseries 30} (1984) 3908--3915}.

\bibitem{Kuniba:1990ci}
A.~Kuniba and T.~Nakanishi, {\slshape {Fusion RSOS models and rational coset
  models},} \href{http://dx.doi.org/10.1007/BFb0101196}{{\em Lect. Notes Math.}
  {\bfseries 1510} (1992) 303--311}.

\bibitem{Chui:2001kw}
C.~H.~O. Chui, C.~Mercat, W.~P. Orrick, and P.~A. Pearce, {\slshape {Integrable
  lattice realizations of conformal twisted boundary conditions},}
  \href{http://dx.doi.org/10.1016/S0370-2693(01)00982-0}{{\em Phys. Lett. B}
  {\bfseries 517} (2001) 429--435},
  \href{http://arxiv.org/abs/hep-th/0106182}{{ arXiv:hep-th/0106182}}.

\bibitem{Chui:2002bp}
C.~H.~O. Chui, C.~Mercat, and P.~A. Pearce, {\slshape {Integrable and conformal
  twisted boundary conditions for sl(2) A-D-E lattice models},}
  \href{http://dx.doi.org/10.1088/0305-4470/36/11/301}{{\em J. Phys. A}
  {\bfseries 36} (2003) 2623--2662},
  \href{http://arxiv.org/abs/hep-th/0210301}{{ arXiv:hep-th/0210301}}.

\bibitem{Aasen:2020jwb}
D.~Aasen, P.~Fendley, and R.~S.~K. Mong, {\slshape {Topological Defects on the
  Lattice: Dualities and Degeneracies},}
  \href{http://arxiv.org/abs/2008.08598}{{
  arXiv:2008.08598~[cond-mat.stat-mech]}}.

\bibitem{Pearce:2025aqh}
P.~A. Pearce, J.~Heymann, and T.~Quella, {\slshape {Unitary and Nonunitary
  A-D-E minimal models: Coset graph fusion algebras, defects, entropies, SREEs
  and dilogarithm identities},} \href{http://arxiv.org/abs/2512.21808}{{
  arXiv:2512.21808~[hep-th]}}.

\bibitem{Cirac:2020obd}
J.~I. Cirac, D.~Perez-Garcia, N.~Schuch, and F.~Verstraete, {\slshape {Matrix
  product states and projected entangled pair states: Concepts, symmetries,
  theorems},} \href{http://dx.doi.org/10.1103/RevModPhys.93.045003}{{\em Rev.
  Mod. Phys.} {\bfseries 93} (2021) 045003},
  \href{http://arxiv.org/abs/2011.12127}{{ arXiv:2011.12127~[quant-ph]}}.

\bibitem{Vancraeynest-DeCuiper:2025msv}
B.~Vancraeynest-De~Cuiper, W.~Wiesiolek, and F.~Verstraete, {\slshape {Les
  Houches Lecture Notes on Tensor Networks},}
  \href{http://arxiv.org/abs/2512.24390}{{
  arXiv:2512.24390~[cond-mat.str-el]}}.

\bibitem{Konechny:2023xvo}
A.~Konechny, {\slshape {RG boundaries and Cardy\textquoteright{}s variational
  ansatz for multiple perturbations},}
  \href{http://dx.doi.org/10.1007/JHEP11(2023)004}{{\em JHEP} {\bfseries 11}
  (2023) 004}, \href{http://arxiv.org/abs/2306.13719}{{
  arXiv:2306.13719~[hep-th]}}.

\bibitem{Delfino:1997ya}
G.~Delfino and G.~Mussardo, {\slshape {Nonintegrable aspects of the
  multifrequency Sine-Gordon model},}
  \href{http://dx.doi.org/10.1016/S0550-3213(98)00063-7}{{\em Nucl. Phys. B}
  {\bfseries 516} (1998) 675--703},
  \href{http://arxiv.org/abs/hep-th/9709028}{{ arXiv:hep-th/9709028}}.

\bibitem{Toth:2004bi}
G.~Z. Toth, {\slshape {A Nonperturbative study of phase transitions in the
  multi-frequency sine-Gordon model},}
  \href{http://dx.doi.org/10.1088/0305-4470/37/41/003}{{\em J. Phys. A}
  {\bfseries 37} (2004) 9631--9650},
  \href{http://arxiv.org/abs/hep-th/0406139}{{ arXiv:hep-th/0406139}}.

\bibitem{Fukusumi:2021qwa}
Y.~Fukusumi and O.~S. Bari\v{s}i\'c, {\slshape {Kubo's response theory and
  bosonization with a background gauge field and irrelevant perturbations},}
  \href{http://dx.doi.org/10.1103/PhysRevB.104.235145}{{\em Phys. Rev. B}
  {\bfseries 104} (2021) 235145}, \href{http://arxiv.org/abs/2106.07339}{{
  arXiv:2106.07339~[cond-mat.stat-mech]}}.

\bibitem{Ando:2026ffy}
T.~Ando and K.~Ohmori, {\slshape {Symmetry Spans and Enforced Gaplessness},}
  \href{http://arxiv.org/abs/2602.11696}{{
  arXiv:2602.11696~[cond-mat.str-el]}}.

\bibitem{Altschuler:1988mg}
D.~Altschuler, {\slshape {Quantum Equivalence of Coset Space Models},}
  \href{http://dx.doi.org/10.1016/0550-3213(89)90320-9}{{\em Nucl. Phys. B}
  {\bfseries 313} (1989) 293}.

\bibitem{Bouwknegt:1992wg}
P.~Bouwknegt and K.~Schoutens, {\slshape {W symmetry in conformal field
  theory},} \href{http://dx.doi.org/10.1016/0370-1573(93)90111-P}{{\em Phys.
  Rept.} {\bfseries 223} (1993) 183--276},
  \href{http://arxiv.org/abs/hep-th/9210010}{{ arXiv:hep-th/9210010}}.

\bibitem{Note14}
We kindly remind again that we studied the dangerously irrelevant perturbation
  in quantum field theory, but it is different from that in statistical
  physics.

\bibitem{Verlinde:1988sn}
E.~P. Verlinde, {\slshape {Fusion Rules and Modular Transformations in 2D
  Conformal Field Theory},}
  \href{http://dx.doi.org/10.1016/0550-3213(88)90603-7}{{\em Nucl. Phys. B}
  {\bfseries 300} (1988) 360--376}.

\end{thebibliography}\endgroup

\end{document}